\documentclass[twocolumn]{emulateapj} 
\newcommand{\ms}{\mbox{m s$^{-1}~$}}
\newcommand{\mse}{\mbox{m s$^{-1}$}}

\newcommand{\msun}{M$_{\odot}~$}
\newcommand{\msune}{M$_{\odot}$}

\newcommand{\lsune}{L$_{\odot}$}

\newcommand{\muas}    {{\mu\mbox{as}}}
\newcommand{\uas}     {{\mu\mbox{as}}}

\newcommand{\muasyr}  {{\mu\mbox{as\,yr}^{-1}}}

\newcommand{\kms}     {{\mbox{km\,s}^{-1}~}}

\newcommand{\kmsMpc}  {{\mbox{km\,s}^{-1}\,\mbox{Mpc}^{-1}}}

\newcommand{\Msun}    {{M_{\odot}}}

 \def\Msun{\hbox{$\dot {M}$}}

 \def\Lsun{\hbox{\it L$_\odot$}}
 
 \def\Msun{\hbox{\it M$_\odot$}}

\def\rel{{\rm rel}}
\def\e{{\rm E}}
\def\au{{\rm AU}}
\def\muas{{\mu\rm as}}
\def\kms{{\rm km}\,{\rm s}^{-1}}

\def\rel{{\rm rel}}

\def\MEarth{M_\oplus}

\def\MJup{M_{\rm Jup}}

\def\gapp{\lower 3pt\hbox{${\buildrel > \over \sim}$}\ }
\def\lapp{\lower 3pt\hbox{${\buildrel < \over \sim}$}\ }
\def\proptosim{\lower 3pt\hbox{${\buildrel \propto \over \sim}$}\ }

\begin{document}

\submitted{Accepted for publication in PASP, January 2008 issue} 
\shorttitle{Precision Astrometry with SIM}
\shortauthors{Unwin {\it et al.}}
\title{Taking the Measure of the Universe:\\Precision Astrometry with SIM PlanetQuest}

\author{Stephen C. Unwin\altaffilmark{1},
Michael Shao\altaffilmark{2},
Angelle M. Tanner\altaffilmark{2},
Ronald J. Allen\altaffilmark{3},
Charles A. Beichman\altaffilmark{4},
David Boboltz\altaffilmark{5},
Joseph H. Catanzarite\altaffilmark{2},
Brian C. Chaboyer\altaffilmark{6},
David R. Ciardi\altaffilmark{4},
Stephen J. Edberg\altaffilmark{2},
Alan L. Fey\altaffilmark{5},
Debra A. Fischer\altaffilmark{7},
Christopher R. Gelino\altaffilmark{8},
Andrew P. Gould\altaffilmark{9},
Carl Grillmair\altaffilmark{8},
Todd J. Henry\altaffilmark{10},
Kathryn V. Johnston\altaffilmark{11,12},
Kenneth J. Johnston\altaffilmark{5},
Dayton L. Jones\altaffilmark{2},
Shrinivas R. Kulkarni\altaffilmark{4},
Nicholas M. Law\altaffilmark{4},
Steven R. Majewski\altaffilmark{13},
Valeri V. Makarov\altaffilmark{2},
Geoffrey W. Marcy\altaffilmark{14},
David L. Meier\altaffilmark{2},
Rob P. Olling\altaffilmark{15},
Xiaopei Pan\altaffilmark{2},
Richard J. Patterson\altaffilmark{13},
Jo Eliza Pitesky\altaffilmark{2},
Andreas Quirrenbach\altaffilmark{16},
Stuart B. Shaklan\altaffilmark{2},
Edward J. Shaya\altaffilmark{15},
Louis E. Strigari\altaffilmark{17},
John A. Tomsick\altaffilmark{18,19},
Ann E. Wehrle\altaffilmark{20},  and
Guy Worthey\altaffilmark{21}
}
\altaffiltext{1}{Jet Propulsion Laboratory, California Institute of Technology, 4800 Oak Grove Drive, Pasadena, CA 91109; stephen.unwin@jpl.nasa.gov}
\altaffiltext{2}{Jet Propulsion Laboratory, California Institute of Technology, 4800 Oak Grove Drive, Pasadena, CA 91109}
\altaffiltext{3}{Space Telescope Science Institute, 3700 San Martin Drive, Baltimore, MD 21218}
\altaffiltext{4}{Michelson Science Center, California Institute of Technology, 770 S. Wilson Ave., Pasadena, CA 91125}
\altaffiltext{5}{United States Naval Observatory, 3450 Massachusetts Avenue NW, Washington DC 20392}
\altaffiltext{6}{Department of Physics and Astronomy, 6127 Wilder Laboratory, Dartmouth College, Hanover, NH 03755}
\altaffiltext{7}{Department of Physics and Astronomy, San Francisco State University San Francisco, CA 95064}
\altaffiltext{8}{Spitzer Science Center, 1200 E. California Blvd., Pasadena, CA 91125}
\altaffiltext{9}{Department of Astronomy, The Ohio State University, 140 W. 18th Avenue, Columbus, OH 43210}
\altaffiltext{10}{Department of Physics and Astronomy, Georgia State University, Atlanta, GA 30303}
\altaffiltext{11}{Van Vleck Observatory, Wesleyan University, Middletown, CT 06459}
\altaffiltext{12}{Columbia University, Pupin Physics Laboratories, 550 West 120th Street, New York, NY 10027}
\altaffiltext{13}{University of Virginia, Department of Astronomy, P.O. Box 400325, Charlottesville, VA 22904-4325}
\altaffiltext{14}{University of California, Berkeley, 417 Campbell Hall, Berkeley, CA 94720}
\altaffiltext{15}{University of Maryland, Astronomy Department, 0227 CSS College Park, MD 20742}
\altaffiltext{16}{Landessternwarte Koenigstuhl 12, 69117 Heidelberg, Germany}
\altaffiltext{17}{University of California at Irvine, Department of Physics and Astronomy, Irvine, CA 92697}
\altaffiltext{18}{University of California at San Diego, Center for Astrophysics and Space Sciences, 9500 Gilman Drive, La Jolla, CA 92093}
\altaffiltext{19}{Space Sciences Laboratory, 7 Gauss Way, University of California, Berkeley, CA 94720-7450}
\altaffiltext{20}{Space Science Institute, 4750 Walnut Street, Suite 205, Boulder, CO 80301}
\altaffiltext{21}{Department of Physics and Astronomy, Washington State University, Department of Physics and Astronomy, Webster Hall 1245, Pullman, WA 99164-2814}

\begin{abstract}

Precision astrometry at microarcsecond accuracy has application to a wide range of astrophysical problems.  
This paper is a study of the science questions
that can be addressed using an instrument with flexible scheduling that delivers parallaxes at about 4 microarcsec ($\muas$)  on targets as faint as $V=20$, and differential accuracy of 0.6 $\muas$ on bright targets.  The science topics are drawn primarily 
from the Team Key Projects, selected in 2000, for the Space Interferometry Mission PlanetQuest (SIM PlanetQuest).  We use the 
capabilities of this mission to illustrate the importance of the next level of astrometric precision in modern astrophysics.
 
SIM PlanetQuest is currently in the detailed design phase, having completed in 2005
all of the enabling technologies needed for the flight instrument.
It will be the first space-based long baseline Michelson interferometer designed 
for precision astrometry.  SIM will contribute strongly
to many astronomical fields including stellar and galactic astrophysics, 
planetary systems around nearby stars, and the study of quasar and AGN nuclei.  Using differential astrometry
SIM will search for planets with masses as small as an Earth orbiting in the `habitable zone' around the nearest stars, 
and could discover many dozen if Earth-like planets are common.  It will characterize the multiple-planet systems that are now known to exist, and it will be 
able to search for terrestrial planets around all of the candidate target stars in the Terrestrial Planet Finder and Darwin mission lists.  It will be capable 
of detecting planets around young stars, thereby providing 
insights into how planetary systems are born and how they evolve 
with time.  Precision astrometry allows the measurement of accurate 
dynamical masses for stars in binary systems.  SIM will observe 
significant numbers of very high- and low-mass stars, providing 
stellar masses to 1\%, the accuracy needed to challenge physical models.  
Using precision proper motion measurements, SIM will probe the Galactic 
mass distribution, and through studies of tidal tails, the formation 
and evolution of the Galactic halo.  SIM will contribute to cosmology through improved accuracy of the Hubble Constant.  With repeated astrometric 
measurements of the nuclei of active galaxies, SIM will probe the 
dynamics of accretion disks around supermassive black holes, and the 
relativistic jets that emerge from them.

\end{abstract}

\vskip 2mm


\section{Introduction \label{CHAPTER1}}

Astrometry is perhaps the most fundamental, and oldest of all areas in astronomy, 
and it remains a cornerstone of the field for the twenty-first century. 
Accurate distances to astronomical objects are essential for deriving fundamental 
quantities like mass and luminosity.
Photographic astrometry in the nineteenth and twentieth centuries laid
the foundation for our understanding of local stellar populations by
identifying the inhabitants of the solar neighborhood \citep{gleise1969,luyten1979}.
The Second US Naval Observatory CCD Astrograph Catalog (UCAC2) is a CCD-based survey covering most of the sky, with accuracies of $15-70$ milliarcsec (mas), depending on brightness \citep{Zacharias2004}, and utilizing the Hipparcos and Tycho-2 reference frame.  Recent CCD based astrometry over 
narrow fields has achieved an accuracy of less than 1 mas in a single measurement \citep{pravdo05}.   On still smaller scales, \citet{lane2004} have demonstrated $\simeq 16$\, microarcsec measurements between the components of a 0.25-arcsecond binary, using the Palomar Testbed Interferometer at $2\, \mu$m.
CCD parallaxes now achieve typical errors of 0.5 mas \citep{Harris2005}.  Wide-angle astrometry using ground-based optical and near-IR interferometers now reaches 20 mas \citep{Hummel1994}.  In the radio range, very long baseline interferometry (VLBI) astrometry of quasars has allowed the creation of a quasi-inertial reference frame, the ICRF \citep[International Celestial Reference Frame;][]{Ma1998} with wide-angle accuracy 0.25 mas.

Space-based astrometry has brought about a renaissance in the field. 
The ESA Hipparcos mission, which operated from 1989-1993, yielded an astrometric 
catalog of 118,000 stars down to 12.5 magnitude, with positional accuracy of 1 mas 
for stars brighter than $V = 11$.   The European Space Agency (ESA) is now developing the Gaia mission as a next generation astrometric survey mission 
\citep{Perryman2001,P2002}, which is expected to develop a catalog of $\sim 10^9$ stars, with accuracy $\simeq 20-25$  microarcsec ($\muas$) for stars brighter than $V=15$.

In this paper, we present an overview of the impact of precision astrometry in many fields of astrophysics.  We use NASA's SIM PlanetQuest mission, hereinafter SIM, as a specific example of a space-based facility instrument for astrometry.  This mission has been under active development since 1996, based on concept studies made several years earlier \citep{Shao1993}.   A major objective of this paper is to show how microarcsecond-level astrometry is a powerful tool for 21-st century astronomy.
There are observing opportunities for new experiments with SIM, and this paper is intended as a resource for astronomers using precision astrometry in their research.
Although presented in the context of the specific capabilities inherent to the SIM design, the topics represent very clearly the impact across many areas of astronomy in which precision astrometry plays a fundamental role.  Most of the science investigations described here are drawn from the Key Projects of the SIM Science 
Team, which was selected via a NASA Announcement of Opportunity 
in 2000 \citep[][and references therein]{U2005}.   The SIM Science Team members are co-authors on this paper.

Recommended by the 1990 NRC Decadal Survey \citep{Bahcall1990}, SIM PlanetQuest entered its Formulation Phase (Phase A) in October 1997 and was approved to enter Phase B in August 2003.  SIM was again endorsed by the 2000 NRC Decadal Survey 
\citep{McKeeTaylor2000} wherein it was assumed that SIM would be completed, making it unnecessary to rank it against new mission recommendations in that report.
Technology development was completed in July 2005 and formally signed off by NASA Headquarters in March 2006 after extensive external independent review. Having completed nearly all of the Formulation Phase (Phase A/B), SIM is ready to enter the Implementation Phase, with mature designs, well understood schedule and cost, and low technical and cost risk.  Unfortunately, there is no official launch date, since budget pressures on NASA's Science Mission Directorate have resulted in NASA delaying the Implementation Phase.

This paper covers the expected science contributions of SIM but does not describe any of the technical details of the instrument or mission.  Brief descriptions of the instrument itself and the supporting technologies may be found in several technical papers \citep{Laskin2006,Marr2006,Shao2006}.  A companion paper (Shao \& Nemati, in preparation) explains the SIM instrument design, operation, performance and calibration in more detail.  The astrometric performance of SIM is based on an hierarchical error budget with more than 1000 terms, and with key sets of parameters verified in a series of testbeds developed during Formulation Phase.  Quoted performance numbers are current best estimates from the error budget and detailed instrument design.

The acronym SIM stands for Space Interferometry Mission.  SIM will be the first space-based Michelson interferometer for astrometry.  The instrument will operate in the 
optical waveband using a 9-m baseline between the apertures.  With a global astrometry accuracy of $3\, \muas$  for stars brighter than $V = 20$, 
it will measure parallaxes and proper motions of stars throughout the Galaxy 
with unprecedented accuracy.  Operating in a narrow-angle mode, it will achieve 
a positional accuracy of $0.6\, \muas$ for a single measurement, equivalent to a 
differential positional accuracy at the end of the nominal 5-year mission of 
$\le 0.1\, \muas$.  This performance is about 1000 times better than existing  capabilities on the ground or in space, and about 100 times better than the upcoming Gaia mission, for differential measurements. Such high accuracy will allow SIM 
to detect and measure unambiguous masses of terrestrial planets around 
stars in our Galactic neighborhood. 

SIM is a targeted mission which measures the astrometric positions of stars, referencing the measurements to a grid of 1302 stars covering the entire sky.  Its scheduling is highly flexible, in both the order of observations, their cadence, and the accuracy of each individual measurement.  This contrasts with the Hipparcos and Gaia missions, which scan the entire sky according to a pre-determined scanning pattern.  Many astrometry experiments can make effective use, or in some cases require, this pointing capability -- for instance, searches for terrestrial planets (especially in multiple planet systems), stellar microlensing events, orbits of eccentric binary systems, and variable targets such as X-ray binaries and active galactic nuclei. Currently, the ICRF, defined 
by the locations of 212 extragalactic radio sources 
\citep{johnston95,Ma1998} with most having errors less than 1 mas, is the standard frame for astrometry.  SIM is expected to yield an optical reference frame at a level of about $3\ \muas$; it will be `tied' to the ICRF by observing a number of radio-loud quasars in common.

This paper is divided into a number of sections, each covering a major area of SIM astrophysics. 
In \S\,\ref{CHAPTER2} we show how SIM can be used to search a large sample of nearby solar-type stars for Earth-like planets orbiting in the ``habitable zone'', and to take a planetary census of an even larger population of stars with a variety of spectral types, ages and multiplicities. Section \ref{CHAPTER3} describes a search for planetary systems around young stars with ages of 1-100 Myr, which will provide knowledge of the evolutionary history of planetary systems.  In $\S$\,\ref{CHAPTER4} we explore how combining SIM with existing datasets extends our sensitivity to long-period planets and allows a very complete picture of planetary systems to be made.
Section \ref{CHAPTER5} shows how microarcsecond astrometry allows us to  make fundamental advances in understanding the stellar mass-luminosity relation.  In 
$\S$\,\ref{CHAPTER6} we show that SIM  contributes to a range of problems dealing with the physics of `exotic' stellar objects such as neutron 
stars and black holes, and stars with circumstellar maser emission.

Sections~\ref{CHAPTER7}, \ref{CHAPTER8}, and \ref{CHAPTER9} cover stellar evolution, Cepheids, and  the luminosity-age
relation in globular clusters and constraints on the ages of clusters and 
the Galaxy.   Section \ref{CHAPTER10} explores the dynamics and evolution of our Galaxy using tidal streams, and  $\S$\,\ref{CHAPTER11} explains how astrometric microlensing provides insight into the mass spectrum of dark bodies in the Galaxy.   
$\S$\,\ref{CHAPTER12} and \ref{CHAPTER13} 
cover astrometric studies of the dynamical properties of our Galaxy, galaxies 
out to 5 Mpc, and the structure and properties of active galactic nuclei.
In $\S$\,\ref{CHAPTER14} we present an application of SIM astrometry to cosmology.
Section \ref{CHAPTER15} describes how SIM will be able to make high dynamic 
range and high angular resolution images using the technique of aperture synthesis.  
In addition to its scientific significance, such data represent a demonstration 
of the future of high-resolution optical/IR imaging in space, similar to the way 
that the NRAO VLA revolutionized ground-based radio imaging in the late 1970s. 
In $\S$\,\ref{CHAPTER16} we discuss a fundamental physics experiment with SIM.

Astrometric measurement techniques which support the science objectives are covered in the Appendices.  Appendix~\ref{APPA} discusses SIM's `narrow-angle' precision 
astrometry mode, which is used for discovering planets around 
other stars and measuring their masses, as well as other science.  In Appendix~\ref{APPB} we show how SIM measurements of `grid stars' are used to construct an astrometric reference frame for wide-angle astrometry, and how these measurements are tied to an inertial reference frame defined by ICRF quasars (Appendix~\ref{APPC}).  Defining a non-rotating frame to high precision is essential for some of the science described in this paper.  Appendix~\ref{APPD} explains the astrophysical criteria used to select suitable stars to serve  as reference objects.  

We conclude the paper (\S\,\ref{CHAPTER17}) with a recap of the major science areas, and an indication of new areas where the implications of microarcsecond-accuracy astrometry have yet to be explored in detail.  This paper may serve as a guide to those interested in using precision astrometry to help further their research interests.  SIM is a facility instrument, and there will exist opportunities for the science community to propose new experiments.  About half of the total observing time is assigned to the Science Team, but the remaining time is open and is not yet allocated.  To assist researchers, \S\,\ref{CHAPTER17} includes a table showing the overall assignment of mission time, including open time for new programs.


\section{The Search for Potentially Habitable Planets \label{CHAPTER2}}

Greek philosophers Epicurus, Metrodorus of Chios, and Lucretius
pondered the possibility that many worlds like Earth existed.
Aristotle and Plato argued that our world was unique. We are now in
a position to resolve this 2400-year debate with NASA space missions
such as Kepler, SIM, and the Terrestrial Planet Finder.  SIM
measures three key characteristics of a planet (1) its mass, (2) the
size and shape of its orbit (semimajor axis and eccentricity), (3)
the inclination of its orbit -- if there are multiple planets, this
will tell if their orbits are co-planar. The planet's mass and orbit
determine whether it can retain an atmosphere, develop a molten core
and protective dynamo-generated magnetic field, and harbor oceans of
liquid water. Such planet characteristics are believed to play vital
roles in the formation and evolution of organic life.

Since 1995, over 200 exoplanets have been discovered, most by using
the Doppler technique to monitor the gravitationally-induced radial
velocity (RV) `wobble' induced by a planet \citep{Marcy_Japan_05}.
Precise knowledge of the orbits and minimum planet masses are given
in the {\it Catalog of Nearby Exoplanets} \citep{Butler06} which
provides the physical and observational properties of known
exoplanets orbiting within 3 AU. Twenty multi-planet systems have
been discovered, spawning theoretical studies of the interactions
between planets, their nascent protoplanetary disks, and other
planets \citep{Ford05, Ida05, Kley05, Alibert05, Chiang02,
Tanaka04}.

Most exoplanets found so far are gas giants or ice giants, with
minimum masses greater than $M_{\rm Nep}$ and Jupiter-like radii
gleaned from transit observations. The planet with lowest minimum
mass found thus far by the Doppler technique has $M \sin i =
5.9\,\MEarth$ and a period, $P = 1.94$ d, orbiting the star, GJ\,876
\citep{Rivera05}. This discovery demonstrated that planet formation
yields masses below $10\,\MEarth$ and motivated questions regarding
the occurrence and properties of rocky planets.  Indeed, the
distribution of masses of the well-characterized exoplanets around
nearby stars (within 200 pc) rises steeply toward lower masses, at
least as fast as d$N$/d$M \propto M^{-1.07}$. While planet
detectability in current RV surveys becomes poor for masses below
the mass of Saturn, the rise toward lower masses and the correlation
of exoplanets with metal abundance suggest that planets grow from
rocky/icy embryos toward larger masses.  Such growth suggests that
rocky planets should be at least as common as the giant planets,
forming from the leftover planetesimals in a protoplanetary disk
\citep{Goldreich04, Kenyon06}. The observed semimajor axes span a
range of 0.02--6.0 AU with a rise observed in ${\rm d} N / {\rm d}
\log a$.  This suggests that planets of terrestrial masses may also
be found at a wide range of orbital distances.

Recent simulations of the giant planet formation process
\citep{Benz2006}
produce large populations of low-mass planets whose growth
was halted before they could become giant planets. These planets
orbit beyond 1 AU of the parent star. Their results indicate that
for every currently known exoplanet, there should be many `failed'
giant planet cores with masses smaller than $5\,\MEarth$. Every
solar-type star may have one or two low-mass planets.

Extrapolation of the RV-discovered distribution to 10 AU and
integration over the entire range of semimajor axes indicates that
at least 10\% of all nearby FGK stars harbor gas giants in the inner
few AU.  Given the wide range of
masses and orbital sizes of known planets, we expect that many rocky
planets will orbit between 0.1 and 2 AU, having masses above
$1\,\MEarth$, just the domain in which SIM is uniquely sensitive.
Note that the RV technique, with a precision of 1 \mse, cannot
detect planets of $1\,\MEarth$ orbiting near 1 AU, as the RV
semi-amplitude will be $\sim$0.1 \mse. Moreover, stellar surface
jitter of 1 \ms makes improvement in the Doppler technique unlikely.
Thus SIM offers a unique opportunity to detect Earth analogs,
planets of one Earth mass in the habitable zones of nearby
Solar-type stars.

\begin{figure*}
\epsscale{1.55}
\plottwo{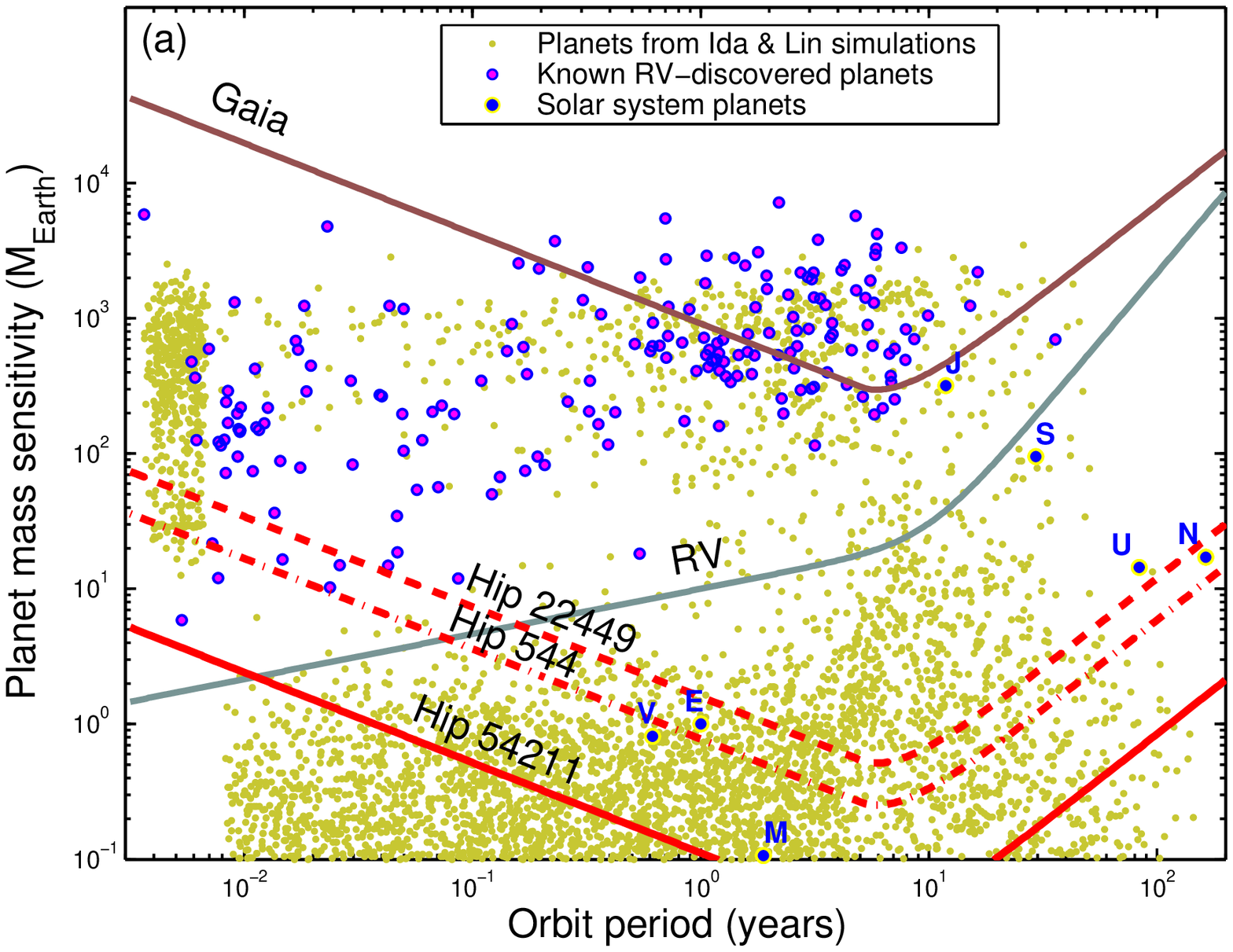}{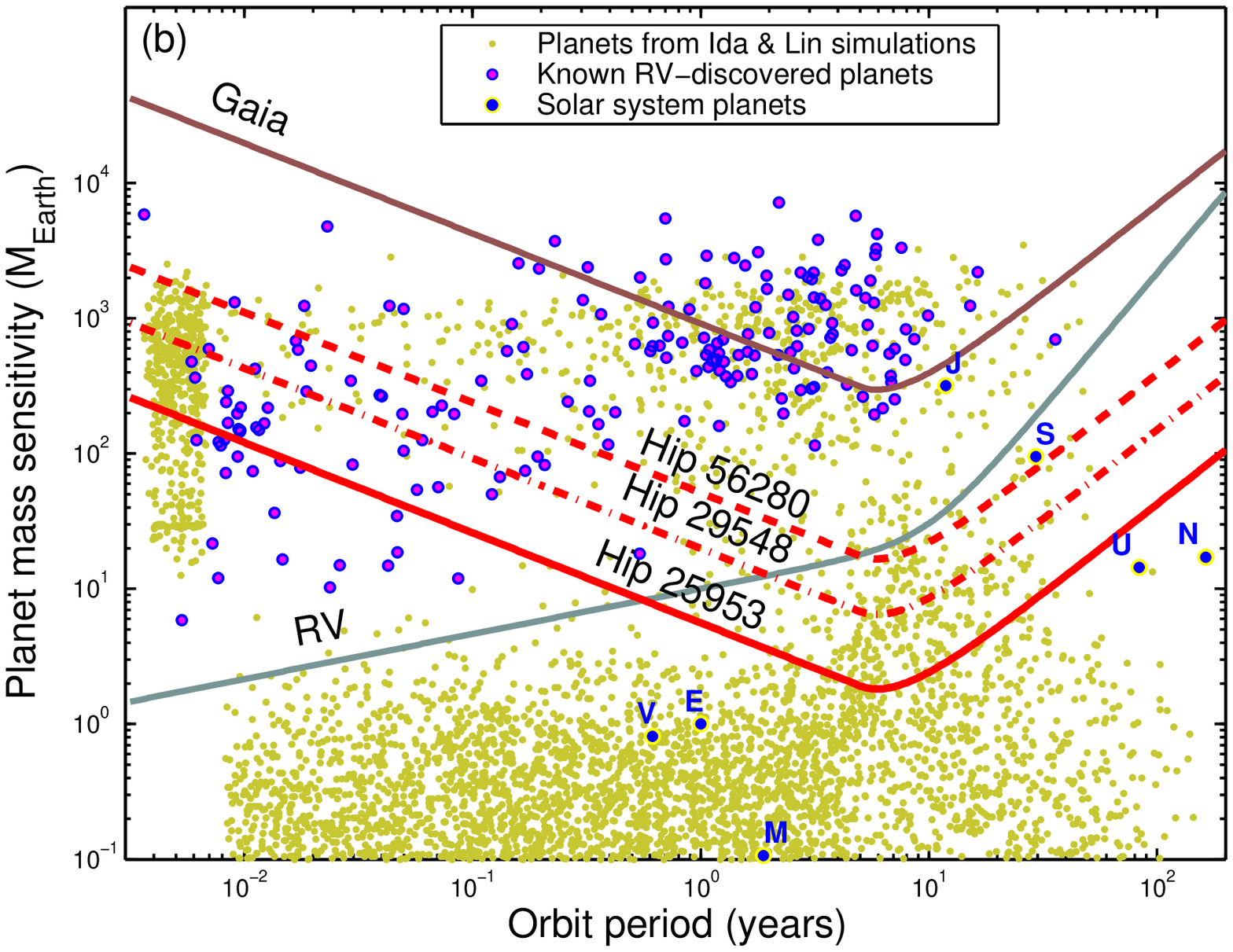}

\caption{(a) Detailed view of the discovery space for rocky Earth-like ($\sim\, 1-10 \MEarth$) planets in the habitable zone ($\sim$ 0.7-1.5 AU for a G star), for the `Earth Analog Survey' of 129 stars described in $\S$~\ref{CHAPTER2}.
The small dots represent a theoretical planet distribution \citep{Ida05} for planets of $0.1-3000 \MEarth$.  In this distribution gas giants have an envelope
mass at least 10 times the mass of the central core; terrestrial planets initially formed within the ice line (2.7 AU for a solar luminosity star);
icy planets formed outside the ice line; and hot Jupiters have periods $\le 0.1$ AU.  Exoplanets discovered as of early 2007 and with semimajor axes $> 0.03$ AU are shown as filled circles.  Planets in our Solar System are labeled with single letters. Labeled curves represent the estimated sensitivity limits of indirect detection methods: for radial velocity method (RV at $1\, \mse$), and astrometry with SIM and Gaia.  
The SIM sensitivity in this space is a broad band, defined by the three lowest curves (labeled with specific Hipparcos star numbers).  The lowest curve shows the `best' star (as computed from star mass and distance); the middle curve represents the median star; and the upper curve shows the least favorable star in the sample.  Also shown is the effective sensitivity of Gaia for stars at 50 pc, a typical distance for Gaia targets.\ \
(b) Detailed view of the discovery space for the SIM `Broad Survey' of 2100 stars, in which a much larger sample of stars is surveyed with less sensitivity than in the `Earth Analog Survey'.  Symbols and curves have the same meaning as in (a).
\label{discovery}}
\end{figure*}

Gaia is expected to detect many exoplanets, but for individual
observations its astrometric precision will be many times lower than
that of SIM, so its main discovery space is that of gas-giant
planets (see Fig.~\ref{discovery}).  Searching the new domain of terrestrial
planets will require the precision and flexibility offered by SIM,
which can select the number and timing of observations, along with
the number of reference stars, allowing a tailored study of each
target.


\subsection{Astrometric Detection of Terrestrial Planets\label{planet-search}}

The angular wobble induced in a star by an orbiting planet is given
by:

\begin{equation}
\alpha = 3.00 \frac{M_{\sun}}{M_{*}} \frac{M_p}{\MEarth}
\frac{a}{1 AU} \frac{1 pc}{D}\  \muas,
\end{equation}

\noindent where $\alpha$ is the angular semi-amplitude of the wobble in $\mu$as,
$M_{\rm p}$ is the mass of the planet, $M_{*}$ is the mass of the star, $a$ is the orbital semimajor axis, and $D$ is the  distance to the system.

SIM's narrow angle observing mode will allow for a single
astrometric measurement precision of 0.6 $\mu$as for stars brighter
than $V = 7$. Narrow-angle astrometry of each target star will be
made relative to at least 3 reference stars selected to evenly
surround the target star within $1\fdg5$. The reference stars are K
giants brighter than $V = 10$, within roughly 600 pc, so that the
astrometric `noise' due to orbiting planets is minimized. Radial
velocity (RV) observations prior to launch will detect brown dwarf
and stellar companions of the reference stars \citep{frink01}. A
ten-chop sequence between a target and a reference star, with 30 sec
integrations per chop, will achieve $0.85\, \muas$ differential
measurement precision for $V=7$ stars, including instrumental and
photon-limited errors; this is more conservative than a scenario individually optimized for each targets, which delivers 0.6 $\muas$.

It is important to clearly define what constitutes astrometric
detection of a planet.  In this paper, for a star of a given mass
and at a given distance, we define the effective mass sensitivity as
the mass of a planet that SIM can detect with false-alarm
probability (FAP) of 1\%, and a detection probability of 50\%.
Effective mass sensitivities can be determined for lists of actual
planet-search target stars using Monte Carlo simulations of detection
of stellar reflex motion due to Keplerian planet orbits. Effective
mass sensitivity provides a good metric of SIM's planet-finding
capability for a given target star, since it depends only on
assumptions about the SIM instrument and the known characteristics
of target stars, without assumptions about the poorly-known
properties of the planets under study, e.g., their mass
distribution, semi-major axis distribution, and frequency of
occurrence.

SIM's capability of detecting planets orbiting in the habitable
zones of nearby stars for several hypothetical planet surveys has
been investigated in detail by \citet{Cat2006}.  In their
simulations, each target was allocated the same amount of observing
time. In this Section we present a different approach, in which each
target star is searched to a specific mass sensitivity. Thus the
simulated observing program computes observing time based on the
mass and distance of each star individually.  This takes advantage
of recent Micro-Arcsecond Metrology (MAM) testbed results at JPL,
indicating that SIM's systematic noise floor is below 0.1 $\muas$
after many repeated measurements, opening up the possibility of
detecting sub-Earth mass planets around the closest stars. The
current best estimate of SIM's single-measurement accuracy is
$0.6\,\muas$.  Both the number of `visits' during the 5-year mission
(nominally $\sim100$ 1-D measurements in each of two orthogonal
baseline orientations) and the spacing of those visits, are
flexibly scheduled, allowing followup of the most interesting
targets during the mission. Approximately 40\% of SIM's five-year
mission time is available for planet searching in narrow-angle mode.

\subsection{The Lowest Mass Planets Detectable by SIM
\label{lowest-mass}}

The threshold planet mass detectable by SIM, for a given orbital
radius around a star of given mass and distance, can be estimated in
several ways.  \citet{Sea2002} used a chi-squared-based test of the
null hypothesis for detection. They derived a detection threshold of
$S = 2.2$, where $S$ is the ``scaled signal'', the ratio of the
angular radius of the astrometric wobble and the astrometric
measurement accuracy. This criterion has been widely quoted in the
planet search community.  But it is overly simplistic to deem a
planet detectable only if the amplitude of the angular wobble is
greater than the astrometric measurement accuracy. Such an estimate
fails to account for both the advantage of large numbers of
observations, $N_{\rm obs}$ and for the temporal coherence of the
orbital position.

In our own Monte Carlo study \citep{Cat2006}, planet detection is
accomplished by using a {\em joint periodogram}, the sum of the
periodogram power in the astrometric measurements along two
orthogonal baseline directions, after fitting out a model of
position offset, proper motion, and parallax.  From the simulations
we derive and validate a more appropriate planet detectability
criterion.

We define SNR as the ratio of the angular wobble to the standard error
of the observation:

\begin{equation}
SNR = \frac{\alpha}{\sqrt{2}\sigma/\sqrt{N_{2D}}}
\end{equation}
where $N_{2D}$ is the number of two-dimensional measurements,
$\sigma$ is the single-measurement accuracy and the factor of
$\sqrt{2}$ is inserted because the measurement is differential. SIM 
measurements are one-dimensional; each target is measured and then
re-observed within a day or so with a baseline orientation that is
quasi-orthogonal to that of the first measurement. So the number of
two-dimension measurements is half the number $N$ of 1-D
measurements. Replacing $N_{2D}$ with $N/2$, we have
\begin{equation}
SNR = \frac{\alpha\sqrt{N}}{2\sigma}\ {\rm\ or\ }\ SNR = 0.5S\sqrt{N},
\end{equation}
in terms of the scaled signal $S$.
It is the SNR rather than ``scaled signal $S$'' which properly
determines detectability; we find that a planet with a SNR of 5.4 is
detectable half the time at the 99\% confidence level. With 200 observations, our detectability criterion is equivalent to S =
0.76, a factor of three below that quoted in \citeauthor{Sea2002}  
With 200 
observations at $0.85\, \muas$ differential accuracy, SIM achieves mass
sensitivities of $\approx 0.2\,\MEarth$ at the mid-habitable zones
of the nearest few targets.

\begin{figure*}[ht!]
\epsscale{0.5}
\plotone{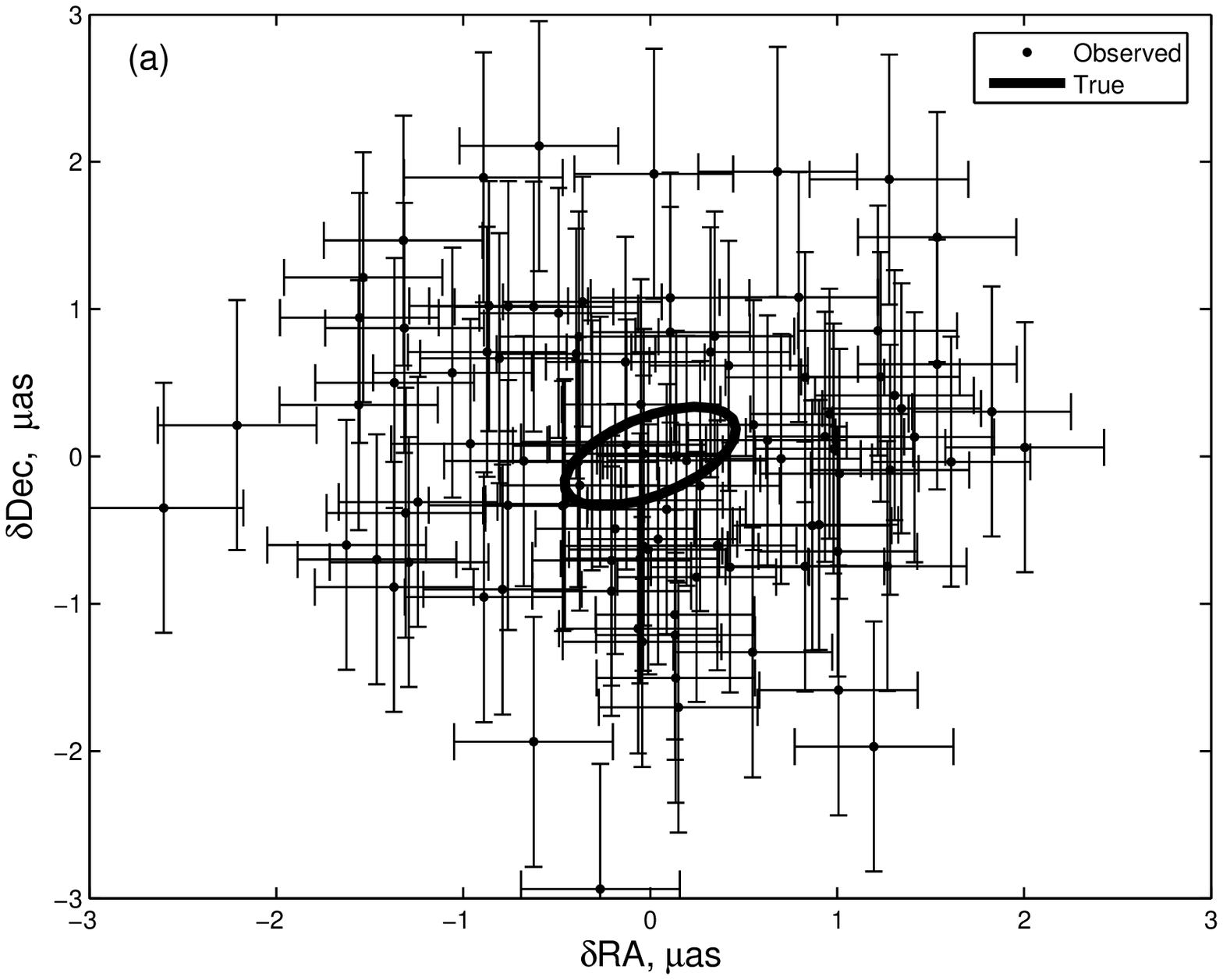}

\noindent
\epsscale{1.0}
\plottwo{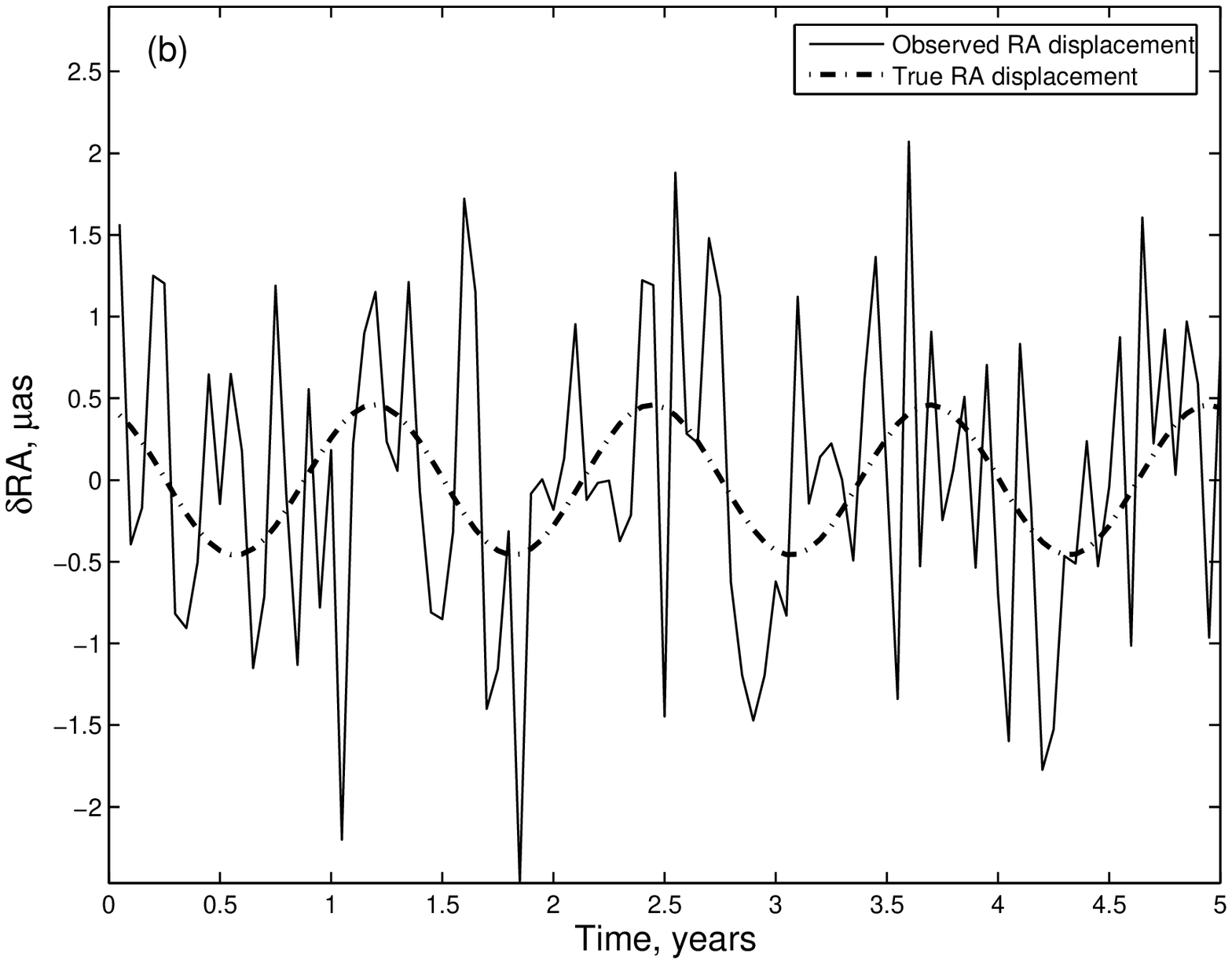}{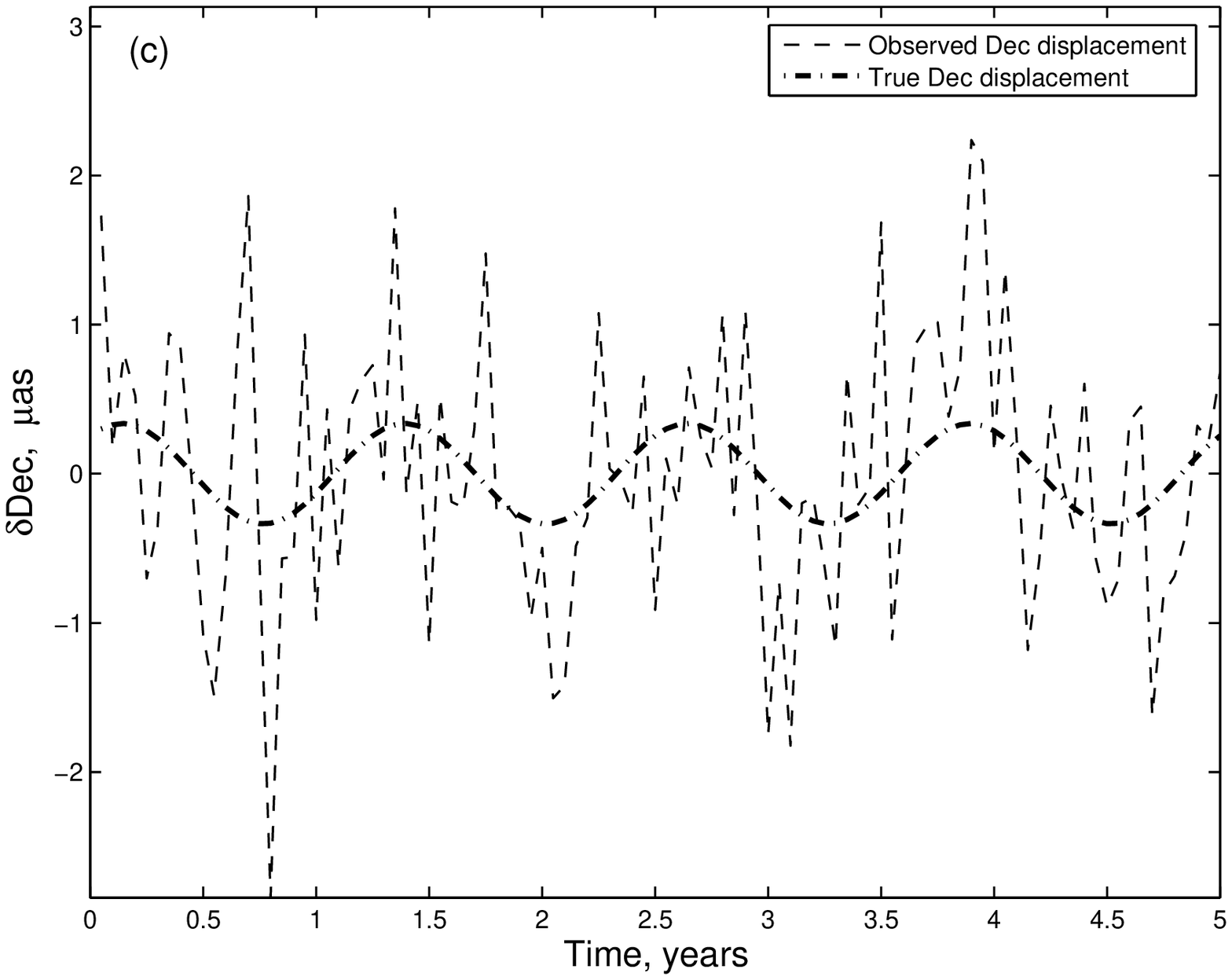}
\plottwo{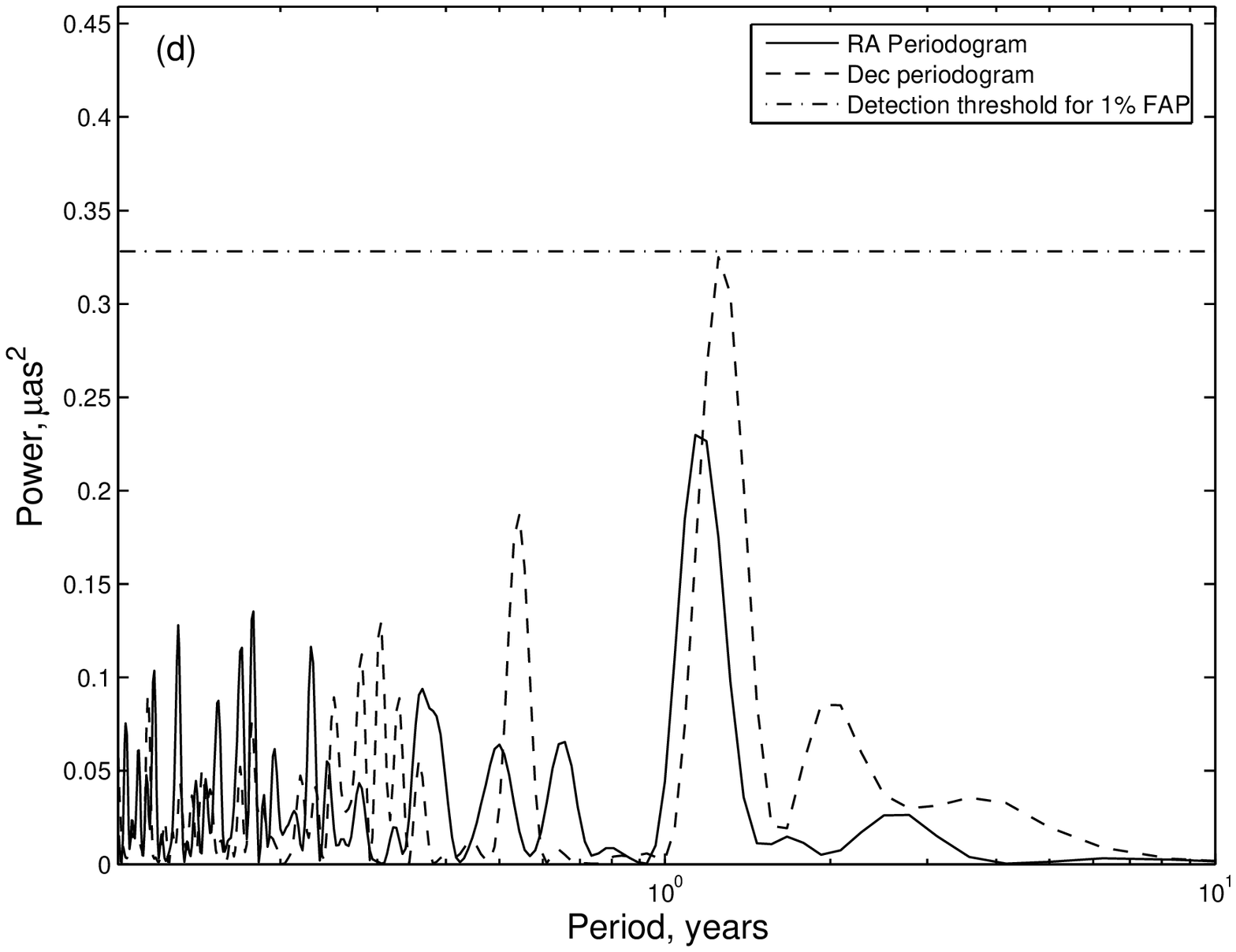}{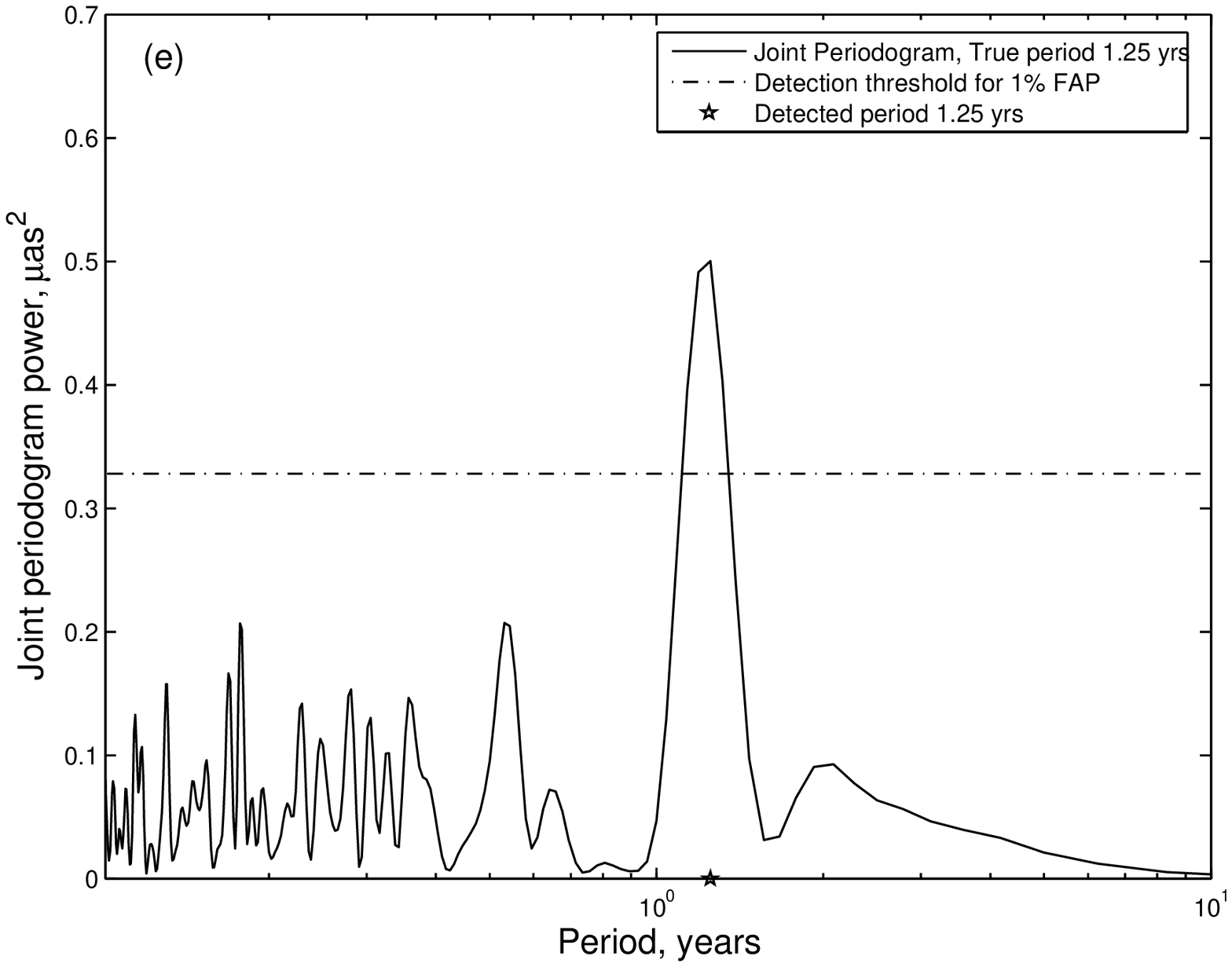}

\caption{Simulation of astrometric detection of a planet with 100 SIM measurements in RA and 100 in Dec, over a 5-year time baseline.   The planet has a mass of 
$1.5\,\MEarth$ and orbits at 1.16 AU from a 1 $M_{\odot}$ star at a distance of 10 pc; it was  chosen to illustrate an orbit close to the limit of detectability.
(a) Sky plot showing the astrometric orbit (solid curve) and the SIM measurements with error bars, for the observing scenario described in \S\,\ref{planet-search}.
(b) and (c), the same data as in (a) but shown as time series along with the astrometric signal projected onto RA and Dec.  (d) Periodograms of the data plotted in (b) and (c). (e) Joint periodogram of data from RA and Dec simultaneously.  Note that the planet is not reliably detected in RA or Dec, but is detected with a false-alarm probability (FAP) of well below 1\% in the joint periodogram.  This illustrates the power of the joint periodogram relative to the $\chi^2$ method which does not use any time information. In this example, the data shown in (a) have (reduced) $\chi^2 = 1.22$, slightly less than the $\chi^2 = 1.25$ required to reject the null hypothesis with $> 99$\% confidence.
\label{astrometry}}
\end{figure*}

To illustrate the detection process we show in Figure
\ref{astrometry} how the joint periodogram, by simultaneously using
data from the two orthogonal baseline directions, is able to
reliably detect a planet in the low signal-to-noise regime where a
simple $\chi^2$ test would reject it.

\subsection{SIM Planet Surveys
\label{planet-surveys}}

Discovery and characterization of many Earthlike planets is one of
SIM's most important scientific objectives.   Up to half of the SIM
observing time will be devoted to three planet surveys, each with
distinct science objectives:

\begin{itemize}

\item A ``Deep Survey'' of up to several hundred stars located within 30 pc, for the lowest-mass planets detectable by SIM.  The Deep Survey is expected to yield a significant number of Earth-like planets.

\item A ``Broad Survey'' of $\sim$\,2100 stars over a variety of spectral types, ages, and multiplicities, for planets with masses of a few $\MEarth$ and greater.  It will explore the
diversity of planetary systems, providing a more complete picture of
planetary systems than is possible with, say, RV or direct imaging
surveys alone.

\item A ``Young Planet Survey'' of $\sim$\,200 stars with ages in the range $1-100$\,Myr.  This survey, when combined with the results of planetary searches of mature stars, will allow us to test theories of planetary formation and early Solar System evolution.
\end{itemize}

In this Section we discuss the objectives of the Deep Survey and Broad Survey.
See \S\,\ref{CHAPTER3} for details of the Young Planet Survey.

The most effective observing strategy  for detecting low-mass planets will depend on the fraction of stars expected to have terrestrial planets ($\eta_\earth$).  We expect results from the Kepler mission (planet detection via transits) to inform that strategy. The basic argument is simple: if Earths
are rare, then SIM should concentrate on a larger sample of hundreds of
stars to get as much information on as many systems as possible.
If on the other hand, terrestrial planets are common, we would like
to probe as many systems as possible for potentially habitable planets.  With
this goal in mind, it makes sense to search each target to the same
mass sensitivity, instead of measuring each target to the same
accuracy, as was adopted in our previous work \citep{Cat2006}.

\begin{deluxetable}{lcc}
\tablecaption{Planet Mass-Limited Surveys with SIM
\label{three-planet-cases}
} 
\tablehead{
\colhead{}& \colhead{Mass sensitivity}& \colhead{Number of stars surveyed}} 
\startdata
Survey 1 & $1.0\ M_{\earth}$           &  129\\
Survey 2 & $2.0\ M_{\earth}$           &  297\\
Survey 3 & $3.0\ M_{\earth}$           &  465\\
\enddata
\tablecomments{In a mass-limited survey, observing time on each star is calculated to yield the given mass sensitivity for a planet at the center of the `habitable zone', computed from the star's distance, estimated mass, and spectral type (see \S\,\ref{planet-surveys}).  Stars are rank-ordered in observing time, and the resulting number of stars which can be surveyed to a given mass sensitivity is shown as a function of planet mass, for three different sensitivity levels.}
\end{deluxetable}

To emphasize this `mass-limited' approach to the search, we term
this survey the ``Earth Analog Survey''.  We allocate to each target
enough integration time to allow a planet of $1\,\MEarth$ to be
detected at the radius of its mid-habitable zone (as determined from
its spectral type). We find that with an assignment of 40\% of a
5-year mission, and a single-measurement accuracy of $0.6\,\muas$,
SIM can probe the mid-habitable zone of 129 stars for $1.0\,\MEarth$
planets (Table~\ref{three-planet-cases}). Essentially, one can
regard this as the survey yield for the idealized case of
delta-function distributions for planet mass and planet orbit radius
($1\,\MEarth$ at 1 AU for a G2V star).  Although unrealistic, this
measure of performance avoids having to make assumptions about the distributions.
(In \S\,\ref{planet-yield} below, we show a second simulation of the planet yield, this time basing it on distributions from \citealt{cumming2007}.)

The discovery space (planet mass vs.~orbit radius) for the ``Earth Analog Survey'' is shown in Figure \ref{discovery}a, with the 129 stars
filling a band in the lower portion of the plot.
Distributing the observing time over a larger target list allows one
to detect more terrestrial planets, albeit at higher masses.
Table~\ref{three-planet-cases} shows the expected SIM yield for
three different values of the search depth. A survey to a
sensitivity of $3 M_{\earth}$ would encompass more nearby stars than
would likely be observed by the Terrestrial Planet Finder (TPF)
Mission.  In each survey, the mass sensitivity improves with orbit
radius, out as far as orbits with periods $\lesssim 5$ yr (see
Fig.~\ref{discovery}a). Note that these surveys are intended to be
illustrative; at the time of SIM launch, the best available data
from all sources will be used to design a survey which might
represent a combination of the approaches in Table
\ref{three-planet-cases}.

In the ``Broad Survey", SIM will probe 2100 stars for planets.  As its name implies, this planetary census includes stars of all spectral types (including
O,B,A and early F, which are not accessible to RV measurements), binary stars, stars with a broad range of age and metallicity, stars with dust disks, evolved stars, white
dwarfs, and stars with planets discovered with RV
surveys.  Each class addresses specific features of the
planet-formation process: {\it Are metals necessary for giant planet
formation?  Does the number of planets decline slowly with time due
to dynamical evolution?  What is the relation between dust disks and
planets?}

Using about 4\% of a 5-year mission, each star will be measured 100 times
at $4\, \muas$ per measurement.  Figure \ref{discovery}b shows the discovery space for the Broad Survey, which is expected to yield a
large sample of hot, cold, rocky, ice giant and gas giant planets,
as well as multiple-planet systems for tests of planet-formation
theories.  Orbit solutions will determine masses and inclinations,
and elucidate planetary system architecture for multiple-planet
systems.

SIM's discoveries will complement future exoplanet missions. SIM
will complete the planetary system architecture for stars with
planets identified by Kepler and COROT. Where Kepler and COROT find
the rocky planets SIM will find the gas giant planets. Furthermore, it will
provide high quality parallaxes (and thus accurate angular diameters) of
stars around which planets have been detected by transits. SIM
resolves the uncertainty in determining planetary orbit radii from
transits. SIM determines the orbit, so that it can provide the
time-dependent location of the planet in the sky, which is critical
for any follow-up program, such as TPF.

\subsection{Expected Planet Yields for SIM Surveys
\label{planet-yield}}

We estimate the likely yield of planets from SIM observations under
plausible assumptions regarding their frequency of occurrence and
distributions as a function of mass and orbit radius.  Tentative
target lists have been selected for the survey of nearby main
sequence stars \citep{Marcy_Japan_05,Shao2006}. Our simulations use
actual star lists, since catalogs of nearby stars are almost
complete, except for late-type stars \citep{DM1991}. Target lists
for the simulations are derived from an initial list of 2350 stars
taken from the Hipparcos catalog, with distances of less than 30 pc
\citep{turnbull2003}.  We excluded stars with luminosity greater
than 25 $L_{\sun}$, thereby eliminating giants from our sample.  To
eliminate the possibility of fringe contamination from a binary
companion, we applied the following selection rules: stars with a
companion closer than 0\farcs4 were excluded; for stars with a
companion that was separated by 0\farcs4 to 1\farcs5, both 
were included as target-star candidates if the magnitude
difference was greater than 1; otherwise, both companions were
excluded. If the target-star candidate had a wide binary companion
that was separated by more than 1\farcs5, the companion was added to
the list of target-star candidates. Surviving candidates were
rank-ordered by effective mass sensitivity.

Although the sensitivity for planet detection at each target is of
primary significance in assessing the capability of any proposed
planet survey, it is also important to understand the yield: how
many planets we do we expect to find and what is their expected mass
distribution?  As discussed in \S\,\ref{planet-surveys}, planet
detection sensitivity may be derived from knowledge of instrument
performance, the target list, the observing scenario, and the
available observation time for each star. To predict the expected survey
yield requires knowledge (or plausible assumptions) of mass and orbit distributions of planets and their occurrence frequency,
for solar-type stars.

Discoveries from the golden age of radial velocity (RV) surveys
\citep{Butler06} have given us robust knowledge of planetary
statistics for orbits out to 3 AU and masses down to a few Saturns.
But the surveys are incomplete for planets on more distant orbits;
and though a handful of planets with masses in the terrestrial range
have been discovered, these are in very close-in orbits. Though RV
is advancing toward detection of Earth mass planets orbiting
M-stars, terrestrial planets in the habitable zones of solar-type
stars will remain beyond its capability, except possibly for a
handful of nearby stars with extremely low variability such as Alpha
Cen\,B. On the other hand, information on orbital and mass
distributions and occurrence frequency of terrestrial planets around
Sun-like stars will be forthcoming from the Kepler mission in a few
years; and COROT will very soon yield statistics of Neptune-class
planets.

At the present time we can only estimate planetary statistics in the
terrestrial mass regime by extrapolation from observational results
and the expectations of planet-formation theorists.  To this end we
created a simple hybrid model based on the power-law mass and period
distributions derived from the RV observational data, obtained from
surveys of solar-type stars \citep{cumming2007}.  For simplicity, in
our model we assume that each star has a maximum of one planet.  We
extrapolated these power laws to orbits out to 10 AU, and to masses
down to the terrestrial mass regime. To account for the prevalence
of `failed cores' expected by many theorists \citep[e.g.,][]{Ida05}, 
we increased the
occurrence frequency of terrestrial planets by a factor of five.  The model distributions are depicted graphically in Figure
\ref{discovery}.  Some recent studies suggest that the planet
occurrence rate is lower in low-mass stars
\citep{butler04,Butler06,endl06,johnson07}, and \citet{gould06}
deduce that about a third of low-mass stars may have cold Neptunes,
whereas extrapolation from \citet{cumming2007} indicates that only
5\% of solar-type stars have Neptunes at all separations.

Our hybrid power-law model has the following properties for
solar-type stars: 73\% of stars have terrestrial mass planets ($0.3
- 10 \MEarth$); $\simeq\,10$\% of stars have terrestrial-mass
planets in the habitable zone ($0.7 - 1.5$ AU); 5\% of stars have
Neptune-class planets ($10 \MEarth$ to $0.1 \MJup$); and 16\% of
stars have Jupiter-class planets ($0.1 - 10\MJup$). According to
this model, the overall occurrence frequency of planets is 95\%. Our
predictions of Neptunes and Jupiters are probably close to reality,
since they involve little extrapolation from observational data, but
the terrestrial mass planet prediction is sensitive to our
extrapolation.

Using this hybrid power-law model, we estimate planet yields for the
`Earth Analog' and `Broad surveys' via Monte Carlo simulation.  For
each survey target star, we generate 1000 planets, with masses and
periods drawn randomly from the model described above. For each
planet we generate a circular orbit, with parameters other than
mass and period randomized.  We calculate the reflex motion
trajectory of each target star due to its planet, and sample it 100
times uniformly over a time baseline of 5 years.  This results in a
time series of 100 pairs of simulated RA and Dec true star
positions.  This database of planets and orbits is then stored away.
Next we create 1000 `sky realizations'; each realization results
from assigning to 95\% of the target stars a planet randomly drawn
from the database; according to the statistics of our model, 5\% of
the targets have no planet. Finally, we generate a simulated survey
for each sky realization by perturbing each
stellar reflex motion trajectory with parallax, and
single-measurement error of $0.6 \muas$.  We pre-process the
simulated observational data by fitting out a model of position,
parallax and proper motion, running the fit residuals through the
joint periodogram (see \S\,\ref{lowest-mass})  with the detection
threshold set to allow only a 1\% chance of false detections. Each
simulated survey therefore has a set of `input' planets, and for
each, a subset of those are detected with SIM.  The most useful
representation of the results are histograms of the ensemble
averages of input and detected masses for the simulated surveys and
planets.

\begin{figure}[ht!]
\epsscale{1.15}
\plotone{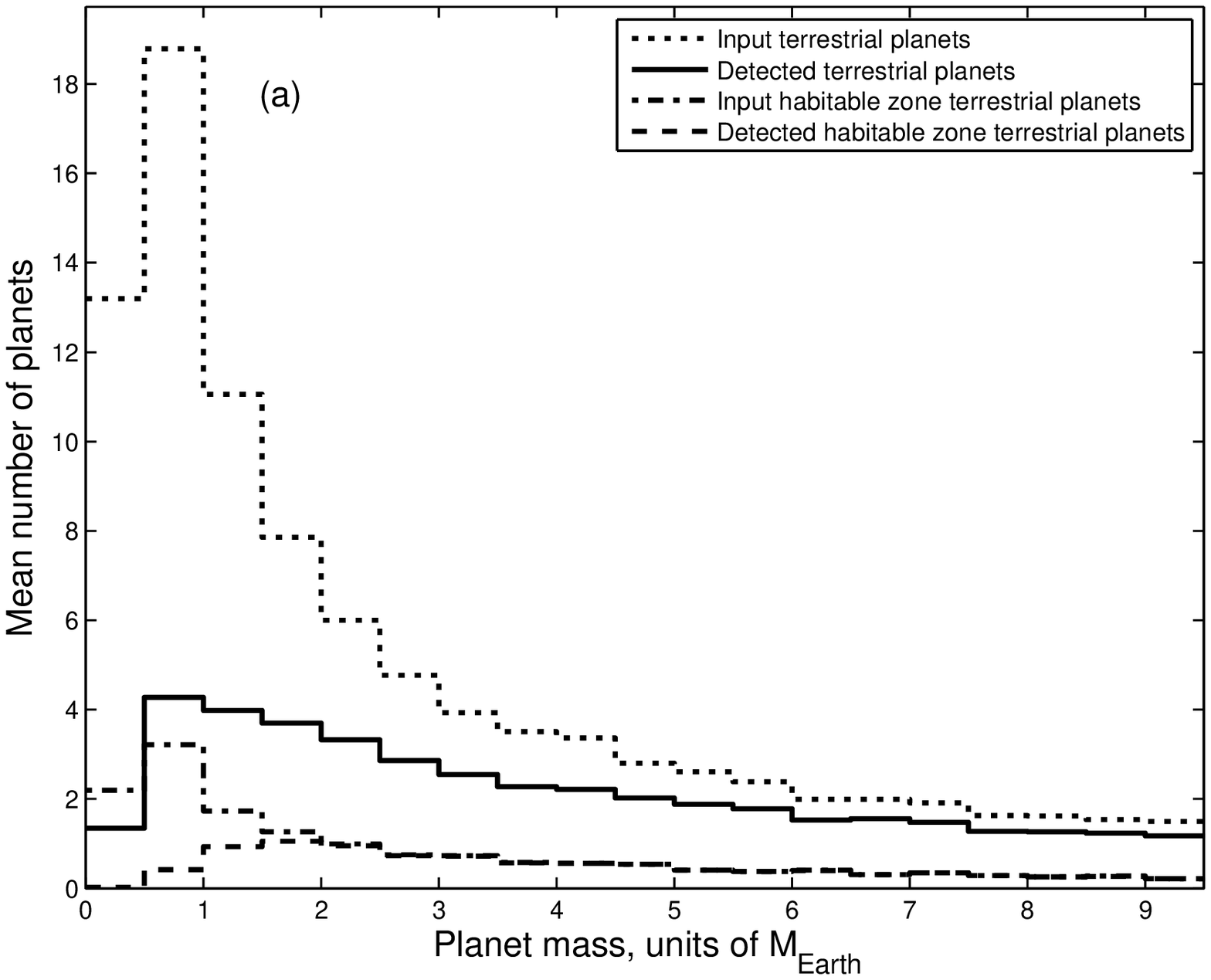}
\vskip 3mm
\epsscale{1.15}
\plotone{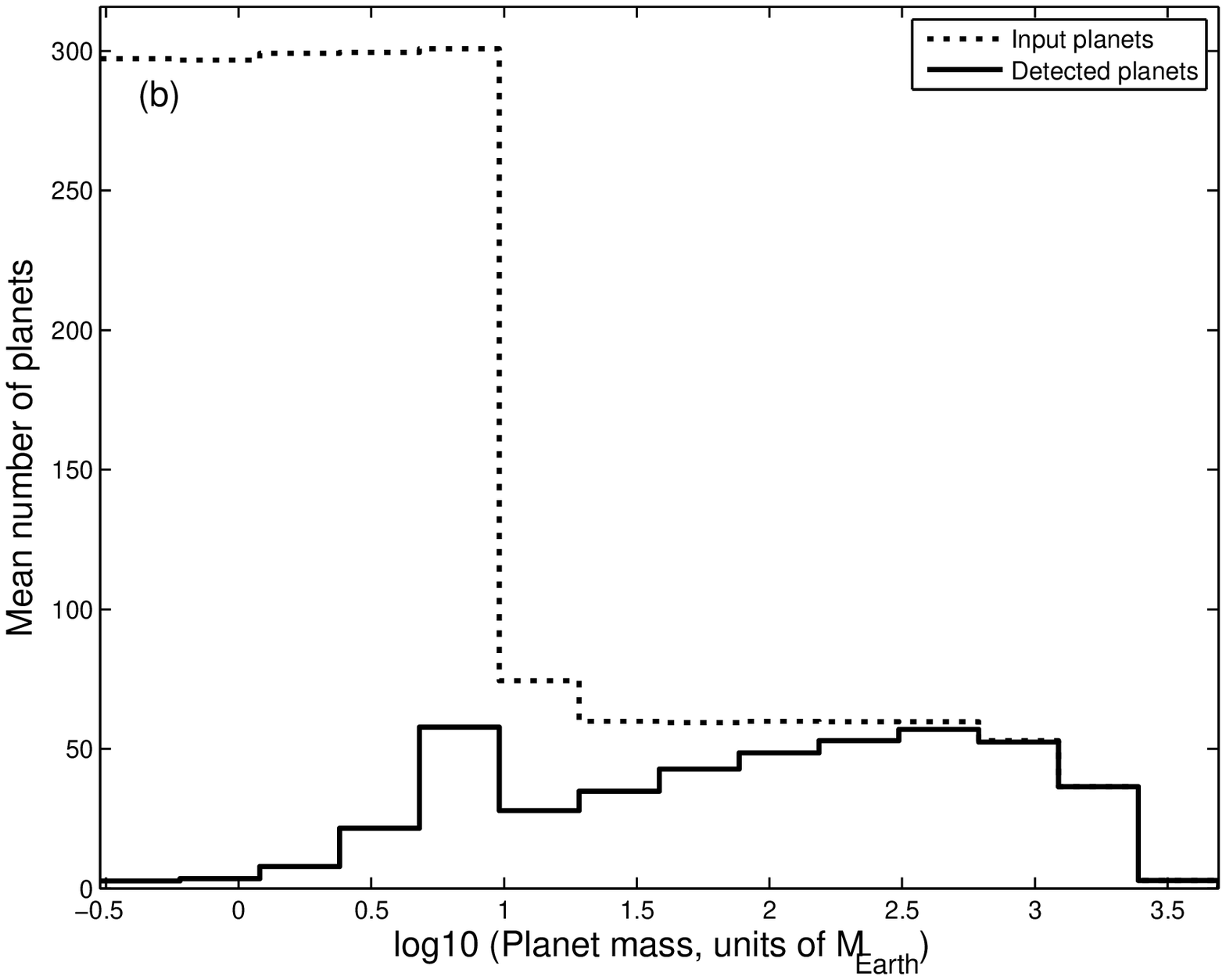}
\caption{(a) Histogram of the expected yield of terrestrial planets for the SIM `Earth Analog Survey' observing program and assumed planetary orbit and mass distributions, and a normalized planet occurrence rate, described in \S\,\ref{planet-yield}.  This histogram is a mean of 1000 simulated surveys in which geometric parameters of the model orbits were randomized.  The mean input distribution of terrestrial ($M < 10 \MEarth$) planets is shown as the dotted curve, 
and the mean number of terrestrial planets detected in the survey by SIM as the solid curve.  For terrestrial  planets in the habitable zone only, the corresponding curves are shown as dash-dotted and dashed respectively.  The integral planet counts are summarized in Table \ref{mass-sens-table}.
Note that the yield of planets, especially at the low-mass end, depends sensitively on both the assumed planetary model parameters and the observing strategy.  (b) The same as in (a) but for the 2100 star `Broad Survey', which includes stars spanning the entire main sequence.  This survey would discover planets over a wide range of masses and orbit radii which are largely unexplored by other detection methods. 
The integral planet counts are summarized in Table \ref{mass-sens-broad-table}.   
\label{planet-histogram}}
\end{figure}

Figure \ref{planet-histogram}a shows the expected histogram for
input vs detected terrestrial planet masses in the `Earth Analog'
survey. The histogram shows fractional counts because it is a mean
over 1000 simulated surveys.  Results for the complete range of
planet masses are shown in Table \ref{mass-sens-table}. In the
habitable zone, SIM would detect 61\% of all the terrestrial
planets, including almost half all planets with masses in the range
$1 - 1.5\MEarth$, and nearly every planet of higher mass.

We repeated the methodology described above, with the same hybrid
power-law distributions for the input planets, on the SIM `Broad
Survey' of 2100 stars. In Figure \ref{planet-histogram}b we show the
mean of the mass histograms (logarithmic mass bins extending over
entire mass range) for 1000 simulated surveys. Table
\ref{mass-sens-broad-table} shows that we expect SIM to find 7\% of
the terrestrial planets, 2\% of all terrestrial planets in the
habitable zone, 47\% of the Neptune-class planets and 87\% of the
Jupiter-class planets.

\begin{deluxetable*}{lccc}
\tablecaption{Expected yield of SIM Earth-Analog Survey of 129
stars
\label{mass-sens-table}
} 
\tablewidth{0pt} 
\tablehead{
\colhead{Planet type}& \colhead{Number} & \colhead{Total Number} &
\colhead{Fraction}\\
\colhead{} & \colhead{Detected} & \colhead{in Sample} & \colhead{Detected}
}
\startdata
Terrestrial, $0.3$  to $ 10 M_{\earth}$           &  43.0 ${\pm}$ 5.0  &   94.0 ${\pm}$  5.1 &  0.46\\
Terrestrial (habitable zone)                      &   9.6 ${\pm}$ 2.9  &   15.7 ${\pm}$  3.7 &  0.61\\
Ice giant, $10 M_{\earth}$  to $ 0.1 \MJup$ &   5.3 ${\pm}$ 2.2  &    6.1 ${\pm}$  2.4 &  0.87\\
Gas giant, $0.1$ to $10 \MJup$              &   21.1 ${\pm}$ 4.2  &   21.3 ${\pm}$  4.3 &  0.99\\
\enddata
\tablecomments{Based on 1000 Monte Carlo survey
realizations, assuming the distribution of planets from the hybrid
model discussed in \S\,\ref{planet-yield}.}
\end{deluxetable*}

It is important to realize that the planet yields predicted by these
simulations depend on many parameters, e.g., the SIM single
measurement accuracy, the observing scenario and time devoted to
each target, the mass and orbit radius distributions of the planets,
and of course, the frequency of occurrence of those planets.  In
particular, we note that the fractions of habitable-zone terrestrial
planets which are input to the simulations are different in Tables
\ref{mass-sens-table} and \ref{mass-sens-broad-table}, due to
different characteristics of the survey stars. The `Broad Survey'
target list includes a larger number of low-mass stars; about half
have masses $< 0.5\Msun$.  Though our hybrid power-law model is
derived from observations of solar-type stars, we have assumed that
it also applies to low-mass stars. One feature of the model is that
there is a decrease in the number of planets per dex as the orbit
radius becomes smaller than about 0.7 AU. Since the habitable zones
of low-mass stars are entirely within 0.7 AU, these stars will
accordingly have fewer habitable zone planets than solar-type stars,
and this is reflected in the tables.

\begin{deluxetable*}{lccc}
\tablecaption{Expected yield of SIM Broad Survey of 2100 stars
\label{mass-sens-broad-table}
}
\tablewidth{0pt} 
\tablehead{
\colhead{Planet type}& \colhead{Number} & \colhead{Total Number} &
\colhead{Fraction}\\
\colhead{} & \colhead{Detected} & \colhead{in Sample} & \colhead{Detected}
}
\startdata
Terrestrial, $0.3$  to $ 10 M_{\earth}$           & 98.3  ${\pm}$ 9.5  & 1511.6 ${\pm}$ 20.4  &  0.07\\
Terrestrial (habitable zone)                      & 2.4   ${\pm}$ 1.6  & 154.2 ${\pm}$ 12.0  &  0.02\\
Ice giant, $10 M_{\earth}$  to $ 0.1 \MJup$ & 47.1  ${\pm}$ 6.6  & 99.8  ${\pm}$ 9.6   &  0.47\\
Gas giant, $0.1$ to $10 \MJup$              &303.5  ${\pm}$ 16.6 & 347.4 ${\pm}$ 17.4  &  0.87\\
\enddata
\tablecomments{Based on 1000 Monte Carlo survey
realizations, assuming the distribution of planets from the hybrid
model discussed in \S\,\ref{planet-yield}.}
\end{deluxetable*}

To summarize, SIM will be capable of detecting a significant
fraction of the expected population of planets
for a large sample of
stars within 30 pc.   As the first planned instrument capable of
detecting terrestrial planets around nearby stars, the planet yield
from SIM will in fact test the degree to which the above model
assumptions are valid.  SIM's scientific discoveries will likely
reveal the erstwhile hidden regime of rocky planets, and make
possible the first thorough checks of the predictions of current
theories of planet formation.

\subsection{Physical Parameters of Habitable Planets}

SIM provides a wealth of planetary astrophysics, including the
masses, orbital radii, and orbital eccentricities of rocky planets
around the nearest stars.  It will also find correlations between
rocky planets and stellar properties such as metallicity and rotation.

SIM and TPF/Darwin together, along with Kepler, provide a valuable
combination of information about rocky planets.
Each mission brings results that illuminate a different portion of the multidimensional space that represents the field of exoplanet research.
Kepler offers the occurrence rate of small
planets.  SIM provides the masses and orbits of planets around
nearby stars, identifying the candidate Earths. TPF/Darwin measures
radii, chemical composition, and atmospheres. In some cases, images
from TPF/Darwin may provide feedback which allows re-analysis of old SIM data,
helping orbit determination, especially for multiple planet
systems.

Imaging surveys (such as TPF and Darwin) require lists of
target stars for observation, ideally those for which rocky planets have been detected.  Assuming that the fraction of stars with Earths in the habitable
zone, $\eta_{\earth}$, is 0.1, SIM will produce a list of target stars for TPF
enriched by a factor of at least 2 in rocky planets between
0.5 and 2.0 AU relative to a TPF-only sample.   For many of
these stars, SIM's orbital solution will be precise enough to predict
the best timing for a direct observation. This information is
crucial for direct imaging, since a planet in the habitable zone can
spend much of its time hidden in the glare of the parent star.
Indeed, habitable rocky planets detected by SIM will likely reside
at angular separations of at least 100 mas from the host star.
Such tantalizing rocky planets will become high priority targets for
those instruments, both on the ground and in space, that can perform
high contrast imaging.  With sufficiently long integration times and
on-band, off-band filters, early imaging of Earthlike planets
around the very nearest stars may be achieved in advance of TPF and
Darwin. SIM also identifies those stars that TPF and Darwin should
avoid, notably those with large planets near the habitable zone that
render any Earths dynamically unstable.  Of course, SIM also detects
those Saturn or Neptune-mass planets located at 2 AU, valuable in
themselves for planetary astrophysics.

As a benchmark, one may assume that at least 10\% of stars have a
rocky planet between 0.5 and 2.0 AU. If so, Kepler will find them in
its transit survey of stars 400-1000 pc away; and SIM is likely to
find the first rocky planets orbiting in the habitable zones of
Sun-like stars closer than 30 pc. Although no rocky planets will be
detected in common between the two missions, SIM could detect gas
giants orbiting Kepler target stars for which rocky planets have
been detected via transits.  (Multiple planet detections by Kepler will likely
be rare due to the very stringent coplanarity requirement).

Detections of rocky planets will spawn
theoretical work about geophysically plausible interior structures
for such planets. SIM measures planet masses, which is {\em the}
basic physical parameter for any planet. Imaging of Earth-mass planets
around all stars within 30 pc remains beyond current technical
ground-based capabilities as such planets are 10$^{10}$ times
fainter than the host star and will be 28th magnitude, comparable to
the background patchwork of high redshift galaxies. Adding to the
challenge, planets in somewhat edge-on orbital planes will spend a
significant fraction of the time located within the
diffraction-limited angle of the host star.  The next generation of space-based telescopes, represented by TPF and Darwin, will have a rich discovery space to explore.  SIM will pave the way by conducting an inventory of rocky planets around nearby stars.

\subsection{The Impact of Starspots on Astrometric Planet Detection}

Stellar variability manifests itself in different ways in photometric,
astrometric, and radial velocity data.  In this subsection we estimate the expected
astrometric centroid jitter due to variability of the
planet-search target stars, and assess the impact on astrometric
planet detection.

The 30 year record of satellite observations of the Sun's
photometric variability shows an RMS of 0.042\%
\citep{Frohlich2006}. Variations on timescales of days to decades
can be attributed to the evolution and rotational modulation of
magnetic surface phenomena, e.g., sunspots and faculae
\citep{Wenzler2005}. In general, photometric variability in a star
due to starspots introduces noise in measurements of both its
photocenter and radial velocity. This noise, in turn, imposes limits
on the mass of a planet detectable by these types of measurements.

To investigate the size of the effect, we developed a simple dynamic sunspot
model that accurately captures the known behavior of Sun's photometric
variations in both time and frequency domains (Catanzarite, Shao \& Law, in preparation).  Starspot noise has a `red' power spectral density (PSD), showing strong variation with frequency, and our model takes account of this. The important frequencies are those associated with the duration of a
measurement (about an hour), an observing campaign (up to a few years) and
with the orbital period of the planet one is trying to detect.

We used our dynamic sunspot model to characterize the jitter in the radial velocity and in the astrometric centroid.  For the Sun, we find typical RMS jitter of $7 \times 10^{-7}$\,AU in the astrometric centroid, and 0.3 \ms in the radial velocity.  Because of the shape of the PSD, a simple RMS does not adequately represent the noise contribution to planet detection.  To gauge the impact on planet detection, we sampled the centroid and the RV signal from our sunspot model once every 11 days (100 epochs)
over three years.  From the PSD of the
resulting time series, we found that the noise level in the centroid
due to starspots is $4 \times 10^{-7}$\,AU for orbit periods longer than 0.6
years,  equivalent to the astrometric signal (at 10 pc) of a $0.1 \MEarth$ planet in a 1 AU orbit, and well below the sensitivity of SIM at this distance.

This level of centroid jitter translates (at 10 pc) to an astrometric noise of
$0.04\,\muas$, substantially below SIM's noise floor of $0.085\,\muas$ achieved with 100 observations with a differential accuracy of $0.85\,\muas$  (see \S\,\ref{planet-search}).  We therefore conclude that if the Sun were at 10 pc, starspot noise would not impact the astrometric detection of terrestrial planets with orbit periods longer than 0.6 years.

Radial velocity measurements of solar-type stars are subject to
variability due to starspots. The PSD in radial velocity is flat in
the same region of frequencies, with a noise level of 0.2 \mse,
comparable to the signal of a $1 \MEarth$ planet in a 1 AU orbit. In
addition, RV measurements may also be subject to astrophysical noise
from other processes involving velocity field fluctuations, such as
p-modes. For this reason, the estimated RV jitter due to starspots
is only a lower bound to the noise in RV measurements. Astrometry is
not affected by these other processes, so our dynamic starspot model
is a good representation of the dominant astrophysical noise source
for astrometric planet detection. A detailed discussion of our
starspot simulations is forthcoming (Catanzarite, Shao \& Law, in
preparation).

\subsection{Detecting Multiple Planets}

Of the planet-bearing stars identified by the RV technique, 20 are
revealed to have multiple-planet systems, comprising 13\% of the
sample. \citet{Sozzetti2003} and \citet{Ford2006} have investigated astrometric
orbit fitting for multiple
planet systems. Our own simulations show that with 200 observations,
SIM can detect and characterize systems with two or three
short-period planets as long as their periods are well-separated,
which should be the case if they are in dynamically stable orbits. Gas giant
companions with periods longer than the mission length are hard to detect, because SIM would detect an acceleration, but not obtain data for a closed orbit \citep{gould01}.  However, SIM can generate valuable statistical data on long-period planets, even if the periods are very uncertain, because these planets are undetectable with RV measurements and too faint for direct imaging.
In intermediate cases, combined RV and astrometric data should
constrain the orbits and make orbit solutions tractable \citep{Eisner2002}.

\subsection{Early-Mission Detection of Planets}

Precision astrometry requires knowledge of the SIM baseline length
and orientation. A set of baseline vectors for each tile is derived
as part of the astrometric grid solution. Early in the mission, the
grid accuracy, and the reconstruction of baseline vectors, is
relatively poor, but it improves rapidly after about 9 months of data
have been taken. An observing and analysis technique termed
grid-based differential astrometry (GBDA) has been developed to make
effective use of early observations of planet-search targets.
Details of the method are given in Appendix \ref{APPA}.  To
demonstrate the GBDA approach, we modeled the detection of the
planet orbiting Tau Boo, previously detected by the radial velocity
method \citep{butler97}. It has a Jupiter-like planet in a 3-day
orbit, and an expected astrometric signature of 9.0 $\mu$as.  The
model shows that this planet would be readily detected, even with
limited baseline knowledge from the grid.  Thus SIM will be able to
make useful detections of planetary systems very early in the
mission.


\section{Jupiter Mass Planets Around Young Stars \label{CHAPTER3}}

A SIM ``Young Planets'' survey, targeted toward 150-200 stars with ages from 1 Myr to 100 Myr, will help us understand 
the formation and dynamical evolution of gas giant planets.
The host stars of the majority of the more than 200 exoplanets found to date  are mature main sequence stars which were chosen based on their having quiescent photospheres for the successful measurement of small Doppler velocities 
($<$10 m s$^{-1}$).  Similarly, stellar photospheres must be quiescent at the 
milli-magnitude level for transit detections since a Jupiter mass planet transiting a solar type 
star reduces the photometric signal by about 1.4\%. Since young stars have RV 
fluctuations or rotationally broadened line widths of {\it at least} 500 m s$^{-1}$ 
and brightness fluctuations of many percent, optical RV measurements accurate to 
$<$ 100 m s$^{-1}$ or transit observations cannot be used to detect planets around 
young stars.  The near-IR is more promising, and a number of groups are attempting RV observations to improve on these limits and find a 
few ``hot Jupiters'' within 0.1 AU. 

A few potentially planetary mass objects have 
been detected at 20-100 AU from young ($<10$ Myr) host stars by direct, coronagraphic 
imaging, e.g., 2MASSW J1207334-393254 \citep{chauvin05} and GQ\,Lup 
\citep{neuhauser05}. However, these companions are only inferred to be of planetary 
mass by comparison to uncertain evolutionary models that predict the brightness of 
young Jupiters as a function of mass and age \citep{wuchterl03,baraffe02,burrows97}. Since dynamical determinations of mass 
are impossible for objects on such distant orbits, it is difficult to be sure that 
these are planets and not brown dwarfs. Nor is it even clear that the origin of 
these distant young Jupiters is due to the same formation processes as planets 
found closer-in. Multiple fragmentation events \citep{boss01}, rather than core 
accretion in a dense disk \citep{Ida05}, may be responsible for the formation of these objects orbiting so far from their star.

\begin{figure*}[ht!]
\epsscale{0.9}
\plotone{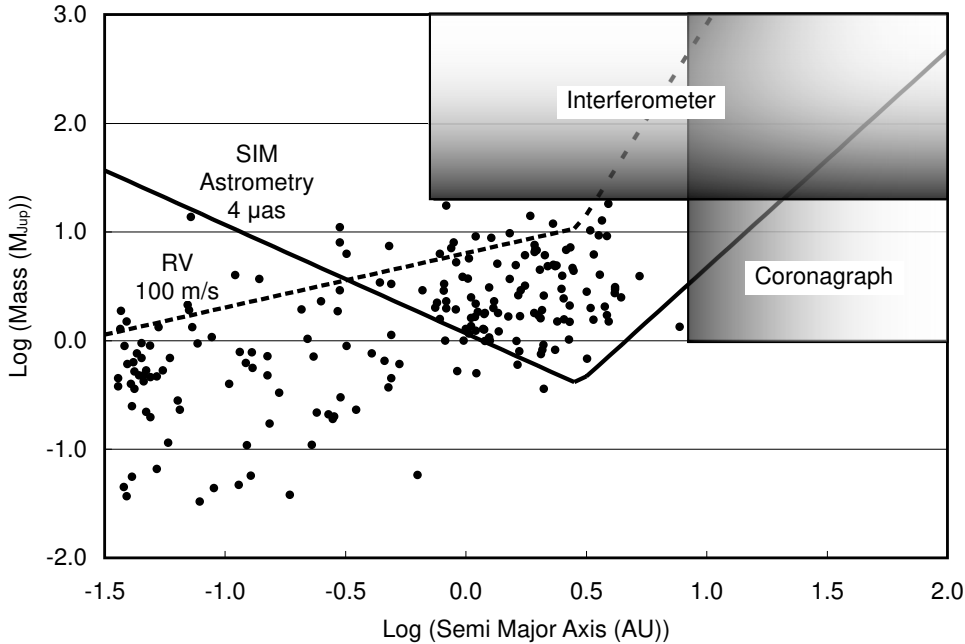} 
\caption{Planet mass detection sensitivity for the SIM-YSO survey (solid curve) in 
M$_J$ versus orbital semi-major
axis.  Estimated capabilities for a large ground-based coronagraph (taken to be a diameter $d = 30$\,m telescope at $\lambda = 1.6\,\mu$m operating an angular distance of $4 \lambda/d$) and a near-IR interferometer (85-m baseline at  
$\lambda = 1.6\,\mu$m and an intermediate stellar distance of 50 pc) are shown as shaded boxes.  Also plotted are the properties of the known radial velocity (RV)  detected planets (dots).  RV accuracy for YSOs (dashed curve) is limited to about $100\,\mse$ (\S\,\ref{CHAPTER3}).  Except for RV, the sensitivity limits assume a distance of 140 pc. 
\label{simperform}
}
\end{figure*}

As a result of the selection biases of the radial velocity, transit and direct imaging 
techniques, we know little about the incidence of planets around young stars in orbits 
close to their stars, leaving us with many questions about the formation and 
evolution of gas giant planets.  

Using Equation 1,  a Jupiter orbiting 5.2 AU away from a 0.8 $M_\odot$ star at the distance of the youngest stellar associations (1-10 Myr) such as Taurus (140~pc) and Chamaeleon would produce an astrometric amplitude of 44 $\mu$as. At the 25-50 pc distance of the nearest young stars (10-50 Myr) such as members of the $\beta$ Pic and TW\,Hya groups, the same system 
would have an astrometric amplitude in excess of 100 $\mu$as. Moving a Jupiter into a 
1 AU orbit would reduce the signal by a factor of 5.2, or 50 $\mu$as for a star at 25 
pc and 8 $\mu$as for one in Taurus. In its narrow-angle mode, SIM will have a Single Measurement Accuracy (SMA) of 0.6 $\mu$as (1$\sigma$); observations made during  wide angle observations (Appendix~\ref{APPB}) will have SMA $\simeq 4\,\muas$.
Thus a search for gas giants falls well within SIM's capabilities 
and forms the core of the SIM-YSO program. Figure~\ref{simperform} shows
SIM's expected astrometric accuracy for the SIM-YSO survey as a function of planet mass and 
semi-major axis. Also plotted is the expected RV accuracy achievable with present day
infrared echelle spectrometers. Unlike RV surveys, SIM will be able to detect Jupiter mass planets at radii out to several AU around young stars.

\subsection{Science Goals}

In a SIM survey of 200 young stars, we expect to find anywhere from 10-20 (assuming that 
only the presently known fraction of stars, 5-10\%, have planets) to 200 (all young 
stars have planets) planetary systems. We have set our sensitivity threshold to ensure 
the detection of Jupiter-mass planets in the critical orbital range of 1 to 5 AU. 
These observations, when combined with the results of planetary searches of mature 
stars, will allow us to test theories of planetary formation and early Solar System 
evolution. By searching for planets around pre-main sequence stars carefully selected 
to span an age range from 1 to 100 Myr, we will learn at what epoch and with what frequency giant planets are 
found at the water-ice `snowline' where 
they are expected to form \citep{pollack96}. This will provide insight into the 
physical mechanisms by which planets form and migrate from their place of birth, and 
about their survival rate. 

With these SIM observations in hand, we will have data, 
for the first time, on a series of important questions:\  What processes affect the 
formation and dynamical evolution of planets? When and where do planets form? 
What is the initial mass distribution of planetary systems around young stars? How 
might planets be destroyed? What is the origin of the eccentricity of planetary 
orbits? What is the origin of the apparent dearth of companion objects between 
planets and brown dwarfs seen in mature stars? How might the formation and 
migration of gas giant planets affect the formation of terrestrial planets?

Our observational strategy is a compromise between the desire to extend the 
planetary mass function as low as possible and the essential need to build up 
sufficient statistics on planetary occurrence. About half of the sample will be 
used to address the ``where" and ``when" of planet formation. We will study 
classical T Tauri stars (cTTs) which have massive accretion disks as well as 
post-accretion, weak-lined T Tauri stars (wTTs). Preliminary studies suggest the 
sample will consist of $\sim$30\% cTTs and $\sim$70\% wTTs, driven in part by 
the difficulty of making accurate astrometric measurements toward objects with 
strong variability or prominent disks. The second half of the sample will be 
drawn from the closest, young clusters with ages starting around 5 Myr, to the 
10 Myr thought to mark the end of prominent disks, and ending around the 100 
Myr age at which theory suggests that the properties of young planetary systems 
should become indistinguishable from those of mature stars. 
The properties of the planetary systems found around stars in these later age bins will be used to address the effects of dynamical evolution and 
planet destruction \citep{lin2000}.  Since we will also measure accurate parallaxes, we will have good luminosities for the host stars, and will use these to help estimate  ages.

\subsection{Astrophysical Noise}

The photospheric activity that affects radial velocity and transit measurements 
affects astrometric measurements, but, as we will now show, at a level consistent 
with the secure detection of gas giant planets. From measurements of photometric 
variability \citep{bouvier89,bouvier95} plus Doppler imaging 
\citep{strassmeier98}, T Tauri stars are known to have active photospheres 
with large starspots covering significant portions of their surfaces 
\citep{schussler96} as well as hot spots due to infalling, accreting material 
\citep{mekkaden98}. These effects can produce large photometric variations which can 
significantly shift the photocenter of a star.  Using a simple model for the 
effect of starspots on the stellar photocenter  \citep{Tanner2007}, 
for a typical T Tauri star radius of 3 $R_\odot$, we see that the astrometric 
jitter is less than 3 $\mu$as for R-band variability less than 0.05 mag. Thus, 
the search for Jovian planets is plausible for young stars less variable than 
about 0.05 mag in the visible even without a correction for jitter that may be 
possible using astrometric information at multiple wavelengths. Note that since 
both the astrometric signal and the astrometric jitter scale inversely with distance, 
there is no advantage (from the jitter standpoint) to examining nearby stars even 
despite their larger absolute astrometric signal. Other astrophysical noise sources, 
such as offsets induced by the presence of nebulosity and stellar motions due to 
non-axisymmetric forces arising in the disk itself are negligible for appropriately 
selected stars. Finally, it is worth noting that searching for terrestrial planets 
will be difficult until stars reach an age such that their photometric variability 
falls below 0.001 mag and the corresponding astrometric jitter below 0.5 $\mu$as. 

\subsection{The Sample}

The youngest stars in the sample  will be located in well 
known star-forming regions such as Taurus, the Pleiades, Sco Cen, and TW Hydra
\citep{Tanner2007} and will be observed in Narrow Angle mode, 
which is 
capable of achieving a single measurement accuracy of 0.6 $\mu$as. 
Somewhat older stars, such as those in the $\beta$ Pictoris 
and TW Hydrae Associations, are only 25-50 pc away and can be observed with less
mission time in Wide Angle mode, capable of 4 $\mu$as single measurement accuracy. We have adopted the following criteria in 
developing our initial list of candidates: a) stellar mass between 0.2 and 
2.0 M$_\odot$; b) $R < 12$ mag for reasonable integration times; c) distance 
less than 140 pc to maximize the astrometric signal to be greater than 
6 $\mu$as; d) no companions within 2$^{\prime\prime}$ or 100 AU for instrumental 
and scientific considerations, respectively; e) no nebulosity to confuse the 
astrometric measurements; f) variability $\Delta R<$0.1 mag; and g) a spread of ages 
between 1 Myr and 100 Myr to encompass the expected time period of planet-disk and 
early planet-planet interactions.

A literature search and precursor observing program  
\citep[described in][]{Tanner2007} was carried out to identify 
and validate stars according to these criteria. The observing program included 
adaptive optics imaging with the Palomar 5m, VLT 8m, and Keck 10m telescopes to 
look for M star or brown dwarf companions; RV measurements 
with the McDonald  2.7 m and HET telescopes, as well as the Magellan 
telescope to look for M star or brown dwarf  companions; and photometric 
observations with smaller telescopes to look for variability. The variability 
program proved to be the most stringent filter with roughly 50\% of the sample 
showing photometric dispersion in excess of 0.1 mag. We now have a validated 
list of 75 stars meeting all of the above criteria. More stars will be added to the 
precursor program to bring the total up to the desired number of $150-200$ stars. 
With the available observing time allocated to this program (1,600 hours), we will 
be able to make 75-100 visits to each star (up to 50 2-D visits) which, spread 
over 5 years, will be enough to identify and characterize up to 3 planets per 
star having periods ranging from less than a year up to 2.5 years. For narrow angle 
targets we will take advantage of the natural clustering of young stars to share 
requisite observations of $\sim5$ reference stars, typically $R=10-12$ mag K giants, 
with multiple (2-8) science stars within a 1$^{\circ}$ radius. With additional observations 
during a 10 year extended mission, it will possible to find planets out to 5 AU. 

A secondary goal of the program is put our knowledge of stellar evolution on a 
firmer footing by measuring the distances and orbital properties of $\sim$100 
stars precisely enough to determine the masses of single and binary stars to an 
accuracy of 1\%. This information is required to calibrate the pre-main sequence 
tracks \citep[e.g.,][]{baraffe02} that serve as a chronometer ordering the events 
that occur during the evolution of young stars and planetary systems. To 
accomplish the goals of this program, we will observe a few dozen binary T Tauri 
stars as well as stars with gas disks observed (in millimeter lines of CO) to 
exhibit Keplerian rotation. With accurate orbits and distances for these systems, 
it will be possible to determine accurate stellar properties for comparison with 
stellar evolution models.


\section{How Unique is the Solar System? \label{CHAPTER4}}

SIM is most sensitive to orbits with periods in the range of $\sim
1-5$ years (Fig.~\ref{discovery}), with the most sensitivity to
periods close to the mission length.  SIM is well-suited to detect Earth-like planets in the habitable zone around nearby stars \citep[see $\S$~\ref{CHAPTER2} and][]{Cat2006}.  For periods up to $\sim 10$
years, estimates of orbital parameters, including period, can be
made, but as the period lengthens, the uncertainties grow quickly.
In the limiting case, SIM can only make a detection of the acceleration due to a companion \citep{gould01}.  Even though the parameters of any one such target may not be well-determined, important statistical conclusions can be
drawn from an ensemble of long-period systems.  Independent data, especially over a long time baseline, can greatly improve our knowledge of long-period companions.  For many targets, there will be a 10-15 year baseline of RV measurements to draw on.  

Combining astrometry from a ``SIM quick look'' (SQL) survey with data from the Hipparcos astrometric mission \citep{ESA97} would constrain orbits of 100 years or more, and this could be done for several thousand Hipparcos stars.  Below, we show results from a simulation of the extraction of planets from the combined dataset.
Orbits of a $1\, \MJup$ planet can be reliably characterized up to periods of about 10 years, $10\, \MJup$ planets up to 80 years, and stellar companions up to 320 years.

\subsection{Masses and Periods of Long-Period Planets}
\label{sec:Long-Period-Planets}

Long-period extrasolar giant planets appear to
be rare: only 25 have periods above 5 years and just one
has a period slightly longer than that of Jupiter. Taking the
selection effects into account, \citet{TT2002} estimate that 3\% of
Sun-like stars should have a planet with a period between 2 days and 10
years and a mass of $1-10\, \MJup$.   
For our simulation, we adopt the normalization of \citet{S2005}, which is 1.62 times that computed by \citet{TT2002}.

We define a Solar System Giant Analog (SSGA) as a planet (or planets)
whose mass and period fall within the range of the giants of our
Solar System (i.e., with a period between 12 and 165 yr and mass of $0.05 - 1\MJup$).  Such systems may or may not contain lower-mass planets, in closer orbits, but the astrometric signatures of SSGAs would normally dominate, and would remain detectable in distant systems for which terrestrial planets are below the detection limit.  Because the extrasolar giant planet period distribution function increases with period, systems dominated by giant planets should be rather
common, and we estimate that 12.6\% of Sun-like stars could harbor SSGAs. 
 
We also define a more massive version of the Solar System Giant Analog, with mass between 1 and $13\, \MJup$, as a Massive Solar System Giant Analog or MSSGA.  These are predicted to be quite abundant, and of course are easier to detect:  20\% of the total number of planetary systems with periods up to 165 years, and occurring around 7.9\% of single stars.  

\subsection{A Survey for Solar-System Analogs}
\label{sec:Finding_Solar-System_Analogs}

To identify likely long-period planetary systems we combine
data from a ``SIM Quick Look" (SQL) survey with Hipparcos data
\citep{ESA97}.  This method uses the astrometric parameters as
determined from a fit to the SIM data to predict the position at the
Hipparcos epoch. Differences between the observed and predicted
positions indicate the presence of a companion \citep{O_EXOPTF_2007}. 
SIM data allow for the determination of the seven astrometric
parameters of an ``acceleration" solution (in addition to the two
positions, two proper motions, and parallax).  Additional SIM observations would help with reliable extraction of accelerations. In any case, the aim of the survey would primarily be to reject the main-sequence (MS) binaries that have huge signals.   Either way, a SIM survey would produce a sample rich in planetary and/or brown-dwarf (BD) companions.

For truly single stars, the SIM data will be an excellent predictor of
positions recorded in the twentieth century. However, if the star
has a companion, the short-term proper motion determined from a SQL survey can be very
different from the center-of-mass motion. For a face-on, circular
system, the semi-major axis of the orbit, orbital speed, acceleration, and the derivative of the acceleration are all substantial for nearby MSSGA systems. For a
1\, \msun star at a distance of 20 pc, and a $1 \MJup$ planet, we can show that MSSGAs with periods in the range of 5 to 160 years are readily detectable.  This issue has been well-studied, in the context of FK5 and Hipparcos astrometry, by 
\citet{Wielen2001} and references therein.

Due to the short observing span with respect to the orbital period,
SQL data effectively determine the instantaneous proper motion
and acceleration due to orbital motion.  The long time baseline $\tau$ between the SQL and Hipparcos epochs allows us to compute a metric,  $\Delta_{xy}(\tau) \equiv \sqrt{\Delta_x^2(\tau) + \Delta_y^2(\tau)}$ which
is independent of phase for circular, face-on orbits
\citep{O_EXOPTF_2007}.  The $\Delta_{xy}(\tau)$ diagnostic is
useful when it exceeds the astrometric error.

The $\chi^2_\nu$ figure of merit is useful in revealing MSSGA, BD
and MS companions. For orbital periods between 5 and 320 years, the
(reduced) $\chi^2_\nu$ values uncover 11\%, 39\%, and 73\% of the companions
in the MSSGA, BD, and MS mass range respectively, if we use only the SIM data to
compute $\chi^2_\nu$.  Here we ignore the effects of inclination and  eccentricity, which complicate the characterization of the companion, although they will not lower the $\chi^2_\nu$ and $\Delta_{xy}$ values very much 
\citep{MK2005,O_EXOPTF_2007}.
Including the available non-SIM astrometry
significantly increases the yield to 46\%, 90\%, and 99.8\% for the three 
mass ranges respectively.  These results indicate that low-mass companions can be efficiently detected by combining SQL and Hipparcos data. 
We find that the orbits of a 1 or $10 \MJup$ MSSGA can be characterized up
to periods of 10 or 80 years, respectively, and for stellar
companions up to 320 years \citep{O_EXOPTF_2007}. The
$\Delta_{xy,\mu}$ values are significant up to 1,200 or 5,000 years for
companions with mass 0.08 or 1.0 M$_\odot$, respectively.

\subsection{Very Long-period Companions}
\label{sec:SQL_Follow-up}

There is strong evidence that the intrinsic multiplicity rate due to
either stars or planets is close to 100\% among Hipparcos MS stars
\citep{O2005}. The lack of cataloged companions is most likely due to
selection effects. Thus, those systems without signs of binarity in
a SQL+Hipparcos survey are likely to have either sub-stellar
companions with an unknown period, or very long-period stellar
companions.  An extended SIM survey would further explore these poorly-characterized systems.

An extended SIM astrometric survey would be significantly more sensitive than the initial Quick Look Survey.  Applying the $\Delta_{xy}$ analysis presented above to the extended SIM survey indicates that the MSSGAs can be detected with
masses as low as $0.1 \MJup$ in 10-yr orbits. The maximum period for which a 
$1 \MJup$
planet can be reliably detected is extended by a factor four (to 40 years) and for $10 \MJup$ by a factor two (to 160 years).  Thus SQL+Hipparcos plus an extended SIM survey can uncover a very significant part of the MSSGA parameter space. 

Given the importance of accurate pre-SIM astrometry, we note
that large-scale ground-based photometric
surveys such as Pan-STARRs will also provide astrometry at the
required (sub-mas) level \citep{PANSTARRS}. Also, data from the Gaia mission
\citep{P2002} will help explore and characterize SSGAs more fully.


\section{Precision M-L Relation for Extreme Stellar Types \label{CHAPTER5}}

Mass is the most fundamental characteristic of a star.  It governs a
star's entire evolution --- determining which fuels it will burn, what
color it will be, and how long it will live.  It is crucial to our
understanding of stellar astrophysics that we determine stellar masses
to high accuracy.  Knowing the masses of main sequence stars answers
basic stellar astrophysics questions such as, {\it What is the mass-luminosity relation for the highest mass and also the lowest mass stars?  What is the initial mass function?  What is
the mass content of the Galaxy and how does it evolve?}  In fact, the
dependence of luminosity upon mass --- the mass-luminosity relation
(MLR) --- is one of the few stellar relations sufficiently fundamental
to be applicable to many areas of astronomy.  With the exception of
the H-R Diagram, it is the single most important ``map'' of stellar
astronomy.  To answer truly fundamental astrophysical questions about
stars, the ultimate goal is to determine masses to 1\% accuracy, which
will allow us to challenge stellar models more severely than ever
before.  Because of SIM's exquisitely accurate astrometric
capabilities, coupled with its faint magnitude limit, we can develop a
well-stocked ``toolbox'' of MLRs that can become the standards against
which all stars are measured.

Here, we consider the extreme ends of the main sequence,
where SIM will be crucial in making real progress in defining the MLR.
In the case of the most massive stars, SIM's extreme accuracy will
allow us to reach further across the Galaxy to pick up the rare O and
B type binaries needed for mass determinations.  For their much less
massive cousins, the red M dwarfs, SIM's faint limit provides the ability to measure the orbital motions of objects all the way
to the end of the stellar main sequence, and into the regime of the
substellar brown dwarfs.

There are two tactics that can be used to pin down the most massive
and least massive stars --- measurements of individual systems in the
field, and the calibration of the so-called third and fourth
parameters, metallicity and age, by targeting stars in clusters for
which those quantities are known.

\begin{figure}[ht!]
\epsscale{1.15}
\plotone{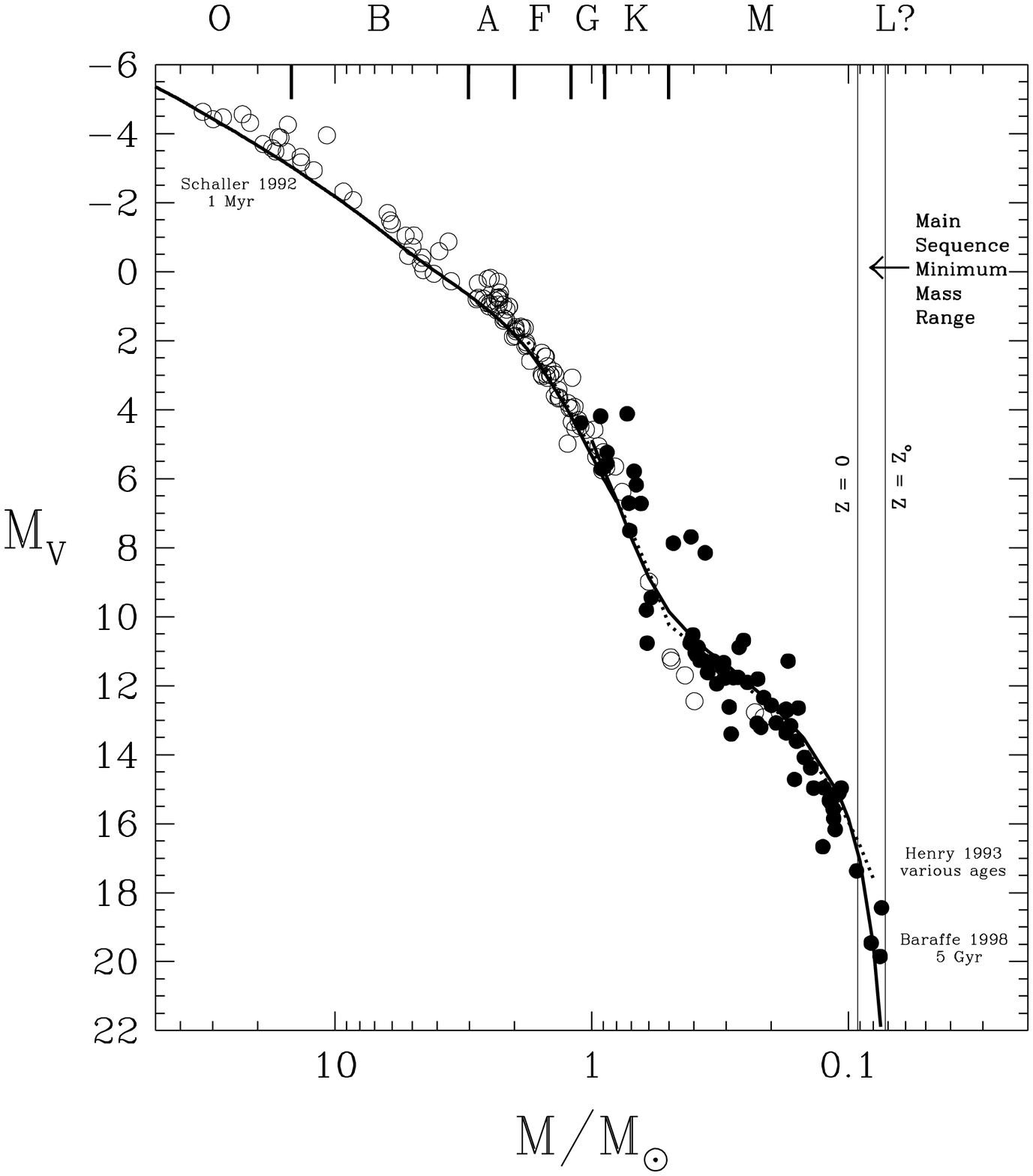}
\caption{The mass-luminosity relation in 2007, using eclipsing binary data 
(open circles) from \citet{Andersen1991} and others, supplemented with visual, speckle and interferometric binary data (filled circles).  
Model curves for the mass-luminosity relation at the indicated ages and solar 
metallicity are shown, from \citet{Schaller1992} at the higher masses and 
\citet{Baraffe1998} at the lower masses.  The empirical fit of \citet{Henry1993} for stars with masses 0.08 to 2.0 M$_\odot$ is indicated with a dotted line.
\label{henry1}
}
\end{figure}

\subsection{Massive O and B Stars}

Massive stars are key contributors to the energy budget and chemical
enrichment of the Galaxy, but little is presently known about their
masses (see Figure~\ref{henry1}).  There are only five known eclipsing binaries among the O stars that have reasonably well established masses 
\citep{harries98}, and this lack of data has seriously hindered our
understanding of the evolution of massive stars.  One unknown, for
example, is the maximum mass possible for a star.  Interior models for
massive stars predict that stable stars can exist with initial masses
of 120 M$_\odot$, but the most massive object among the five eclipsing
binaries is only $33 M_\odot$.  Furthermore, indirect methods of
estimating mass by comparison with model evolutionary tracks and
through spectroscopic diagnostics lead to discrepancies as large as a
factor of two \citep{Herrero2000}.  SIM will record the
photocentric and/or absolute orbits of many binaries and by combining
this information with spectroscopic data it will be possible to
determine accurate distances, inclinations, and masses.  An excellent
example is the massive binary HD\,93205, which consists of an O3V +
O8V pair in a 6.08 day orbit.  SIM observations will show a 45 $\mu$as
photocentric variation that will yield the first accurate mass for a
star at the top of the main sequence \citep[only known to be in the range of
32--154 M$_\odot$;][]{Antokhina2000}.

\begin{figure}[th!]
\includegraphics[scale=0.33,angle=-90]{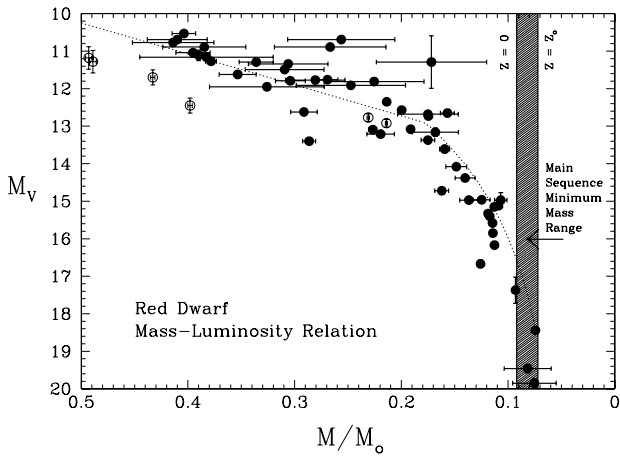}
\caption{ A zoom in of the mass-luminosity relation in 2007, focusing on red dwarfs. 
Eclipsing binary data are represented by open blue points, and visual binary data 
from the MASSIF Team (solid red points) and others (open red points).  
The empirical fit of \citet{Henry1999} is indicated with a dotted line, 
as well as the range of minimum masses for main sequence stars, depending on 
metallicities.  Note the disconnect between some of the eclipsing and visual 
binary points, as well as the need for a revision to the fit (toward higher masses) 
for the lowest mass stars.
\label{henry2}
}
\end{figure}

SIM will also provide the first accurate masses of the evolutionary
descendants of massive stars.  The most massive stars develop strong
outflows later in life and appear as Wolf-Rayet (WR) stars.  SIM
measurements of the WR binary, WR 22 
\citep[WN7 + O9III;][]{Schweickhardt1999}, will show a 250 $\mu$as astrometric variation through
the 80.3 day orbit. These measurements will provide the mass of
this extraordinary object, currently estimated to be $55 \pm 7$
M$_\odot$, the most massive star known.
Intermediate-mass B stars in close binaries are believed to suffer
extensive mass transfer and mass loss during the Roche lobe overflow
phase.  The best example of this evolutionary stage is the enigmatic
binary, $\eta$ Lyr \citep{Bisikalo2000}, which consists of a
bright, 3 M$_\odot$ star losing mass to a 13 M$_\odot$ star hidden
in an extensive accretion disk.  The astrometric orbital motion of the
bright component will amount to 820 $\mu$as, and SIM will provide
accurate mass estimates at this key evolutionary stage.  Finally, two
other examples of massive stars with longer periods include HD 15558
(O5e) and HD 193793 (WR), with periods of 1.2 and 7.9~years,
respectively.  At distances of 1.3 and 2.6 kpc, each system has a
semimajor axis of 5--10 mas, easily within reach of SIM.

\subsection{Low Mass M Stars and Brown Dwarfs}

Red dwarfs dominate the solar neighborhood, accounting for at least
70\% of all stars, and represent nearly half of the Galaxy's total
stellar mass \citep{Henry1999} and Figure~\ref{henry2}.
These stars have spectral type
M, $V = 9-20$, and masses 0.08--0.50\,$\Msun$ \citep{Henry1993,Henry1994}. 
The MLR remains ill-defined for M
dwarfs, so their contribution to the mass of the Galaxy is a guess at
best, and the conversion of a luminosity function to a mass function
is problematic.  At masses less than $\sim\,0.20 \Msun$ an accurate
MLR can provide a strict test of stellar evolutionary models that
suggest the luminosity of such a low-mass star is highly dependent
upon age and metallicity.  Finally, the MLR below 0.10 $\Msun$ is
critical for brown-dwarf studies because accurately known masses can
convincingly turn a candidate brown dwarf into a {\it bona fide} brown
dwarf.

In recent decades, the masses of red dwarfs have been determined using
a combination of infrared speckle interferometry and HST-FGS, and
occasionally via radial velocity efforts.  The number of red dwarfs
with accurate mass measurements less than 0.20 $\Msun$ has
increased from four in 1980 \citep{Popper1980} to 22 
\citep[][and unpublished]{Henry1999}.
The sample of more massive red dwarfs in the
range $0.50 \Msun > M > 0.20 \Msun$ has also increased,
with particular improvement in the quality of the available masses.

SIM is critical for M dwarf systems because they are typically faint
and do not allow high-precision RV measurements due to
their slow orbital motion and poorly separated spectral lines.  In
addition to accurate orbital monitoring, SIM will provide two crucial
pieces of information required to reduce mass errors to the 1\% level,
where they become astrophysically interesting: parallaxes and mass
fractions.  As an example, we examine the nearby binary Gl 748, which
represents the current state-of-the art accuracies for red dwarf
masses (2.4\%).  SIM can improve the mass by reducing the error in
the semi-major axis of the absolute orbit ($147.0 \pm 0.7$ mas) by a
factor of 18 (to 0.04 mas, or 10 times the nominal astrometric
accuracy of SIM for Global Astrometry) and the error in the parallax
($98.06 \pm 0.39$ mas) by a factor of 10.  The result would be mass
errors of only 0.1\%.

Mass is the best discriminator between stars and brown dwarfs.  An
object's mass determines whether or not temperatures in the object's
core are sufficiently high to sustain hydrogen fusion --- the defining
attribute of a star.  L dwarfs are objects with smaller masses and
cooler temperatures ($\sim$1500--2000K) than those of M dwarfs, but no
accurate masses of L dwarfs have yet been measured, so the models of L
dwarfs are completely untested by data.  Several hundred L dwarfs have
been discovered to date, and appropriate systems observable at SIM's
faint limit are being found.  One example is GJ 1001 BC at a distance
of 13 pc.  With an orbital period of $\sim$4 years, this system is
ideally suited to the nominal SIM mission lifetime.


\section{The Late Stages of Stellar Evolution
\label{CHAPTER6}}

The first discovery of an X-ray binary occurred over four decades ago
when Scorpius X-1 was detected during a New Mexico rocket flight
\citep{giacconi62}.  Although the nature of Sco~X-1 was not immediately
clear, it was not long before \cite{shklovskii67} suggested the correct
explanation that Sco~X-1 is a neutron star accreting from a stellar
companion.  Over the years, in excess of 25 X-ray satellites have found
hundreds of X-ray binaries with neutron star or black hole accretors
exhibiting a rich variety of physical phenomena.  Studies of these systems
allow us to probe the most extreme physical conditions in the universe,
including magnetic fields at the surfaces of neutron stars that can be in
excess of $10^{12}$~Gauss \citep{coburn02}, densities in neutron star cores
that may be as much as an order of magnitude above nuclear densities
\citep{lp04}, and gravitational fields near black holes and neutron
stars can provide tests of strong gravity \citep{psaltis04}.

Studies of accreting stellar mass black holes also improve our understanding
of Active Galactic Nuclei (AGN) and quasars.  X-ray binaries with relativistic
jets are often called microquasars \citep{mirabel92} because of the
similarities between these systems and quasars.  However, detailed
comparisons between the two populations are hampered by uncertain distances
to the microquasars, making parameters like total luminosity and jet velocity
uncertain.

While much has been learned about the physics of X-ray binaries, it
is evident that more precise measurements of physical properties are
required to make further progress in testing theory.  Some of the
properties that are the least accessible using current instrumentation,
such as the source distance ($d$), proper motion, and binary inclination
($i$), will be readily measured using SIM.  Here, we discuss some of the
issues related to the physics of X-ray binaries that SIM will help to
address.

For these sources as well as for other radio-emitting stars, it
will be possible to combine SIM's observations with Very Long Baseline
Interferometry (VLBI) observations to place the radio components within an
absolute reference frame that is accurate to 3 $\mu$as.

\subsection{Masses of Neutron Stars}

Neutron stars (NSs) provide a unique opportunity to understand what happens
to matter as densities are increased beyond the density of nuclei.  Thus,
measuring the NS equation of state (EOS) has important implications for
nuclear physics, particle physics, and astrophysics, and measuring NS
masses, radii, or both provide constraints on the EOS.  The NSs for which
accurate mass measurements have been made lie very close to the canonical
value of 1.4 \Msun\ \citep{tc99}, and EOSs
with normal matter (neutrons and protons) as well as exotic matter (e.g.,
hyperons, kaon condensates, and quark matter) can reproduce this mass
for a large range of radii \citep{lp04}.  However, more recently, there
are indications that some systems may harbor higher mass, 1.8--2.5\Msun,
NSs \citep{barziv01,clark02,nice05} and confirming these high NS
masses by reducing the uncertainties would lead immediately to ruling out
a large fraction of the proposed EOSs.

\begin{figure}[ht!]
\epsscale{1.15}
\plotone{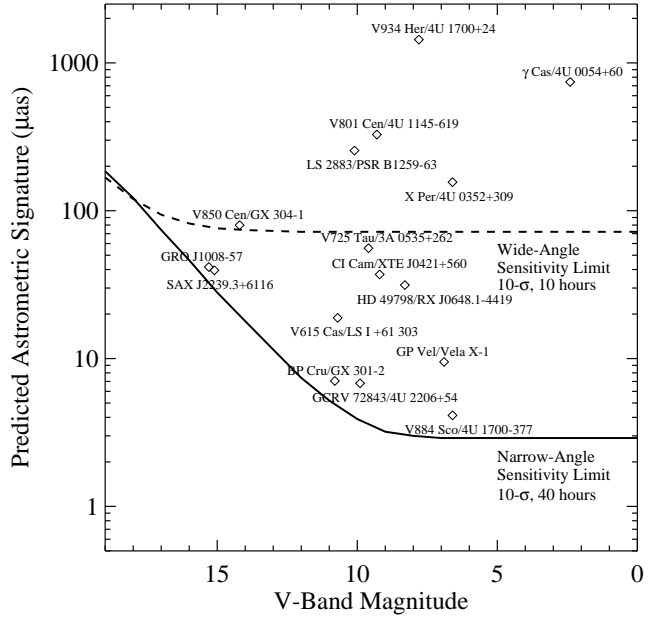}
\caption{The expected astrometric signature from orbital motion vs. V-band magnitude
for the 16 neutron star and neutron star candidate High Mass X-ray Binaries
which we expect to have large enough signatures to be detected by SIM.
The solid line shows the 10-$\sigma$ narrow-angle SIM sensitivity limit
found from simulations \citep{tqr05}  where
40 hour-long observations are made, and we have a priori knowledge of
the binary orbital period.  The dashed line shows the 10-$\sigma$ wide-angle
SIM sensitivity limit for 20 half-hour-long observations.
\label{hmxbs}
}
\end{figure}

SIM will be capable of making precise orbital measurements for
a large number of High-Mass X-ray Binary (HMXB) systems.  These systems
typically have O- or B-type companions with $\sim$25 HMXBs being
brighter than $V\sim 15$.  They also have orbital periods ($P_{\rm orb}$)
of days to a couple of years, and their wide orbits give large astrometric
signatures.  Taking estimates of HMXB parameters ($P_{\rm orb}$, $d$, and
the component masses) from \cite{lvv00} as well as more recent literature,
we find that 21 likely NS HMXBs have orbital signatures (the
semi-major axis of the optical companion's orbit) of $a_{\rm sig}\ge 5\,
\mu$as and 8 HMXBs have $a_{\rm sig}\ge 40\, \mu$as.  Detailed simulations
that account for the optical source brightnesses \citep{tqr05} show
that SIM is expected to be capable of detecting orbital motion for 16
NS HMXBs (see Figure~\ref{hmxbs}).

The most interesting among these 16 systems are those for which
the projected size of the NS's orbit ($a_{\rm x}\sin{i}$) has already
been measured \citep{bildsten97}.  The five sources for which this is
the case are Vela X-1, X Per, 3A~0535+262, GX 301--2, and PSR B1259--63.
SIM measurements, along with $a_{\rm x}\sin{i}$, will immediately yield
a NS mass measurement.  Perhaps Vela X-1 is the most tantalizing as it is
suspected of having an over-massive NS.  The current NS mass measurement
for Vela X-1 is $M_{\rm x} = 1.86\pm 0.16$\Msun~(1-$\sigma$ errors)
\citep{barziv01}.  In 40 hours of narrow-angle SIM observations of
Vela X-1, it will be possible to measure $M_{\rm x}$ to 3.9\%
\citep{tqr05}.  This is a major improvement over the current mass
measurement and will be sufficient to determine if Vela X-1 harbors an
over-massive NS.  As our estimate of the astrometric signature for
Vela X-1 is 9.5 $\mu$as, the microarcsecond measurement accuracy provided
by SIM is critical.

\subsection{Masses of Stellar Black Holes}

Black holes (BHs) are among the most fascinating celestial objects.
Stellar BHs in our galaxy accreting from a normal star facilitate
investigation of disk accretion and relativistic jets. The mass is a
fundamental property of a stellar BH and has critical implications
for the evolution of BH binaries.  Although BHs and NSs are often
difficult to distinguish, it is believed that a compact object that
is more than 3 $\Msun$ is probably a BH while objects less than
3 $\Msun$ could be either a BH or a NS.  Present measured masses of
twenty confirmed BHs have possible masses ranging from 3 to 18
$\Msun$ \citep{Rem06}.  The masses of stellar BHs have large
uncertainties due to the unknown orbital inclination, parallax,
and other systematic errors.

As an example, our understanding of the nature of the galactic BH
Cyg X-1 could change substantially depending on its true distance
and the companion's mass.  The mass of the BH in Cyg X-1 is estimated
as 10 $\Msun$ \citep{Herrero1995}.  However, the companion star of
Cyg X-1 might be undermassive for its early spectral type.
Additionally, there is a huge range of distance estimates for the
system:  Hipparcos measurements place it at $1724 \pm 1000$ pc; VLBI
estimates $1400 \pm 900$ pc; spectral analysis places it at
$2000 - 2500$ pc. If the lower companion mass and nearest distance
are adopted, the mass of Cyg X-1 could be as low as 3 $\Msun$. SIM
can refine the mass measurements for X-ray binaries that are thought
to harbor BHs. Currently, the main uncertainty in the component
masses arises from uncertainty in the binary inclination and distance
to the system, quantities that will be measured accurately by SIM.
For the case of Cyg X-1, the $\sim$20 $\Msun$ supergiant companion
has an orbital astrometric signature of 27 $\mu$as at a distance of
2.5 kpc, or 34 $\mu$as at 2.0 kpc. Thus, the orbit of Cyg X-1 can be
easily resolved by SIM, allowing the binary inclination to be
determined accurately and the semi-major axis to be determined to
better than 2\% \citep{ps05}. SIM's measurements, combined with X-ray
spectroscopic and photometric observations and VLBI observations,
will give us a complete physical picture of stellar BHs for the
first time.

SS~433 is an HMXB and microquasar with unique relativistic baryonic
jets that precess with a 162 day period \citep{margon84}.  Although
there is evidence to support the presence of a BH in the system
\citep{hillwig04}, the nature of the compact object is still debated.
Due to the uncertainty in the compact object's mass, the orbital
astrometric signature is also uncertain, but could easily be
$10-30 \mu$as.  Because the object is bright ($V=14$), SIM
narrow-angle measurements are feasible, and a detection of its
orbital motion would allow for a definitive answer regarding the
nature of the compact object.

\subsection{Formation and Evolution of Black Holes and Neutron Stars}

The evolutionary endpoints of massive stars result in compact objects,
such as white dwarfs, NSs, or BHs. It is of great importance to
investigate their birth place, asymmetric birth kicks, and the path
of formation and evolution of these compact objects.  In theory, a
BH can be formed in two different ways: a supernova (SN) explosion or
collapse of a massive star without an energetic explosion.  These
formation mechanisms can be distinguished by unique information
contained in the kinematic history.  If a galactic BH or a NS is
formed in a supernova event, very often a supernova remnant is
nearby.  It is crucial to use 3-dimensional space velocity measurements
to determine the object's runaway kinematics and the galactocentric
orbits \citep[see][and references therein]{mr03a}. Unfortunately, the
current measurement precision of transverse velocities is limited.
The image superposition technique with HST has a precision of
$\approx 1$ mas for proper motions and an error of $16\%$ for
velocities \citep{mirabel92}. SIM can provide at least two orders of
magnitude improvement for kick velocity measurements, and can identify
associations between a compact object and supernova remnant for many X-ray binaries.

For the scenario where a BH is formed without a supernova event, the
most important issues include determination of its birth place,
measurements of its galactocentric orbit, and a thorough investigation
of its space environment \citep{ps05}.  So far, only Cyg X-1 provides
observational evidence of this outcome, because it appears to have
proper motion in common with the association  OB-3, and there is no
nearby supernova remnant.  For this type of BH formation, key parameters, such as
inclination, kick velocity, distance and masses, are either indirectly
known or lack sufficient precision from current observations.

\subsection{LMXB Distances and Constraints on Physical Parameters}

As most X-ray binaries are too far away for parallax measurements,
distance measurements for these systems are, for the most part, highly
uncertain.  It is not unusual for an X-ray binary's only distance
estimate to be based on a companion's spectral type and a system
brightness, and these estimates can be uncertain by a factor of
2 or more.  This leads to uncertainty about many basic system
parameters such as luminosities, mass accretion rates, radii of
NSs \citep{rutledge02}, sizes of accretion disks, and jet velocities
\citep{fender06}.

SIM will be able to measure the distances of Low-Mass X-ray Binaries
(LMXBs), and we have used \cite{lvv01} as well as more recent literature
to compile a list of LMXBs.  As LMXBs tend to be optically faint, the
main selection criterion is V-band magnitude.  Although the optical
brightness can be strongly variable for transients, these systems spend
most of their time in quiescent (i.e., low flux) states.  There are 27
LMXBs with $V < 20$ for which SIM parallax measurements will be feasible.
For the brightest few sources ($V = 12$--13), it will be possible to
measure distances to accuracies as high as 2\% using 1 hour of SIM time.
While more time will be required for the fainter sources, accurate
distance measurements will still be feasible.  For the 19 sources on
our list with $V > 17$, we typically expect to obtain distance
measurements to 5\% accuracy using, on average, 5 hours of SIM time
per source.  Thus, these systems make excellent use of SIM's ability
to observe fainter targets.

The LMXBs on our target list include microquasars such as V4641\,Sgr
and GRO\,J1655--40 for which accurate distance measurements will
provide a test of whether their jet velocities actually exceed 0.9\,$c$.
Observations of Cen X-4 will allow improved constraints on NS radius
measurements, and we will be able to determine if the brightest
persistent NS systems (such as Sco X-1 and Cyg X-2) and the brightest
X-ray bursters (such as 4U\,1636--536) reach the Eddington limit.

\subsection{Active Stars and Micro-quasars}

There are various types of stars that produce continuum emission
at radio wavelengths including:  RS CVn binaries, eclipsing Algol-type binaries,
X-ray binaries, novae, pre-main sequence stars, and micro-quasars.
Two areas in which a SIM astrometric mission would have a significant impact 
in the study of radio stars are: (1) establishing a link between the ICRF and the 
optical reference frame and (2) in the study of the stars themselves, 
specifically, the location of the radio emission and the mechanism
by which it is generated.

Traditionally, links between the radio and optical frames have been determined
through observations of radio stars.  At optical wavelengths, the Hipparcos Catalogue
currently serves as the primary realization of the celestial reference system.
The link between the Hipparcos Catalogue and the ICRF was accomplished through
a variety of ground and space-based efforts \citep{KOVALEVSKY:97, LESTRADE:99}.  
The standard error of the alignment was estimated to be 0.6~mas at epoch 1991.25, 
with an estimated error in the system rotation of 0.25 mas yr$^{-1}$ per axis 
\citep{KOVALEVSKY:97}.  For future astrometric missions such as SIM, the link between
the ICRF and the optical frame will be established through direct observations
of the quasars.  However, observations of a number of radio stars should
provide a useful check on this important frame tie.

In a series of radio observations made with connected element interferometers
\citep{JOHNSTON:03, BOBOLTZ:03, FEY:04, BOBOLTZ:07}, positions and proper 
motions of $\sim$50 radio stars were determined in the ICRF.   One goal of 
this program was to investigate the current accuracy of the ICRF-Hipparcos frame
tie.  Most recently, \cite{BOBOLTZ:07} compared radio star positions and proper
motions with the Hipparcos Catalogue data, and obtained results consistent 
with a non-rotating Hipparcos frame with respect to the ICRF.  These studies 
demonstrate the methods by which the optical and radio frames can be 
linked on levels of a few milliarcsec using radio stars.
Such a connection between a future SIM optical frame and the ICRF will
require much more accurate VLBI observations in the radio, and will
take into account phenomena related to the orbits of the close
binary companions.

In addition to establishing a link between frames, observations of
active radio stars performed with ground-based VLBI
and SIM will greatly enhance our understanding of these objects.  
Many of the stars emitting in the radio are close
RS CVn and Algol-type binaries with separations $<$20~mas and
orbital periods $<$20 days.  SIM will provide unprecedented insight into
the process of radio emission and mass transfer for such stars.  For example,
the prototype radio star, Algol, is a triple system.  From VLBI
measurements, it was concluded that the two orbital planes of the
close and far pairs are perpendicular to each other, rather than
being co-planar \citep{Pan1993}.  The cause of such perpendicular
orbits in stellar evolution theory is on-going \citep{Les93}.

In both RS~CVn and Algol-type binaries it is unclear where exactly the
radio emission originates relative to the two stars in the system.
Competing mechanisms for generating radio emission are reviewed in 
\cite{RANSOM:02} and include phenomena such as:  gyrosynchrotron radiation from
polar regions of the active K-giant star \citep{MUTEL:98},
emission from coronal loops originating on the K-giant \citep{FRANCIOSINI:99}, and 
emission from active regions near the surface of both stars with possible
channeling of energetic electrons along interconnecting magnetic field lines
\citep{RANSOM:02}.   An astrometric mission such as SIM should provide
stellar positions on the 10 $\mu$as level, and the full three-dimensional orbits
required to distinguish between the various emission mechanisms.

SIM will also have the flexibility to coordinate observations with ground-based 
instruments such as the VLBA to allow the location of the radio emission 
relative to the stars as a function of time, even for the shortest period 
($\sim$1 day) binaries.  In addition, studies of the dynamics of 
radio jets from micro-quasars will take advantage of SIM's flexible `Target of 
Opportunity' scheduling.  Most micro-quasars are X-ray transients, and when 
they undergo their month- to year-long outbursts they become millions of times
brighter in X-rays, thousands of times brighter in the optical,
and they often produce observable radio jets.  SIM observations of an outburst 
will provide the precise absolute location of the compact object and accretion disk, 
which is critical for interpretation of the locations and velocities of the jets.

A final issue regarding radio stars and micro-quasars is the
establishment of the linear scale sizes of the systems through
accurate parallax measurements.  Existing parallax measurements
are sometimes in conflict.  For example, with Hipparcos, the distance to the
micro-quasar LS\,I\,+61\,303 was found to be 190 pc; however, VLBI
observations place it at a distance of 1150 pc \citep{L2000}.  Through
accurate parallax measurements, SIM will provide the linear scale sizes
necessary to relate the radio emission to stellar positions and to
constrain theoretical models of radio star and micro-quasar emission.

\subsection{Late-type Stars with Maser Emission\label{masers}}

The evolution of stars along the asymptotic giant branch (AGB), including Mira variables, 
semi-regular variables, and supergiants, is accompanied
by significant mass loss to the circumstellar envelope  (CSE).  The nature of this 
mass-loss process and the mechanism by which spherically 
symmetric AGB stars evolve to form axisymmetric planetary nebulae (PNe) is 
not well understood.

\begin{figure}[ht!]
\epsscale{1.1}
\plotone{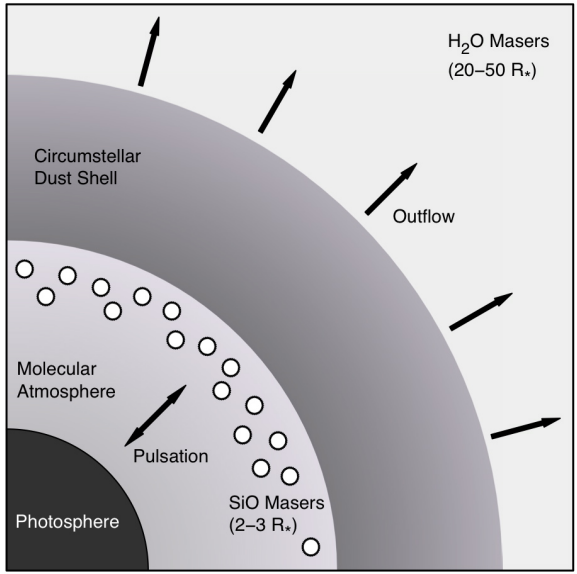}
\caption{A schematic view of the radial structure of the envelope 
of a typical AGB star 
with circumstellar maser emission.  Interferometry in the optical, near-infrared, and 
mid-infrared can be used to study the photosphere, the molecular atmosphere and the 
circumstellar dust.  VLBI at radio wavelengths can be used to study the circumstellar 
SiO and H$_2$O masers.
\label{AGB_SCHEMA}}
\end{figure}

The circumstellar maser emission (OH, H$_2$O and SiO) associated with 
many AGB stars provides a useful probe of the structure and kinematics of the 
nearby circumstellar environment.  Figure~\ref{AGB_SCHEMA} shows a schematic 
view of the inner CSE of a typical AGB star with masers.  The various
maser regions can be studied at radio wavelengths with VLBI,
while the star itself, the molecular atmosphere, and the circumstellar dust
can be studied using long baseline interferometry in the optical and infrared.

While ground-based techniques provide a powerful tool to study
AGB stars, there are still unanswered questions for which SIM could
provide crucial information.  For example:  (1) What is the underlying cause of the
transition of symmetrical AGB stars to asymmetrical PNe (e.g., unseen binary
companions, non-radial pulsations)? (2) What are the positions of AGB stars relative
to the circumstellar masers within the CSE?  (3) What is the linear scale size of the CSE?

A review of the current research regarding PNe shaping is presented in \citet{BF:02}.  
Theoretical models involve interacting stellar winds, 
magnetic field shaping, astrophysical jets, and unseen companions.  Observational 
radio/IR/optical interferometric studies probe the inner regions of the progenitor AGB 
stars and provide evidence for asymmetry \citep{MONNIER:04, BD:05}, significant 
magnetic fields \citep{VLEMMINGS:06}, and highly collimated astrophysical jets 
\citep{IMAI:02, BM:05}.  Whatever the mechanism for shaping PNe 
from AGB stars, it must be operating at the innermost scales of the CSE.  This is just the regime that SIM will be able to probe.  

\begin{figure}[ht!]
\epsscale{1.3}
\plotone{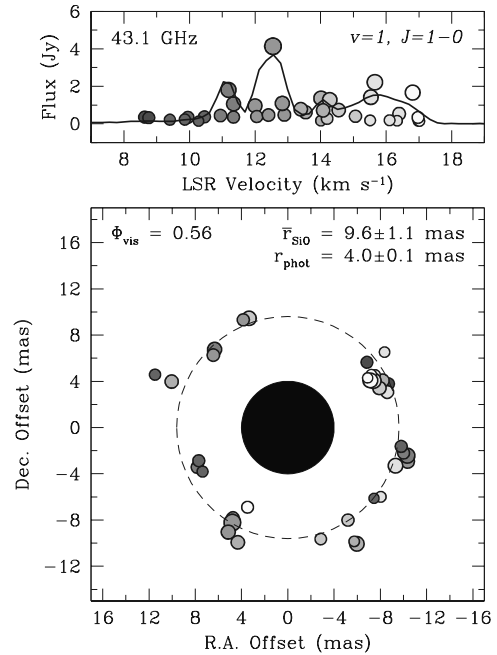}

\caption{The $v=1, J=1-0$ SiO maser emission toward the Mira variable S Ori
at a stellar phase $\Phi_{\rm vis} = 0.56$, as measured by the VLBA.  
The top panel shows the spectrum 
formed by plotting maser intensity versus velocity.  The bottom panel plots 
the spatial and velocity distribution of the masers with point shading
representing the corresponding velocity bin in the spectrum and point size 
proportional to the logarithm of the flux density.  The dashed circle is based 
on the mean angular distance of the SiO masers from the center of the 
distribution.  The dark circle in the center illustrates the angular size of the 
continuum photosphere as determined from VLTI measurements \citep{WITTKOWSKI:07}. 
\label{SORI_VEL}}
\end{figure}

Figure~\ref{SORI_VEL} illustrates the problem of referencing optical/IR interferometry 
to radio interferometry results.  Shown are the results of a joint
Very Large Telescope Interferometer (VLTI) and Very Long Baseline Array (VLBA)
study of the Mira variable S~Ori \citep{WITTKOWSKI:07}.  The stellar diameter,
represented by the dark circle in the center, was measured with the VLTI while
the circumstellar SiO masers were imaged concurrently with the VLBA.  The 
referencing of the star to the masers is purely conjectural, however, 
with additional astrometric information from SIM, this assumption would become 
unnecessary.

A similar astrometric problem is demonstrated by recent H$_2$O maser observations
of disks toward silicate carbon stars.  In the case of the star V778~Cyg,
\citet{SZCZERBA:06} were able to use Tycho astrometric data to associate the
H$_2$O masers with an unseen companion orbiting the carbon-rich AGB star.
A similar disk has been observed for the carbon star EU And, also traced by
H$_2$O masers; however, it is impossible to determine whether the disk 
is associated with the AGB star with the available astrometric data.  With 
SIM astrometry, such a determination would be routine.

Finally, SIM will greatly improve the study of AGB stars by
providing the precise parallax distances essential to establishing a linear 
scale size for the star and the various regions of the CSE.  Knowledge of these 
scale lengths is important for theories relating to the chemistry of CSEs, the formation 
points of circumstellar masers and dust, and the strength of the stellar magnetic field.  
Furthermore, the linear velocity of circumstellar gas as traced by maser proper motions 
has yet to be accurately determined for many stars without precise distances.


\section{Stellar Evolution, Extragalactic Distances, and Galaxy Formation \label{CHAPTER7}}

The study of normal galaxy evolution is greatly enhanced not only by SIM projects that
target dynamics and dark matter, but also by those that target the distance
scale. We show in this Section that better distances translate to much more
precise information on the chemical and age structure of the various stellar
populations that make up the Galaxy and external galaxies. SIM can be used to
obtain parallax distances to Galactic (disk; Population I) clusters. This will
complement SIM's Population II (halo and thick disk) distance scale investigations
described in $\S$ \ref{CHAPTER9}.

A critical step in studying ages and chemical compositions of stellar populations is to establish a collection of standard clusters, mostly Galactic clusters, for which distances, reddenings, abundances, and ages will be derived with unprecedented accuracy.  SIM is critical to this task by providing accurate parallax distances.
These standard clusters can then be used
to tightly constrain theoretical isochrone sets more stringently than ever
before. The isochrones, in turn, give ages for clusters and also for galaxies
via integrated-light models, and precision studies of galaxy evolution are the
ultimate aims enabled by this SIM study. Morphological lookback studies
\citep[e.g., GEMS;][]{rix04} and spectroscopic surveys 
\citep[e.g., COMBO-17;][]{wolf04} dovetail nicely with stellar populations 
studies that earlier
predicted in a broad way what the direct observations are finding. That is, spiral galaxies look as if they have had quasi-continuous star formation
for long epochs, as expected, but elliptical galaxies, while mostly dead
today, have also had much more complex star formation histories than one would
suppose, not too drastically different from spirals \citep{worthey1998}. 

A plethora of questions regarding field versus cluster environment, chemical
evolution, morphological evolution remain, and are likely to remain for many
years. Present age errors intrinsic to isochrone-based models are of order
30\% \citep{charlot96}, but the clusters studied by SIM should allow
for increased precision for better understanding of galaxy evolution at
all redshifts. A reasonable goal in this regard is 5\% age precision for
favorable, well-observed extragalactic stellar populations from the next
generation of large, ground-based spectroscopic telescopes.

But an absolute 5\% age precision requires a much better grip on the
systematics of stellar populations than presently exists.   With precise distances from SIM, distance will no longer contribute significantly to the uncertainties, and instead other effects will dominate.  For example, the uncertainty in heavy
element abundance ($Z$) propagates approximately as $\delta {\rm
log\ age} = -3/2\ \delta {\rm log}\ Z$ \citep{worthey1994} using stellar
temperatures as age indicators, as one is forced to do for integrated-light
applications. This implies that overall heavy element abundance uncertainty be
less than 0.02 dex, a goal reached only rarely at present, but which should be
very common in the near future. The detailed, element-by-element composition
also matters.  \citet{worthey1998} estimates that abundance ratio effects need
to be tracked and calibrated if they induce more than a 7 K shift in stellar
temperature. Progress on such detailed effects is underway and should be
available in a few years. Progress on bolometric corrections, absolute flux
scale, and stellar color-$T_{\rm eff}$ relations can also be expected
shortly. What a standard cluster set does is to provide tie-down points
for stellar modelers, which relates intimately to the interpretation of
high-redshift stellar population studies.

SIM will measure parallax distances to the Galactic clusters
(supplemented by the globular cluster distances
described in $\S$ \ref{CHAPTER9}). The luminosity of the main sequence turnoff in the
color-magnitude diagram is the best age indicator: the one with the
smallest errors \citep{chaboyer1995} and the one that ties most
directly to the ``fusion clock'' of the hydrogen-burning
star. Distance uncertainty is currently the dominant uncertainty, and
that will be removed by SIM (to less than 1\% for most individual
Galactic clusters, and perhaps 1\% for the Globular clusters in
aggregate). After abundance effects, the remaining uncertainty is that
of interstellar extinction, which may prove to be the dominant
uncertainty in the end, although progress is being made in that area
as well.

The target clusters were chosen to fulfill the following science
goals. (1) We would like to see extragalactic stellar population age
estimates with 5\% absolute precision, at least for ``red envelope''
galaxies. This requires that isochrone sets be calibrated to the
standard cluster set, and that the clusters themselves have well-determined
ages (Table \ref{wortheytable}). (2) Of the many distance-scale issues that are benefitted by SIM,
another important one is the surface brightness fluctuation method
\citep{tonry1988,mei2005} that can be
used to chart local galaxy flows and matter distributions. This method
depends directly on the isochrone sets and is thus tied to the
standard cluster set. (3) In our Galaxy, the standard clusters can
be tied in to the photometry of the rest of the globular and open
cluster system in order to investigate the chemical and dynamical
history of the Galaxy.  In external galaxies this is filtered through the isochrone sets. (4) The cluster ages themselves are important. 
(5) Finally, the clusters were chosen to be as massive as
possible in order to attempt to populate the rarer,
post-hydrogen-burning phases of evolution. These phases are the
current frontier of stellar evolution, and the more constraints we can
place, the better off we are.

\begin{deluxetable}{lllrr}
\tablecaption{Clusters Selected for Population Studies\label{wortheytable}}
\footnotesize
\tablehead{
\colhead{Cluster} & \colhead{Distance (kpc)} & \colhead{E(B-V)} & \colhead{[Fe/H]} & \colhead{Age (Gyr)} 
}
\startdata
NGC 6528    & 9.1 & 0.6 & -0.2 & 12 \\
Palomar 6   & 7.3 & 1.5 &  $\sim$0.0 & 12 \\
NGC 6440    & 8.4 & 1.1 & -0.3 & 12 \\
Collinder 261 & 2.2 & 0.27 & -0.2 & 9 \\
NGC 6791    & 4.2 & 0.1 & 0.4 & 8 \\
Melotte 66  & 2.9 & 0.2 & -0.4 & 7 \\
NGC 6253    & 1.5 & 0.2 & 0.4 & 5 \\
Messier 67  & 0.8 & 0.02 & -0.1 & 4 \\
NGC 2420    & 2.2 & 0.02 & -0.4 & 4 \\
Berkeley 18 & 5.8 & 0.46 & 0.0 & 4 \\
NGC 6819    & 2.4 & 0.05 & 0.1 & 2 \\
NGC 7789    & 1.9 & 0.22 & -0.2 & 1.7 \\
IC 4651     & 0.9 & 0.15 & 0.1 & 1.5 \\
NGC 2243    & 4.5 & 0.05 & -0.5 & 1.1 \\
NGC 2477    & 1.2 & 0.3 & 0.0 & 1.0 \\
NGC 6134    & 0.9 & 0.4 & 0.3 & 0.9 \\
Messier 44  & 0.2 & 0.0 & 0.2 & 0.7 \\
NGC 1817    & 2.0 & 0.33 & -0.3 & 0.4 \\
NGC 2324    & 3.8 & 0.11 & -0.8 & 0.4 \\
NGC 2099    & 1.4 & 0.3 & 0.1 & 0.4 \\
\enddata
\tablecomments{This list is given in order of decreasing age. All parameters given are approximate.}
\end{deluxetable}

The primary selection criterion is to cover as much age versus
metallicity parameter space as the Galaxy allows. This makes
``oddball'' clusters with atypical abundances very important. For
instance, young and metal-poor clusters are rare in the Galaxy, so NGC
2243 becomes a very important cluster. The globular clusters do not
reach to supersolar abundance, so the old, metal-rich cluster NGC 6791
becomes a valuable tie-point. The clusters to be observed with SIM cover about
1.5 dex in age, and are listed in Table \ref{wortheytable}.


\section{Cepheids in the Milky~Way \label{CHAPTER8}}

SIM's contributions to Cepheid
science are at least fourfold: 1) approximate Cepheid distances are
fairly easily estimated, so that once a variable star is identified as
a Cepheid it will be useful for Galactic rotation-curve studies if a
SIM-based parallax and proper motion are available, 2) an accurate
distance calibration allows for an accurate determination of
extinction and metallicity effects on the inferred absolute
luminosity, 3) the physics of the pulsation mechanism (including the
mysterious amplitude decline of Polaris) can be studied in great
detail for those nearby Cepheids where the extinction is small and/or
well measured (such as in clusters), and for Cepheids that are members
of binary systems where accurate mass measurements can be made, and 4)
the changing of the color of the Cepheids during its pulsation phase
($\Delta (V-I) \sim 1.1 \, \Delta V$; \citet{O_EXOPTF_2007}, private
communication) could help calibrate SIM's color-dependent astrometric
terms, while on average, $\Delta V$ increases with period: $\Delta V
\sim\,\log{ P_{\rm days} }$ \citep[e.g.,][]{BCM2000}.

The second and third points are essential for our understanding and
usage of Cepheids as extra-galactic distance indicators. Currently,
our lack of detailed understanding of the physics of the pulsation
mechanism (i.e., the calibration of the period-luminosity-color
relation) yields galaxy distances which carry systematic uncertainties
of order $\pm 5$\% \citep{P2006_NGC55, M2006_NGC4258}. A
better understanding of the physics would likely result in smaller
systematic errors, and hence, in an easier method for determining accurate
distances to a large number of galaxies. For example, it has been
claimed that `bump Cepheids' can be used to determine distances
below the 2\% level. This method is based on a detailed analysis of
the light-profile and nonlinear pulsation models \citep[e.g.,][]{KW2006}. 

The Milky~Way is the only galaxy for which we can perform detailed
three-dimensional dynamical studies because all six phase-space
parameters can be determined for a number of tracers of the gravitational
potential. Such studies are essential for the interpretation of
velocity fields of external galaxies, especially those at high
redshift which are used to infer galaxy-formation scenarios. Young
stars such as Cepheids (age $\sim$50 Myr) are very sensitive to small-
and large-scale perturbations of the potential \citep{M1974}, and are
thus very useful to study the dynamical effects of, for example: 1)
the bar, 2) spiral structure, 3) the Gould Belt, and 4) the warp.  For
such studies, the apparent magnitude is not important, just the
distance and space velocity. 
Cepheids are useful for these kind of studies because they
can be identified based on their periodic signal \citep{MCS1998}. A
total of about 900 Galactic Cepheids are currently known
\citep{W1998}, while only the 200-odd nearest of these stars are typically used
in studies of Galactic dynamics \citep{Z2000, MCS1998, FW1997,
Pea1997, Pea1994, CC1987}. The Cepheid sample provides a unique
opportunity to perform very detailed studies of the dynamics of disk
galaxies. Many of these Cepheids are too distant for Gaia, but are easy targets for SIM. Because the binarity rate amongst
Cepheids is large \citep[$\gtrsim$ 80\%;][]{Sz2003}, it is crucial to
monitor the Cepheids astrometrically throughout the SIM mission.

Cepheids in the Milky~Way suffer a significant amount of extinction
($A_V$).  For example, the nearest 180 stars in the sample of Pont and
collaborators \citep{Pea1994,Pea1997} have $A_V = 1.7 \pm 1$ mag, where the
extinction correction is uncertain by about 0.1 magnitude. In general,
it is hard to determine extinction better than to $\pm$0.05 mag for
stars with Cepheid colors, even with the Gaia instrument suite
\citep{GAIA_Phot2006}. Currently, the extinction is estimated from an
intrinsic period-color relation \citep[e.g.,][]{LS1994, CC1986} which
is calibrated on Cepheids in open clusters. If more accurate methods
become available to determine extinction, the Galactic relation
between period, luminosity, color, metallicity, etc., would be
very-well calibrated. Cross-validation of such a calibration would be
available via the Cepheids in galaxies with rotational-parallax
distances (see \S\ref{sec:Realistic_Rotational_Parallaxes}): M\,31 and M\,33 (employing SIM) and the LMC (from Gaia data).

The `expanding photosphere' or `Baade-Wesselink' or
`Barnes-Evans' or `infrared surface brightness' method has been
used for many years to yield `geometric' distances for Cepheids. In
this method, one can equate the integral of the changing radial
velocity of the stellar envelope during the pulsation cycle to observed changes in
radius. In principle, this method is very accurate because the radial
velocities can be measured very accurately, while the radii of nearby
Cepheids can be measured employing ground-based interferometry or via
a surface-brightness color relation. However, this method also suffers
from zero-point issues \citep{G2005_CephISB} that may depend on, for
example, metallicity, period and pulsation mode. Thus, a large sample
of Cepheids with a range of physical properties is required to
establish this relation firmly. SIM could provide a much better
calibration of this method than Gaia because SIM can reach the
required distance accuracies at both faint and bright magnitudes.

A final, perhaps philosophical, point is that Cepheids are variable
stars, and it is through this variation that we can learn much more
about the internal structure and atmospheric physics than for
normal stars. For example, the confirmation by helioseismology of
the standard solar model firmly established neutrino oscillations,
and hence proved that neutrinos are massive.


\section{Accurate Ages and Distances for Population II Objects \label{CHAPTER9}}

The metal-poor stars in the halo of the Milky Way galaxy were among
the first objects formed in our Galaxy.  These Population II stars are
the oldest objects in the universe whose ages can be accurately
determined.  Age determinations for these stars allow us to set a firm
lower limit to the age of the universe and to probe the early
formation history of the Milky Way.  The age of the universe
determined from studies of Population II stars may be compared to the
expansion age of the universe and used to constrain cosmological
models.  Globular clusters (GCs) provide the best opportunity to determine
ages of Population II (hereafter Pop II) stars, as it is easy to
identify the various evolutionary sequences in a GC color-magnitude
diagram.  The main sequence turnoff (MSTO) luminosity is the best
stellar `clock' which can be used to determine the absolute ages of
GCs \citep[e.g.][]{demarque, rood, vandenberg,renzini, CH96}.

\begin{figure}[b!]
\epsscale{0.9}
\plotone{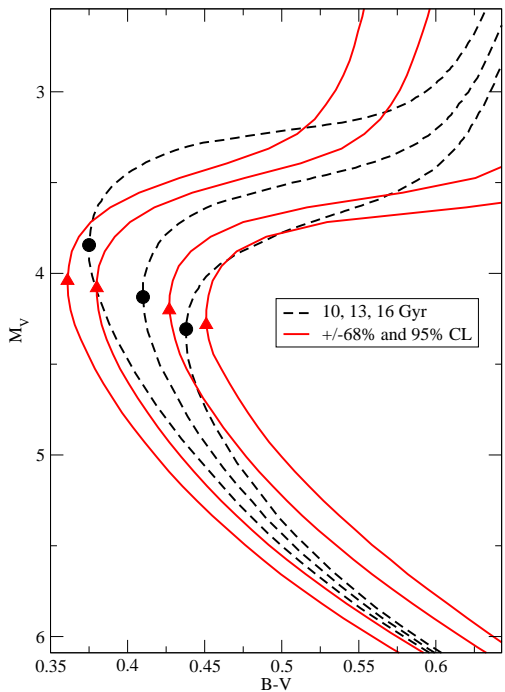}
\caption{Theoretical isochrones with [Fe/H] $= -1.6$ calculated using 
the stellar evolution code described in \cite{bjork}, showing the change in the position of the MSTO (solid circles) when the age of the standard 13 Gyr isochrone is changed by $\pm 3\,$Gyr (dashed curves).  The effects of changes in other parameters (e.g., nuclear reaction rates, opacities, treatment of convection, and oxygen abundance) were evaluated using the Monte Carlo simulation described in \S\,\ref{CHAPTER9}, and are shown as the $\pm 68\%$ and $\pm 95\%$ confidence levels in a standard isochrone (solid curves), with corresponding MSTO (triangles).
\label{iso2}
}
\end{figure}

The theoretical isochrones in Figure \ref{iso2} demonstrate how age
affects the color-magnitude diagram for a cluster of stars with uniform age and 
metallicity.  It is immediately apparent that the
MSTO and sub-giant regions are most sensitive to age differences.  The
MSTO becomes redder and fainter as a cluster of stars gets older.
Thus, in principle one could determine the age from the color of the
turn-off, independent of distance. 

The predicted colors of MSTO stars are subject to a great deal of
uncertainty.  To quantify this, we ran a Monte Carlo simulation to determine the uncertainty
in the calculation of isochrones.  First,  distribution functions for the various input parameters used in a stellar evolution code (such as the opacities, nuclear
reaction rates, treatment of convection, oxygen abundance, etc) are
determined based upon the known uncertainties in the determination of
each of the various quantities.  Isochrones are then calculated for a
given set of input parameters which are drawn at random from the
specified distribution function.  This procedure was then repeated 1119
times in order to determine how the known uncertainties in the input
parameters required for stellar evolution calculations affect the
theoretical isochrones.  The results may be represented as confidence contours in the color-magnitude diagram.  Full details are given in \cite{bjork}.
Figure \ref{iso2} illustrates the large uncertainty in the predicted colors
of the MSTO for a given age and metallicity, by comparing the Monte Carlo isochrones to standard isochrones with different ages.  The uncertainty in the theoretical 
calculation of the MSTO color leads to an error of approximately $\pm 2\,$Gyr
($1\,\sigma$).  In contrast, theoretical uncertainty in the MSTO luminosity
is considerably lower, of order $\pm 1\,$Gyr.

The largest uncertainty in the determination of globular cluster ages
based upon the MSTO luminosity is the distance scale for Pop II
objects.  A 1\% error in the distance leads to a $\simeq$2\% error in
the derived age \citep[e.g.,][]{CH96}.  SIM will be able to
determine distances to globular clusters and other stars in the halo
with unprecedented accuracy, thereby significantly reducing the
uncertainty in the derived ages of metal-poor stars.  Table \ref{gctable}
provides
basic data on the 21 globular clusters that will be observed with SIM.  These clusters were chosen based upon the following
properties: (a) distance from the Sun, (b) metallicity, 
(c) reddening, and (d) whether the cluster is thought to
belong to the Old Halo (OH), Young Halo (YH) or Thick Disk (TD)
(column 3). This grouping of globular clusters is based upon their
kinematics, metallicity and horizontal branch morphology \citep{dacosta}.

\begin{deluxetable*}{cccrcccc}
\tablewidth{0pt}
\tablecaption{Target Globular Clusters \label{gctable}}
\tablehead{
\colhead{NGC}   &   
\colhead{[Fe/H]}   &   
\colhead{Group}   & 
\colhead{Nrr}   &   
\colhead{E(B -- V)} &
\colhead{$\rm D_\sun$ (kpc)} &
\colhead{$\sigma_D$ (\%)} &
\colhead{$\sigma_{\mathrm{age}}$ (\%)}
}
\startdata
  6341&   $-2.29$&   OH& 25 &  0.02&     8.1&    4&    8\\       
  7099&   $-2.12$&   OH& 10 &  0.03&     7.9&    4&    8\\       
  4590&   $-2.06$&   YH& 41 &  0.04&    10.1&    5&    10\\
  6397&   $-1.95$&   OH& -- &  0.18&     2.2&    1&    5\\       
  6541&   $-1.83$&   OH& -- &  0.12&     7.4&    3&    8\\       
  6809&   $-1.81$&   OH& 10 &  0.07&     5.3&    2&    6\\       
  5139&   $-1.62$&   OH&152 &  0.12&     5.1&    2&    7\\       
  5272&   $-1.57$&   YH&260 &  0.01&    10.0&    5&    10\\
  6752&   $-1.55$&   OH& -- &  0.04&     3.9&    2&    5\\       
  6205&   $-1.54$&   OH& 3  &  0.02&     7.0&    3&    7\\       
  6218&   $-1.48$&   OH& -- &  0.19&     4.7&    2&    8\\       
  3201&   $-1.48$&   YH& 85 &  0.21&     5.1&    2&    8\\       
  5904&   $-1.29$&   OH&123 &  0.03&     7.3&    3&    8\\       
   288&   $-1.24$&   OH& -- &  0.03&     8.1&    4&    8\\       
   362&   $-1.16$&   YH& 13 &  0.05&     8.3&    4&    8\\       
  6723&   $-1.12$&   OH& 29 &  0.05&     8.6&    4&    9\\
  6362&   $-1.06$&   OH& 33 &  0.09&     7.5&    3&    8\\       
  6652&   $-0.85$&   YH& -- &  0.09&     9.4&    4&    10\\      
   104&   $-0.76$&   TD&  1 &  0.05&     4.3&    2&    5\\       
  6838&   $-0.73$&   TD& -- &  0.25&     3.8&    2&    7\\       
  6352&   $-0.70$&   TD& -- &  0.21&     5.6&    3&    9\\       
\enddata  
\tablecomments{
Values taken from the \cite{harris96} compilation, unless
otherwise noted.\hfill\\
Notes on individual columns:\hfill\\
Group: OH = Old Halo; YH = Young Halo; D = Thick Disk \citep{dacosta}\hfill\\
Nrr -- Number of RR Lyrae stars in the cluster (from compilation by Carney, 
private communication)\hfill\\
$D_\odot$ -- distance from the Sun; these values are uncertain by $\pm
10\%$.\hfill\\
$\sigma_\pi$ --  percent uncertainty in the parallax, assuming
$4.6\,\muas$ accuracy\hfill\\
$\sigma_{\mathrm{age}}$ -- percent uncertainty in the age estimate
including
contributions due to uncertainties in the distance determination, 
reddening, photometric zero-point and metallicity determination.
\vskip 8mm}
\end{deluxetable*}

For each globular cluster, we plan to observe approximately six red
giant branch stars with an accuracy of 7$\mu$as.  We will average
together the parallaxes of all the confirmed members to obtain a final
parallax to the cluster with an accuracy of 4 $\mu$as,
which supported by the expected accuracy of the SIM grid of about 
3 $\mu$as.  This will
determine the distances to the individual globular clusters with an
accuracy of 1\% to 5\% (column 7 in Table \ref{gctable}), 
which is a factor of 2 to
10 times better than currently achieved.  The current
uncertainties in the globular cluster distance scale is dominated by
systematic errors, while our distance scale will be dominated by
random errors.  This will allow us to average together the age of 
the most metal-poor globular clusters in our sample, thereby significantly 
reducing the uncertainty in the determination of the mean age of
the oldest globular clusters.  

In order to determine the expected accuracy in our absolute age
estimate for the oldest, most metal-poor globular clusters, a Monte
Carlo simulation was performed, similar to that outlined in
\cite{bjork}.  In this simulation, we varied all of the sources of
error in our age determinations within their expected uncertainties,
including: the reddening determinations, photometric zero-points,
parallax uncertainties, uncertainties in the exact composition of the
stars (helium abundance, oxygen abundance and iron abundances), and
uncertainties in the stellar models (including nuclear reaction rates,
opacities, treatment of convection, model atmospheres, diffusion,
etc).  The distribution function for each of the individual input
parameters was determined by a careful consideration of the expected
uncertainties in the various quantities when SIM will
deliver its final parallaxes.  The simulation showed that we will determine
the absolute age of the oldest globular clusters to an accuracy of
$\pm 3\%$, or $\pm 0.4\,$Gyr.

To study the relative age distribution of stars in the halo, 
SIM will  observe 60 metal-poor turn-off/subgiant branch stars in the field.  
To illustrate the expected accuracy of our relative age determinations for the field halo stars and the globular clusters in our SIM program, we ran a
Monte Carlo simulation which allowed for the true distance to the
object to vary within its current estimated uncertainties, and which
took into account the uncertainties in the SIM distance
determination, reddening determinations and in the chemical
composition of the stars.  We find that the field stars will have a
typical uncertainty of $\pm 0.6$ Gyr, while the globular cluster ages will
have an error of $\pm 0.9$ Gyr (column 8 in Table \ref{gctable}).  From these simulations we conclude that we will be able to detect age differences of the order of 1 Gyr between various stellar populations.


\section{Exploring Galactic Stellar Populations and Dark Matter on Galactic 
Scales \label{CHAPTER10}}

N-body simulations of the formation of structure in the Universe in the
presence of dark matter (and dark energy) show galaxies (and all large
structures) building up hierarchically.  The active merging history on
all scales demonstrated by high resolution, Cold Dark Matter (CDM)
numerical simulations has had remarkable success in matching the
observed properties of the largest structures in the Universe, like
galaxy clusters, but are a challenge to reconcile with the observed
properties of structures on galactic scales.  The Milky Way and its
satellite system is a particularly important laboratory for testing
specific predictions of the CDM models, most especially because high
accuracy astrometric observations enabled by SIM allow definitive tests
of dynamical effects specifically predicted by CDM.

SIM will make possible unprecedented opportunities to explore stellar
dynamics with a precision that will allow critical measurements of
gravitational potentials from Local Group size (see Section~\ref{CHAPTER11})
to Milky Way, dwarf galaxy and star cluster scales.  We
outline in more detail below several specific important SIM
contributions that will bear directly on tests of dark matter (DM) and the
evolution of galaxies like the Milky Way and its stellar populations.

\subsection{Probing the Outer Halo with Tidal Tails}\label{TidalTails}

\begin{figure*}[!ht]
\epsscale{1.0}
\plotone{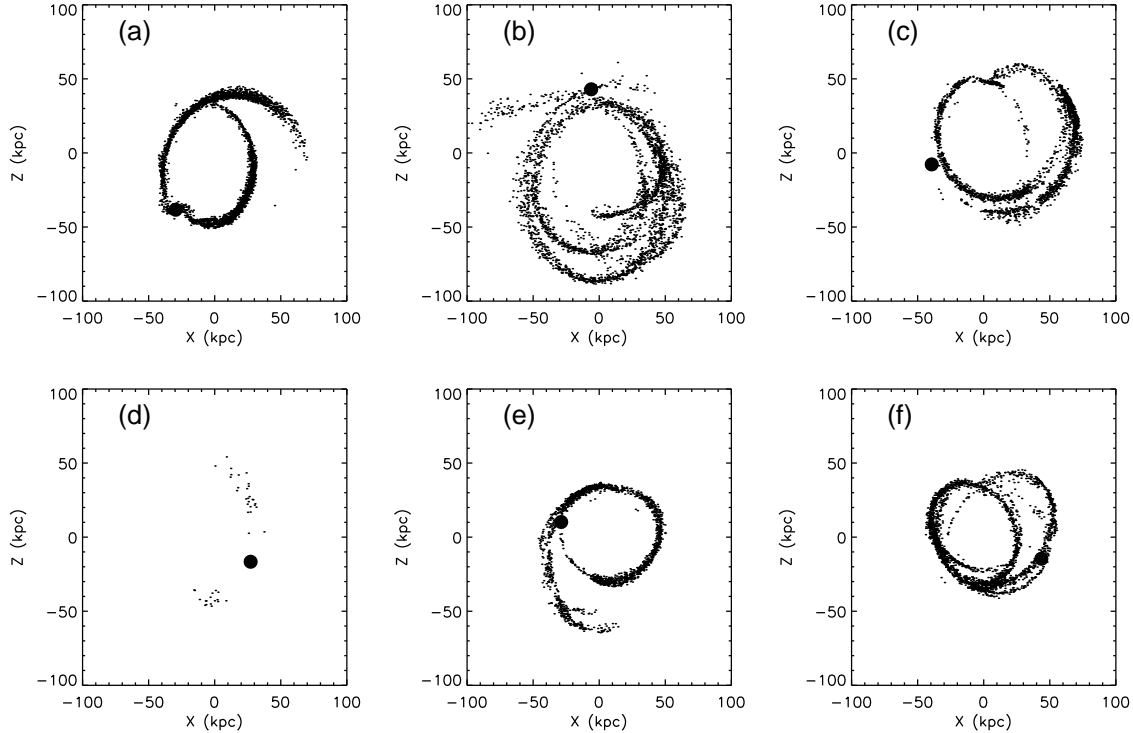}
\caption{A demonstration of the sensitivity of SIM to the Galactic 
  potential using stellar streams.  A Sagittarius-like tidal stream was
  created by the disruption of a dwarf satellite in a time-independent Galactic
  potential through a semi-analytical N-body simulation \citep{Law2005}.  This
  resulted in the tidal stream demonstrated in panel (a).  Complete 6-D
  phase space information on stars in the stream will be derived using
  the astrometric data from SIM.  With such data in hand, guesses may be
  made on the strength and shape of the Galactic potential, and the
  orbits of the individual stars in the tidal streams run backwards
  under these assumed potentials.  Panels (b) and (c) demonstrate what
  happens when the strength of the Galactic potential is underestimated
  by varying degrees: When the orbits are run backwards, the tidal
  stream stars orbit at too large a radius and do not converge to a
  common phase space position.  In panels (e) and (f) the strength of
  the Galactic potential has been overestimated, and when the clock is
  run backwards the tidal stream stars assume orbits that are too small
  and once again do not converge on a common phase space position.  In
  panel (d) a Galactic potential of the correct strength was guessed,
  and when the stream star orbits are run backwards, the tidal stream
  stars collect back into the core of the parent satellite.
\label{backwards_stream}
}
\end{figure*}

At large distances from the Galactic center ($>30$ kpc), the stellar
distribution is far from homogeneous. Standard methods of estimating the
depth of the Milky Way's gravitational potential using a tracer
population whose orbits are assumed to be random and well-mixed would be
systematically biased under these circumstances
\citep[e.g.,][]{Yencho2006}. However, these inhomogeneities themselves are
thought to have formed through the infall and disruption of satellites
and hence we actually have more information about the stars in these
lumps than in a truly random sample.  For example, we know that stars
that are clearly part of a stream of debris were once all part of the
same satellite.  We can use this knowledge to map the mass distribution
in the Galaxy. If we could measure the distances, angular positions,
line-of-sight velocities and proper motions of debris stars, we could
integrate their orbits backwards in some assumed Galactic potential.
Only in the correct potential will the path of the stream stars ever
coincide in time, position and velocity with that of the satellite
(see Fig.~\ref{backwards_stream}).

SIM measurements combined with ground-based 
line-of-sight velocities should provide 
everything needed to undertake the experiment; however, obtaining precision
trigonometric parallaxes of numerous distant debris stars would involve
a significant investment of SIM observing time.  On the other hand,
distances could be estimated either by using accurate photometric parallaxes (calibrated with SIM) for red giant/horizontal branch stars or by
exploiting our expectations for the orbital energy distribution in the
debris. In the latter case, we know the mean offset of the leading and
trailing debris from the satellite's own orbital energy
\citep{Johnston1998} and hence can solve the energy equation for the
distances to stars in each of these groups by assuming each has this
mean energy \citep{Johnston1999}. This should be more accurate than using
a photometric parallax so long as the offset in orbital energy is less
than $\delta d$ $d\Phi/dr$, where $\delta d$ is the uncertainty in
distances and $d\Phi/dr$ is the gradient in the assumed Galactic
potential.

If we find a coherent stream but not the associated satellite, the same
techniques apply, but with the parent satellite's position and velocity
as additional free parameters.  The usefulness of Galactic tidal streams
for probing the Milky Way potential has long been recognized, and the
results of astrometric space missions to provide the last two dimensions of
phase space information for stream stars has been eagerly anticipated
\citep[e.g.,][]{Johnston1999, Penarrubia2006}.

Applying this method to simulated data observed with the $\mu$as
yr$^{-1}$ precision proper-motions possible with SIM and $\kms$
radial velocities suggests that 1\% accuracies on Galactic parameters
(such as the flattening of the potential and circular speed at the Solar
Circle) can be achieved with tidal tail samples as small as 100 stars
\citep{Johnston1999,Majewski2006}.  Dynamical friction is not an important
additional consideration if the change in the energy of the satellite's
orbit in $N_{\rm orb}$ orbits is less than the range in the energies of
debris particles.  For $N_{\rm orb}=3$, this condition is met for all
satellites except the LMC/SMC and Sagittarius.  Evolution of the
Galactic potential does not affect the current positions of tidal
debris, which respond adiabatically to changes in the potential and
therefore yield direct information on the {\it present} Galactic mass
distribution independent of how it grew \citep{Penarrubia2006}.

Ideally, stars from several different tails at a variety of distances
from the Galactic center and orientations with respect to the Galactic
disk would be probed. It is also important to sample the tidal tails out
to points where stars were torn from the satellite at least one radial
orbit ago and hence, have experienced the full range of Galactic
potential along the orbit \citep{Johnston2001}. Finally, the
stars also need to have proper motions measured sufficiently accurately
that the difference between their own and their parent satellite's
orbits are detectable --- this translates to requiring proper motions of
order 100 $\mu$as yr$^{-1}$ for Sgr but 10 $\mu$as yr$^{-1}$ for
satellites that are farther away.

The possible existence of a significant fraction 
of the halo in the form of dark satellites has been 
debated in recent years
\citep{Moore1999,Klypin1999}.  These putative dark subhalos could
scatter stars in tidal tails, possibly compromising their use as
large-scale potential probes, but astrometric measurements of stars in
these tails could, on the other hand, be used to assess the importance
of substructure \citep*{Ibata2002,Johnston2002}.  Early tests of such
scattering using only radial velocities of the Sgr stream suggest a
Milky Way halo smoother than predicted \citep{Majewski2004etal}, but
this represents debris from a satellite with an already sizable
intrinsic velocity dispersion.
Because scattering from subhalos should be most
obvious on the narrowest, coldest tails (e.g., from globular clusters)
these could be used to probe the DM substructures, whereas the
stars in tails of satellite galaxies such as Sgr, with larger
dispersions initially and so less obviously affected, can still be used
as global probes of the Galactic potential.

In the last few years a number of well-defined Galactic tidal tails have
been discovered at a wide range of radii and, with SIM, can be used to
trace the Galactic mass distribution as far out as
the virial radius with an unprecedented level of detail and accuracy.
For example, M-giant stars associated with Sgr have now been traced entirely
around the Galaxy \citep{Majewski2003}, the globular clusters Pal 5 and
NGC\,5466 both have tidal tails traced to over 20 degrees from their
centers \citep{GrillmairJohnson2006,Belokurov2006c,GrillmairDionitas2006a}, two
apparently cold streams over sixty degrees long have been discovered in the
Sloan Digital Sky Survey
\citep{GrillmairDionitas2006b,Grillmair2006a,Belokurov2006a}, and
the nearby ``Anticenter", or ``Monoceros" stream
has been shown to be broken up into dynamically
colder ``tributaries" \citep{Grillmair2006b}.
In the outer Galaxy, there are suggestions of debris associated with the Ursa Minor,
Carina, Sculptor and Leo I dSphs, 
\citep{Palma2003,Majewski2005,Westfall2006,Munoz2006a,Sohn2007}, and 
evidence for outer halo debris from the Sgr dSph \citep{Pakzad2004}.
With SIM, for the first time it will be possible to probe the full
three-dimensional shape, density profile, and extent of (and substructure
within) an individual DM halo.

\subsection{Hypervelocity Stars \label{Hypervelocity}}

Hypervelocity stars (HVS) were postulated by \cite{Hills1988}, who
showed that the disruption of a close binary star system deep in the
potential well of a massive black hole could eject one member of the
binary at speeds exceeding 1000 km s$^{-1}$. HVSs can also be produced
by the interaction of a single star with a binary black hole
\citep{Yu2003}.  These remarkable objects have now been discovered:
\cite{Brown2006} report on five stars with Galactocentric velocities
between 550 and 720 km s$^{-1}$, and argue persuasively that these are
HVSs in the sense that they are ``unbound stars with an extreme velocity
that can be explained only by dynamical ejection associated with a
massive black hole". It is likely that many more HVSs will be discovered
in the next few years, both by ground-based surveys and by the Gaia
mission.

The acid test of whether these remarkable objects are HVSs is whether
their proper motions are consistent with trajectories that lead back to
the Galactic center. The magnitudes of the known HVSs range from 16 to
20, so their proper motions should be measurable by SIM with an accuracy
of a few $\mu$as yr$^{-1}$. At the estimated distances of these stars
(20 to 100 kpc) a velocity of 500 km s$^{-1}$ corresponds to a proper
motion of 1000 to 5000 $\mu$as yr$^{-1}$, so SIM should be able to
determine the orientation of their velocity vectors to better than 1\%.

If these measurements confirm that the HVSs come from the Galactic
center, then we can do more.  For many stars, SIM can measure all six
phase-space coordinates, but for HVSs, the orbits are far more tightly
constrained because we know the point of origin. \cite{Gnedin2005}
have pointed out that the non-spherical shape of the Galactic
potential---due in part to the flattened disk and in part to the
triaxial dark halo---will induce non-radial velocities in the HVSs of
5--10 km s$^{-1}$, corresponding to 10--100 $\mu$as yr$^{-1}$. 
Each HVS thus provides an
independent constraint on the potential, as well as on the solar
circular speed and distance to the Galactic center.

\subsection{Dark Matter Within Dwarf Galaxies}\label{DarkMatter}

Dwarf galaxies, and particularly dSph galaxies,
are the most DM-dominated systems known to exist.
Due to the small scale sizes (about 1 kpc) and large total
mass-to-light ratios (approaching $100$\,\msune/\lsune),
the dSphs provide the opportunity to study the
structure of DM halos on the smallest scales.
The internal structure of the dSphs, as well as their commonality
within the Local Group, make possible a new approach to determining
the physical nature of DM with SIM.

CDM particles have negligible
velocity dispersion and very large central phase-space density,
resulting in cuspy density profiles over observable scales
\citep{NFW1997, Moore1998}.
Warm Dark Matter (WDM), in contrast, has smaller central
phase-space density, so that density profiles saturate to
form constant central cores. Due to the small scale sizes of dSphs,
if a core is a result of DM physics then the cores
occupy a large fraction of the virial radii, which makes
the cores in dSphs
more observationally accessible than those in any other galaxy type.
Using dSph central velocity dispersions, earlier constraints on
dSph cores have excluded extremely warm DM,
such as standard massive neutrinos \citep{Lin1983, Gerhard1992}.
More recent studies of the Fornax dSph provide strong constraints
on the properties of sterile neutrino DM
\citep{Goerdt2006,Strigari2006}.

The past decade has seen substantial progress in measuring radial
velocities for large numbers of stars in nearby dSph galaxies
\citep{Armandroff1995, Tolstoy2004, Wilkinson2004, Munoz2005,
Munoz2006b, Walker2006}.
In all dSphs, the projected radial velocity dispersion
profiles are roughly flat as far out as they can be followed,
with mean values between 7 and 12 km s$^{-1}$.
 From these measurements, the DM density profiles
are obtained by assuming dynamical equilibrium and solving the
Jeans equation \citep{Richstone1986}.   In this analysis,
the surface density of the stellar distribution is required,
and in all dSphs these stellar distributions
are well-fitted by King profiles, modulo slight variations.
In the context of equilibrium models, the
measured velocity profiles typically imply at least an order of magnitude more 
mass in DM than in stars, and 
imply mass-luminosity ratios that increase with radius
  --- in some cases quite substantially \citep[e.g.,][]{Kleyna2002} ---
though at large radii tidal effects may complicate this picture
\citep{Kuhn1993, Kroupa1997, Munoz2005, Munoz2006a, Sohn2007}.
Knowing whether mass follows light in dSphs or the luminous components
lie within large extended halos is critical to establishing the
regulatory mechanisms that inhibit the formation of galaxies in
all subhalos (\S \ref{TidalTails}).

\begin{figure}[ht!]
\vskip 5mm
\includegraphics[scale=0.60,angle=0,bb=80 360 458 680]{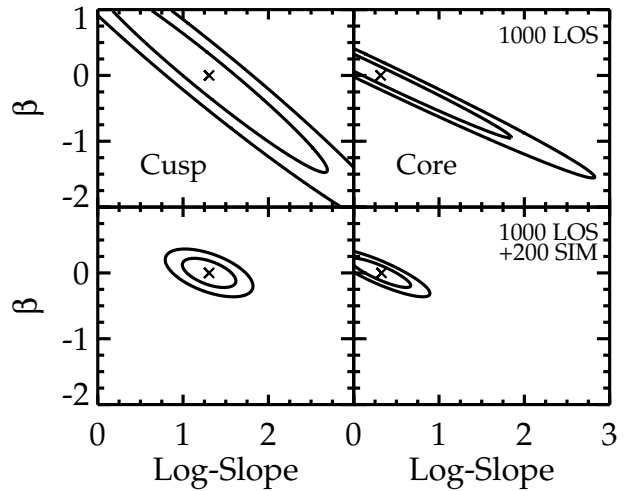}
\caption{A demonstration of the ability to recover information
on the nature of dark matter using observations of dSph stars, 
from analytical modeling by \citet{Strigari2007}.  Each panel shows the 
the $68\%$ and $95\%$ confidence regions for the errors in the 
measured dark halo density profile (log) slope (measured at twice the King core radius) and 
velocity anisotropy parameter $\beta$ for a particular dSph;  small crosses indicate the fiducial input model values.  In the upper plots, derived from 
line of sight (RV) velocities for 1000 stars, the two parameters are highly degenerate, for both a cusp model (left panel), and core model (right panel).
The addition of 200 proper motions from SIM providing
$5\,\kms$ precision transverse velocities (lower panels) dramatically reduces the uncertainty in both parameters. 
\vskip 3mm
\label{BETA_SLOPE}
}
\end{figure}

Unfortunately, for equilibrium models the solutions to
the Jeans equation are degenerate in that the dark
matter density profiles are equally well-fitted by both cores or cusps.
In particular, there is a strong degeneracy between the inner slope
of the DM density profile and the velocity anisotropy,
$\beta$, of the stellar orbits; this leads to a strong dependency
of the derived masses on $\beta$.
Radial velocities alone cannot break this
degeneracy (Fig.~\ref{BETA_SLOPE}),
even if the present samples of radial velocities are
increased to several thousand stars \citep{Strigari2007}.
The problem is further compounded if we add triaxiality, Galactic
substructure, and dSph orbital shapes to the allowable range of
parameters.

The only way to break the mass-anisotropy degeneracy
is to measure more phase space coordinates per star.
The Jeans equation written for the transverse
velocity dispersion probes the anisotropy parameter differently
than that for the radial velocity dispersion.
Thus, combining proper motions with the present
samples of radial velocities will provide orthogonal constraints
and has the prospect to break the anisotropy-inner slope degeneracy.

The most promising dSphs for this experiment
will be the nearby (60-90 kpc away) systems Sculptor, Draco, Ursa Minor, Sextans and Bootes
for which the upper giant branches require proper motions of stars with $V \simeq 19$.
The latter four of these dSphs include the most DM-dominated
systems known \citep{Mateo98,Munoz2006b} as well as a system with
a more modest $M/L$ (Sculptor).  While
Sagittarius is closer still, its strong interaction with the Galaxy and
obvious tidal distortions indicate a system clearly not in
dynamical equilibrium; thus it offers an interesting, possibly
alternative case study for establishing the internal dynamical effects
of tidal interaction. 
To sample the velocity dispersions properly will require
proper motions of $> 100$ stars per galaxy with accuracies of
7 km s$^{-1}$ or better (less than 15 $\mu$as yr$^{-1}$).
Detailed analysis shows
that with about 200 radial velocities and 200 transverse velocities
of this precision,
it will be possible to reduce the error on the log-slope of the dark
matter density profile to about 0.1 \citep{Strigari2007}.  This is
an order of magnitude smaller than the errors attainable from a sample
of 1000 radial velocities, and sensitive enough to rule out nearly
all WDM models (see Figure~\ref{BETA_SLOPE}).
Obtaining the required transverse velocities,
while well-beyond the capabilities of Gaia, is well-matched to the
projected performance of SIM.

\subsection{Galactic Satellite Proper Motions}\label{dSph_mu}

Knowing the bulk proper motions of the dSphs is key to modeling not only
their structure and evolution, but whether they correspond to the
predicted dark subhalos and how they relate to the two primary solutions
to the missing satellites problem, i.e., the ``very massive dwarf''
versus the ``very old dwarf'' pictures \citep*{Mashchenko2006}.  The
derived orbits of the dSphs can be compared directly to predictions for
the orbits of infalling DM substructure \citep{Ghigna1998,
Benson2005, Zentner2005}.  These orbits can also be used to test for
the reality of the purported dynamical families \citep*[e.g.,][]
{Lynden-Bell1995, Palma2002} hypothesized to be the product of the break
up of larger systems 
\citep{Kunkel1979,Lynden-Bell1982}, to determine the possible
connection of their apparent alignment to local filaments
\citep{Libeskind2005}, and to verify whether distant systems like Leo I
\citep[e.g.,][]{Sohn2007} are bound to the Milky Way and can be used as
test particles for measuring the mass profile of the Galaxy.

The proper motions of the more distant satellites of the Milky Way are
expected to be on the order of 100s of $\mu$as yr$^{-1}$, and to derive
transverse velocities good to $\sim10$ km s$^{-1}$ ($\sim10\%$) requires
a bulk proper motion accuracy of $\sim 10\, \muas$ yr$^{-1}$ for the most
distant satellites (Leo I, Leo II, Canes Venatici), in which the
brightest giant stars have $V \sim 19.5$.  The constraints are slightly
more relaxed for satellites at roughly 100 kpc distances, $\sim 20\,
\muas$ yr$^{-1}$ for systems with brightest stars at $V \sim\, 17.5$.
Such measurements for individual stars are well within the capabilities
for SIM, which could derive the desired bulk motions for Galactic
satellites with only a handful of stars per system (suitably placed to
account for rotation and other internal motions, \S \ref{DarkMatter}).

With  much less per star precision, but many  more stars, attempts to
measure  dSph galaxy  proper  motions from the  ground with long time
baselines  \citep*[e.g.,][]{Scholz1994,  Schweitzer1995,  Schweitzer1997},
with  the Hubble  Space Telescope and short  time baselines \citep[et
seq.]{Piatek2002},  and with combinations  of both  ground and  HST data
\citep{Dinescu2004}  still  lead   to  bulk  motions  measurements  with
uncertainties of  order the  size of  the motion of  the dSph,  and with
significantly different results (e.g., two recent measures of the proper
motion  of   the  Fornax  dSph   differ  at  the  2$\sigma$   level  ---
\citealt{Dinescu2004} versus \citealt{Piatek2002}).  Even for the closer
Magellanic  Clouds,  there  still  remain  significant  and  disturbing
variations in  the derived proper  motions \citep*{Jones1994, Kroupa1994,
KroupaBastian1997, Anguita2000,  Momany2005, Kallivayalil2006, Pedreros2006}.
These inconsistencies  not only reflect  the difficulty of  beating down
random uncertainties  with  $\sqrt N$  statistics  in systems  with
limited numbers of sufficiently bright  stars --- a problem that will be
severe in the most recently discovered dSphs, where the populations of giant
stars number at most in the tens  \citep[e.g.,][]{Willman2005b,
Belokurov2006b,  Zucker2006} --- but  in  dealing  with  numerous
systematic problems, 
such as establishing absolute proper motion zero-points  \citep[see discussion
in][]{Majewski1992, Dinescu2004}.   With its $\mu$as-level extragalactic
astrometric reference tie-in, and intrinsic per-star proper motion precision
SIM will make definitive measures of the proper motions of the
Galactic satellites that will  overcome the previous complications faced
by satellite proper motion studies.

The same stellar precisions conferred by SIM to
the study of distant dwarf satellites and star clusters (\S\,\ref{GC_mu})
may also be applied to proper
motions of individual distant field stars, whose full space motions can
be derived and used as additional point mass probes of the Galactic
potential, whether they are part of tidal streams or a well-mixed halo
population.

\subsection{Globular Cluster Proper Motions}\label{GC_mu}

Globular clusters, which have a spatial distribution that spans the
dimensions of the Milky Way and which readily lend themselves to
abundance and age assessments, have long served as a cornerstone stellar
population for understanding galaxy evolution.  Several distinct
populations of globular clusters are known: a disk \citep{Armandroff1989}
and/or bulge \citep{Minniti1996} population and at least two kinds of halo
clusters \citep{Zinn1996}.  Since \citet{Searle1978}, the notion that halo
globular clusters were formed in separate environments ---
``protogalactic fragments" --- 
later accreted by the Milky Way
over an extended period of time has been a central thesis of Galactic
structure studies.  More recently, direct substantiation of the
hypothesis for at least some clusters has come from the identification
of clusters that are parts of the Sagittarius stream \citep*{Ibata1995,
Dinescu2000, Majewski2004etal, Bellazzini2002, Bellazzini2003}
and the Monoceros structure \citep{Crane2003, Frinchaboy2004,
Bellazzini2004}.  These systems represent more obvious
``dynamical families" associated with recent mergers having readily
identifiable debris streams, but presumably such mergers 
were even more common in the early Galaxy.  Ancient mergers
may have tenuous stellar streams today, and their identification
will require 6-D phase space
information to locate objects of common energy and angular momentum.
Tracing ancient mergers via their identified progeny will provide key insights
into the evolution of substructure and star clusters in
hierarchical cosmologies \citep{Prieto2006}.

Of course, halo globular clusters will serve as valuable test particles
for determining the halo potential, but these cluster data will also
play an essential role in understanding clusters as stellar systems. The
dynamical evolution of small stellar systems is largely determined by
external influences such as disk and bulge shocks
\citep*[e.g.,][]{Gnedin1999}, so determining cluster orbits by measuring
their proper motions will dramatically improve our understanding of
their evolution and address the long-standing issue of whether the
present population of these systems is the surviving remnant of a much
larger initial population. The formation of globular clusters remains a
mystery that can be better constrained by 
understanding cluster orbits.
By obtaining definitive orbital data for the
entire Galactic globular cluster system, SIM will clarify the range
of extant cluster/satellite dynamical histories and how the
Galactic ensemble evolved and depopulated.


At present, $\lesssim$ 25\% of Galactic globular clusters have had {\it any}
attempt at a measured proper motion, and reliable data generally
exist only for those clusters closest to the Sun \citep*[see summaries
in][]{Dinescu1999, Palma2002}.  As in the case of the Galactic
satellites (\S\,\ref{dSph_mu}), even in the rare cases when appropriate
data for proper motion measurements exist, analyses are hampered by
critical systematic errors, notably the tie-in to an inertial reference
frame. Galaxies yield unreliable centroids and QSOs have too low of a
sky density at typical magnitudes probed.
Moreover, most of the outer halo globular clusters,
including some newly discovered examples \citep{Carraro2005, Willman2005a} have 
very sparse giant branches, reducing the
effectiveness of averaging the motions of numerous members to obtain a
precision bulk motion. SIM will immediately resolve these problems.


\section{Astrometric Microlensing \label{CHAPTER11}}

What would an unbiased census of Galactic objects, dark and
luminous, reveal?  At a minimum, it would yield the frequency
of black holes, neutron stars, and old brown dwarfs, which are
either completely dark or so dim that they defy detection by
normal methods.  It might also find a significant component of
the dark matter, although the majority of dark matter cannot
be in the form of compact objects \citep{alcock00,tisserand07}.  The
only known way to conduct such a census is to put a high-precision
astrometry telescope in solar orbit.

Masses of astronomical bodies can be measured only by the
deflections they induce on other objects, typically
planets and moons that orbit Solar System bodies and
binary companions that orbit other stars.  Masses of
luminous isolated field stars can be estimated from their
photometric and spectroscopic properties by calibrating
these against similar objects in bound systems.  Hence,
photometric surveys yield a reasonably good mass census of 
luminous objects in the Galaxy.

Dark objects like black holes are another matter.  Mass measurements
of isolated field black holes can be obtained only by their
deflection of light from more distant luminous objects.  Indeed,
it is difficult to even detect isolated black holes by any other
effect.  However, to go from detection to mass measurement
(and therefore positive identification) of a black hole is quite
challenging.

Gravitational microlensing experiments currently detect about
500 microlensing events per year.  The vast majority of the
lenses are ordinary stars, whose gravity deflects (and so magnifies)
the light of a more distant `source star'.  As the source gets
closer to and farther from the projected position of the lens,
its magnification, $A$, waxes and wanes according to the \citet{einstein36}
formula
\begin{equation} 
A(u) = {u^2 + 2\over u\sqrt{u^2+4}},
\quad u(t) =\sqrt{u_0^2 + \biggl({t-t_0\over t_\e}\biggr)^2},
\label{eqn:aofu}
\end{equation} 
where $u$ is the source-lens angular separation (normalized to the
so-called Einstein radius $\theta_\e$), $t_0$ is the time of maximum 
magnification (when the separation is $u_0$) and $t_\e$ is the Einstein radius
crossing time, i.e., $t_\e = \theta_\e/\mu$, where $\mu$ is the lens-source
relative proper motion.  The mass $M$ cannot be directly inferred from
most events because the only measurable parameter that it enters is $t_\e$,
and this is a degenerate combination of $M$, $\mu$ and the source-lens
relative $\pi_\rel$: 

\begin{equation} 
t_\e = {\theta_\e\over\mu},\quad
\theta_\e = \sqrt{\kappa M\pi_\rel},
\label{eqn:te}
\end{equation}
where $\kappa\equiv 4 G/(c^2\au)\sim 8\,{\rm mas}\,M_\odot^{-1}$.

It follows immediately that to determine $M$, one must measure {\it three} parameters,
of which only one ($t_\e$) is routinely derived from microlensing events.
Another such parameter is $\theta_\e$, which could be routinely measured from
the image positions, if it were possible to resolve their $O({\rm mas})$
separation.  A third is the ``microlens parallax'' 
$\pi_\e = \pi_\rel / \theta_\e$.
Hence the lens mass can be extracted from $\theta_\e$ and $\pi_\e$ alone
\citep[see e.g.,][]{natural}:

\begin{equation} 
M = {\theta_\e\over\kappa\pi_\e}.
\label{eqn:mval}
\end{equation} 

Just as $\theta_\e$ is the Einstein radius projected onto the plane
of the sky, $\pi_\e$ is related to $\tilde r_\e\equiv\au/\pi_\e$, the
Einstein radius projected onto the observer plane.  And just as $\theta_\e$
could, in principle, be measured by resolving the two images on the sky,
$\pi_\e$ could be routinely measured by simultaneously observing the
event from two locations separated by $O(\tilde r_\e)$
\citep{refsdal66,gould95}.  ``Routine'' measurement of both $\pi_\e$ 
and $\theta_\e$ is essential.  As of today, there have been a few dozen
measurements of these parameters separately
\citep[e.g.,][]{poindexter05}, 
but only one very exceptional
microlensing event for which both were measured together with sufficient
precision to obtain an accurate mass \citep{gould05}.

In fact, such routine measurements are possible by placing an accurate
astrometric and photometric telescope in solar orbit.  For current
microlensing experiments carried out against the dense star fields of
the Galactic bulge, $\pi_{\rm rel} \sim$40 $\mu$as, so for stellar masses,
$\theta_\e\sim 500\, \mu as$ and $\tilde r_\e\sim 10$ AU.  Hence,
a satellite in solar orbit would be an appreciable fraction of an
Einstein radius from the Earth. As a result, the photometric event described
by equation~(\ref{eqn:aofu}) would look substantially different than
it would from the ground.
From this difference, one could infer $\tilde r_\e$ (and so $\pi_\e$).

Determining $\theta_\e$ is more difficult.  As mentioned above, this
would be straightforward if one could resolve the separate images, but
to carry this out routinely (i.e., for small as well as large values of
$\theta_\e$) would require larger baselines than are likely to be
available in next-generation instruments.  Rather, one must appeal
to a more subtle effect, the deflection of the {\it centroid} of the
two lensed images.  This deflection is given by \citep{my95,hnp95,walker95}
\begin{equation} 
\Delta\theta = {u\over u^2 + 2}\theta_\e.
\label{eqn:deltatheta}
\end{equation} 
Simple differentiation shows that this achieves a maximum at $u=\sqrt{2}$, for
which $\Delta\theta = \theta_\e/\sqrt{8}$, roughly 1/3 of an Einstein radius.
Hence, if the interferometer can achieve an accuracy of O(10 $\mu$as)
{\it at the time when this deflection is the greatest}, then $\theta_\e$
can be measured to a few percent.

There are some subtleties as well as some challenges.  Satellite measurements
of $\tilde r_\e$ are subject to a four-fold discrete degeneracy, which
can only be resolved by appealing to higher-order effects \citep{gould95}.  
It is not
enough to measure the centroid location to determine the astrometric 
deflection: one must also know the undeflected position to which the
measured position is to be compared, and this can only be found by
extrapolating back from late-time astrometry.  And the precision of the
mass measurement depends directly on the signal-to-noise ratio of the
underlying photometric and astrometric measurements.  This is important
because space-based astrometric telescopes are likely to be photon challenged
and so to require relatively bright (and hence rare) microlensing events to
provide accurate mass measurements.  
\citet{gs99} estimated that $\simeq 1200$ hours of SIM time would yield 5\% mass measurements for about 200 microlenses.
Most of these lenses will be stars, but at least few percent are likely
to be black holes, and several times more are likely to be other dark
or dim objects like neutron stars, old white dwarfs, and old brown
dwarfs.  Since such a census is  completely new, it may also turn
up unexpected objects.
\vskip 7mm


\section{Dynamics of Galaxy Motions: Numerical Action and SIM \label{CHAPTER12}}

If one could measure the proper motions of galaxies with global 
accuracies of a few $\muasyr$, one could obtain another two 
components of phase space with which to construct flow models and 
determine histories and masses for galaxies and galactic groups.  
For a galaxy 1 Mpc away, $4 \muasyr$  corresponds to $19\, \kms$
transverse motion, which is small compared to the expected transverse 
motions in the field, $\sim100\,\kms$.  With an instrument such as SIM, one 
could measure accurate positions of a few dozen stars, with 
$V < 20$, 
in each galaxy and after $\sim$5  years obtain proper motions with adequate 
accuracy.  For a typical dwarf galaxy, after averaging randomly 
located stars, the contribution to the error from the internal motions 
would be only a few $\kms$.  For larger galaxies, simple rotation models, 
adjusted to the observed velocity profiles, can be removed from the motions 
for $<20\, \kms$ accuracy. There are 27 galaxies known (all within 5 Mpc) that 
have stars sufficiently bright.

Just beyond the outermost accessible 21-cm isophotes of galaxies, 
the dark matter distribution becomes unknown. Within the light-emitting 
parts of galaxies, rotation curves are flat and not falling according to Kepler's 
law, which implies that mass grows roughly linearly with radius.  The total 
mass-to-light ratio depends critically on where this mass growth ends, but 
this is generally not observed.  As a result, critical questions about dark 
matter on scales of galaxies to groups remain:  Do dwarf galaxies have lower 
or higher mass-to-light ratios than regular galaxies? Do the dark matter halos 
of galaxies in groups merge into a common envelope?  How do these mass components 
compare with the warmer dark matter particles smoothly distributed across 
superclusters or larger scales.  At present, we can only detect dark matter 
through its gravitational effects; therefore, a careful study of the dynamics 
of nearby galaxies is one of the few ways to resolve these issues.  

SIM measurements of the deviation from Hubble flow will be of lasting importance
in the modeling of  the formation of the Local Group, several nearby groups and 
of the plane of the Local Supercluster.  Note that with Gaia only M31 and M33 have
sufficient numbers of stars that are bright enough to attempt a proper motion 
determination.  If Gaia achieves $25\, \muasyr$ (the presently stated goal) 
or $\sim100\,\kms$ at 1 Mpc, it will obtain only $\sim$\,1-$\sigma$ detections for these two galaxies.

\subsection{Peculiar Velocities}

Most analyses of peculiar velocity flows have applied linear perturbation 
theory (appropriate for scales large enough that overdensities 
are $\ll 1$) to spherical in-fall \citep{peebles80}.  Peculiar velocity analysis 
\citep{Shaya92,dekel93,ph05} have proven the general concept that 
the observed velocity fields of galaxies result from the summed gravitational 
accelerations of overdensities over the age of the universe.  They also agree 
with virial analyses of clusters and WMAP observations that indicate the 
existence of a substantial dark matter component strewn roughly where the 
galaxies are.  But these studies apply only to large-scales of $>$10 Mpc,  the 
scales of superclusters and large voids.  Spherical infall (including 
``timing analysis" and ``turnaround radius") studies do not presume low 
overdensities and have been applied to the smaller scale of the Local Group 
\citep{belllin77}.  These studies indicate mass-to-light ratios for the 
Local Group of roughly $M/L \sim\, 100\,\Msun/\Lsun$,
but the model is crude; non-radial motions and 
additional accelerations from tidal fields and sub-clumping are expected to be 
non-negligible and would substantially alter the deduced mass.  A more complete 
treatment of solving for self-consistent complex orbits is required.

\subsection{The Numerical Action Method}

The application of the Numerical Action Method   
\citep[NAM,][]{peebles89} allows one to solve for the trajectories that result 
in the present distribution of galaxies (or more correctly, the centers 
of mass of the material that is presently in galaxies).  By making use of 
the constraint that early time peculiar velocities were small, the problem 
becomes a boundary value differential equation with constraints at both early 
and late times. For each galaxy, a position on the sky, either a redshift or 
a distance and an assumed mass (usually taken to be the luminosity times an assumed 
mass-to-light ratio) are required inputs.  In addition, one must presume an 
age of the universe (which is known from WMAP).   Several studies comparing 
NAM with N-body solutions have shown that NAM can recover accurate orbits 
and positions or velocities \citep{branchini02, phelps02, rom05}.  
\citet{Sharpe01}   have used NAM to predict distances from 
redshifts and then compared these to Cepheid distance measures.

\begin{figure}[ht!]

\vskip 2mm
\epsscale{2.45}
\plottwo{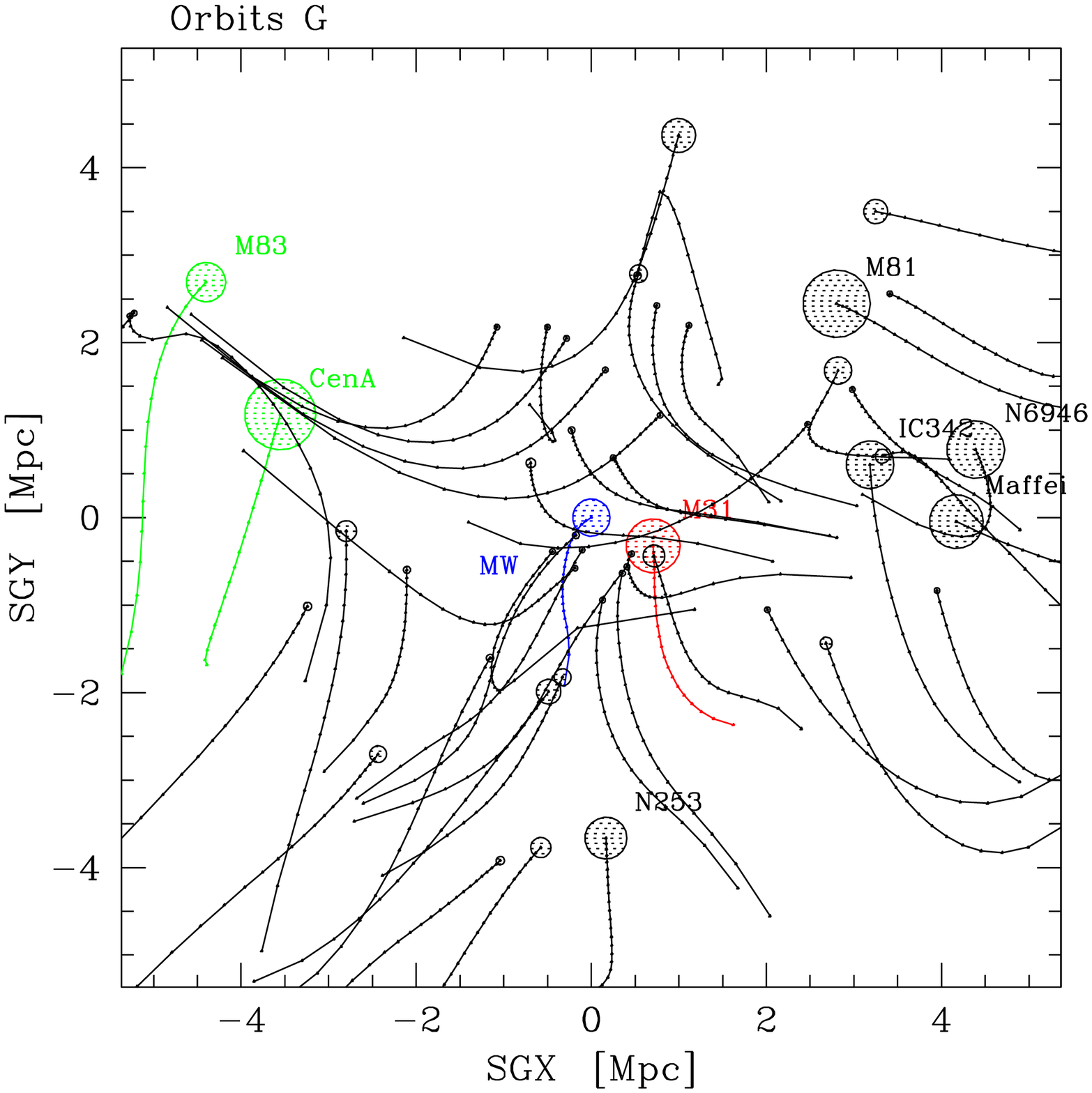}{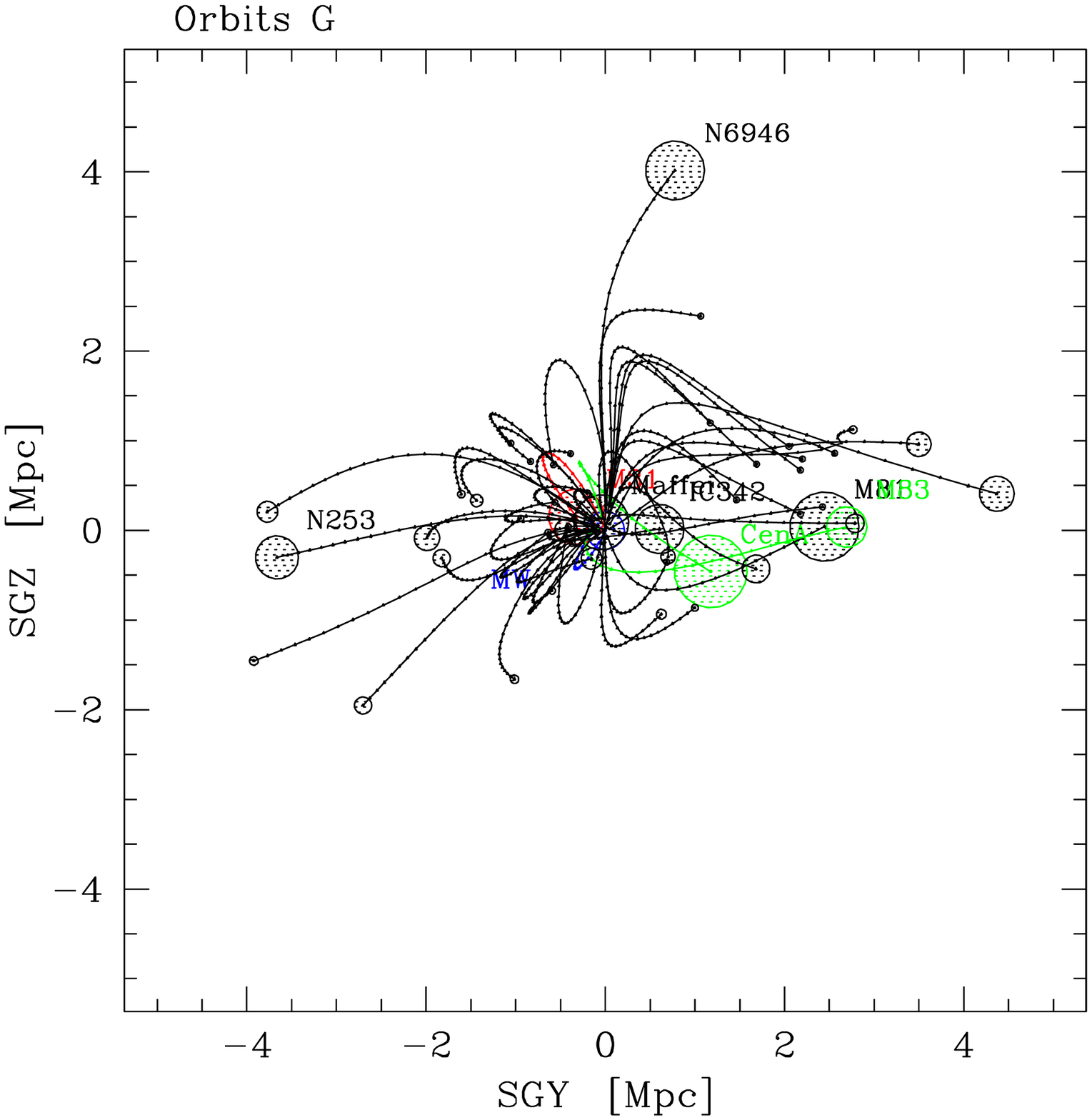}

\caption{Upper -- The trajectories of nearby galaxies and groups going out to the distance of the Virgo Cluster from a Numerical Action Method calculation with parameters M/L = 90 for spirals and 155 for ellipticals, $\Omega_m=0.24$, $\Omega_{\Lambda}=0.76$.  The axes are in the supergalactic plane (SGX-SGY) in comoving coordinates. There are 21 time steps going from $z=40$ to now. The large circle is placed at the present position and the radius is proportional to the square root of the mass. The present estimated distances were fixed and the present redshifts were unconstrained.  Lower -- A view of the galaxy trajectories, except that the coordinates here are real-space (proper) rather than comoving, and the axes are SGY-SGZ allowing a view of the collapse into a plane. 
\label{gal1}
\vskip 5mm}
\end{figure}

Figure~\ref{gal1} (upper) shows the output of a recent NAM calculation for the orbits 
of nearby galaxies and groups going out to the distance of the Virgo Cluster.  
The orbits are in comoving coordinates.  This is just a single solution of a 
set of several solutions using present 3-d positions as inputs.  The four 
massive objects (Virgo Cluster, Coma Group, Cen\,A Group, and M\,31) have been 
adjusted to provide best fit to observed redshifts.  Figure~\ref{gal1} (lower) is the same calculation, but now the coordinates are real-space rather than comoving, and  the supergalactic plane is viewed edge-on.  It appears that the plane of the Local Supercluster was not created by matter raining in from large distances but rather that material has never achieved great heights away from the plane.

Assuming a constant mass-to-light ratio provides only a rough guess for the 
galaxy masses. Adding proper motions measured by SIM will allow us to
solve for the actual masses of the dominant galaxies. 
The dwarf galaxies are essentially massless test particles of the potential.  
Also, the two components of proper motion would be extremely useful for determining 
several supplementary but crucial parameters: the halo sizes and density falloff 
rates at large radii, the mass associated with groups aside from that in the 
individual galaxies, and the amount of matter distributed on scales larger than 
5 Mpc.

\subsection{Preliminary Science Data}

Ongoing ground- and space-based observations should help in the analyses of nearby galaxy 
dynamics.  Accurate relative distances of nearby galaxies, often with accuracy as 
good as 5\%, are being made at a rapid rate using methods such as Cepheids, tip of the red giant branch, 
eclipsing binaries, maser distances, etc.  SIM should help reduce distance 
errors by providing better and therefore more consistent calibrations of these 
techniques.  In addition, measurements of the proper motions of a few galaxies 
(only M\,33 and IC\,10 thus far) using masers supplement our knowledge. For the fortuitous 
cases in which both a maser and SIM measurements are made for the same galaxy, 
the maser information can be averaged to beat down errors from internal motions.  
However, it is unlikely that future maser proper motion measurements, if any, 
will be in the same galaxies as those to be measured by SIM and thus they may 
provide completely complementary information.

\subsection{Future NAM Studies using SIM Data}

It appears that the mass-to-light ratio for systems with early type galaxies is 
higher than systems with late type galaxies.  Cluster mass-to-light ratios and 
elliptical galaxy X-ray data tend to $M/L \sim\, 300\,\Msun/\Lsun$ while group virial masses, turnaround radii, and the timing analysis lead to ratios of 50-100 in the field.   In  NAM calculations, we are beginning to see this trend as well, with 
Virgo and Cen\,A requiring  ratios of $\sim\, $450 and ordinary groups at $\sim\, $120.  
A detailed study of the flow on the 5 Mpc scale with all three components of velocity 
that SIM would allow, will be able to determine if the normal groups have additional 
dark matter at larger radii that could bring the baryon-to-dark matter ratios of the 
two species into better alignment.

The total masses of galaxies and the sizes of their halos are essential parameters for 
an understanding of dark matter and large-scale structure formation.  It may be that 
the mass of the dark matter particle is in the range in which it is neither completely
cold nor hot. If this is so, the clumping scale of the dark matter will provide a 
mass estimate of the dark matter particle.  Or it may be that the dark matter particle is mildly dissipative.  In which case, the halos will be flattened and the orbits of 
dwarf galaxies in different planes about some massive galaxy would be subjected to 
different effective masses. These would be important clues to revealing the physics 
and identity of the dark matter particle.  If we combine the constraints on the 
$1 - 5$ Mpc scale with flow studies on larger scales, it will be possible to strongly 
constrain a constant density component, resulting in a new astrophysical limit to the 
mass of the neutrino.


\section{Quasar Jets and Accretion Disks \label{CHAPTER13}}

Galaxies that possess spheroidal bulges also appear to possess central supermassive
black holes whose mass is of order 0.1\% of the bulge.  Episodic accretion of gas, dust and stars onto the central black hole gives rise to the active galactic nucleus phenomenon.   The potential SIM discovery space for AGN observations is broad because direct measurement 
on scales less than $\sim100\, \muas$ has previously been done only at radio wavelengths.
Most of what is known about the optically-emitting inner regions of
AGN comes from optical variability studies of the Broad Line Region: the intrinsic size-scales are very small (light days to a light year,
or $\sim\,0.1 - 10\, \muas$) in extent.  The ability of SIM to measure
motions and position differences on microarcsecond scales means that
we can study AGN on scales of tens to hundreds of Schwarzschild radii, perfect for studying accretion disks,
jet collimation and possible orbital motion of binary black holes.

Some key questions that SIM can answer are:
(1) Does the most compact optical emission from an AGN come from an
accretion disk or from a relativistic jet?  Does this depend on whether
the AGN is radio loud or radio quiet? (2)
Do the cores of galaxies harbor binary supermassive black
holes remaining from galaxy mergers? (3) Does the separation of the radio core and the optical photocenter
of the quasars used for the reference frame tie change on the timescale of
the photometric variability, or is the separation stable?   The use of quasars in the astrometric grid to approximate a perfectly inertial frame is described in Appendix~\ref{APPC}.
We first briefly review the basic properties of AGN and the physics of the
current accretion disk-jet paradigm, and then describe how 
SIM will answer these questions, including examples of what SIM
should see for specific targets.

\subsection{Basic Quasar Properties: Radio Quiet and Radio Loud Sources}

The observational signature of a quasar is an optically-emitting source
that is very bright intrinsically (up to $10^{13}$ \lsune)
but very small in physical size ($\sim 10^{15} \, {\rm cm}$).
At a 1 Gpc distance, for example, the size of the
central quasar engine is of order $0.1 \, \muas$ and that of the broad
line region of order $1 \, \muas$.
About 90\% of
all quasars are `radio quiet'; their emission is dominated
by optical and X-ray emission.  The optical continuum spectrum
consists of a fairly steep power-law, sometimes with a ``Big Blue
Bump'' in the blue or near ultraviolet region \citep{band89,Zheng95}.
In addition, there are broad emission
lines from highly-ionized elements, which are produced rather close to
the central source ($\sim 10^{16 - 17} \, {\rm cm}$) and ionized by it.

The remaining 10\% of quasars are `radio loud';  they  have strong diffuse radio emission in addition to all the properties of radio quiet quasars, with radio
jets  extending on both sides of the optical source out to the external
radio lobes.  In many cases these jets flow at relativistic speeds  (Lorentz factors of 10 or more).  About 10\% of these (i.e., about 1\% of all quasars)
are blazars --- strong and variable compact radio sources, which also emit in the optical (mainly red and near infrared), and in X-rays and $\gamma$-rays.  They are believed to be normal radio-loud quasars viewed by us nearly end-on to the jet.
Relativistic beaming toward the observer can produce an enhancement of an order of magnitude or more in apparent radio luminosity, as well as apparent proper motions of jet components of up to 1000 $\muas \, {\rm yr^{-1}}$.  If they display
similar internal proper motions in their optical jets, these would be readily detectable with SIM.

\subsection{The Location of the Most Compact Optical Emission in the Accretion Disk-Jet Paradigm\label{astrometric-shift}}

\begin{figure*}[ht!]
\epsscale{0.91}
\plotone{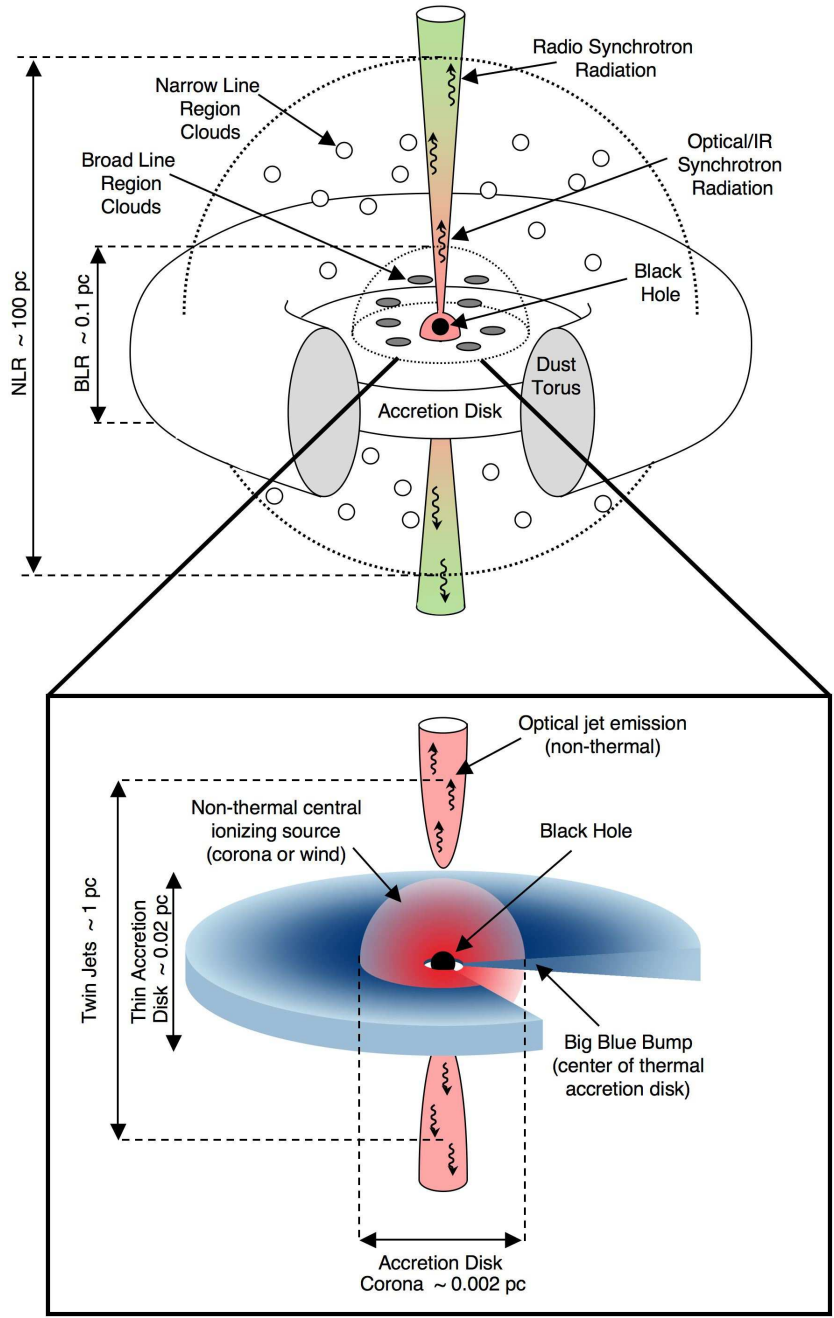}
\caption{Schematic diagram of the structure of a typical quasar on scales from $0.002 - 100$ pc, after \citet{Elvis2000} and Niall Smith (private communication, Cork Institute of Technology). {Upper ---} On scales from $\sim 0.1 - 100$ pc, the main components are the broad- and narrow-line emission clouds and a dust torus that obscures the black hole in certain orientations. In radio-loud quasars, powerful (and often relativistic) jets are ejected from the central engine \citep{konigl81}.  These jets emit over the entire EM spectrum, with optical/IR dominating on scales $\lesssim 0.1$ pc, and radio emission dominating on larger scales.
{Lower ---} On scales $\lesssim 1$~pc, a geometrically thin disk forms around the massive accreting black hole.  In some quasars, the innermost portion of the accretion disk is seen as the `big blue bump' \citep{band89}.  All quasars also have a source of non-thermal emission, a corona, which dominates in the red, and produces enough UV radiation to ionize the broad and even narrow line regions.  In radio-loud quasars, the base of jet also will produce non-thermal, typically highly-variable, red emission that can outshine the central corona.  SIM can probe the sub-parsec structure using time-dependent and color-dependent astrometry (see \S\,\ref{qso-color}). The blue disk and red corona are very compact and spatially coincident, so SIM would not expect to detect time- or color-dependent astrometric shifts in these components.
However, if the jet dominates, then a color shift of $\sim 0.5$~pc (for a $z = 0.6$ quasar; see text) would be seen, and any variability would be along the direction of the jet.  SIM will be able to test these predictions for radio-quiet and radio-loud quasars. \label{fig:jet_disk}
}
\end{figure*}

In order to understand how SIM observations can provide insights into the physical processes in AGN, we need to briefly review the major components and current constraints on their parameters.
A sketch of the canonical AGN model is presented in
Figure~\ref{fig:jet_disk}.   Observations on this size scale are indirect, so this picture is based largely on theoretical models.   There are three
possible sources of compact optical emission: the accretion disk, the
disk corona or wind, and the relativistic jet.  For a number of
nearby AGN, SIM will be able to distinguish which component dominates and to study AGN properties on scales of 1 pc or less.

\subsubsection{The Big Blue Bump and Hot Corona}

In radio quiet quasars, there should be two sources of
optical-UV emission:  1) thermal accretion disk emission, which produces the `big blue bump', and which dominates in the blue; and 2) a nonthermal corona whose origin is a steep power-law ionizing source, which dominates in the red.  Both of these emission regions should be physically centered on the black hole within $\sim 1
\muas$, and both should have a similar $\sim$1 $\muas$ size.

In the high-accretion case, the disk produces a thermal
peak in the near ultraviolet region.  For a typical $10^9 \, M_{\odot}$
black hole system, accreting at 10\% of $\dot{M}_{Edd}$, the diameter
of the portion of the disk that is radiating at a temperature of $10^4
\, K$ or above is $\sim 3.6 \times 10^{16} \, {\rm cm}$, or 0.012 pc 
\citep{shakura73}.  At the distance of M\,87 (16 Mpc) this region would subtend an angular size of $\sim
160 \, \muas$, while at moderate redshift ($z = 0.5$) 
the angular size would be only $\sim 2 \, \muas$.  

The hot corona is believed to be responsible for exciting the observed emission lines of the high-ionization species of carbon, silicon, nitrogen and oxygen, 
\citep{osterbrock86}.  
Coronal emission, therefore, is likely non-thermal ---
either optical synchrotron or inverse-Compton-scattered emission from a radio
synchrotron source.  Models of this ionizing source \citep[e.g.,][]{band89} 
indicate a size of only $\sim 70$ Schwarzschild radii.
At moderate redshift (e.g., 3C\,345, $z = 0.6$) this subtends an angular size of
only $\sim 1\, \muas$, centered on the black hole and comparable in size to the Big Blue Bump.

\subsubsection{The Beamed Relativistic Jet in Compact Radio-Loud Quasars}

Radio-loud quasars
possess  powerful radio-emitting jets, in addition to the central optical-UV source seen in radio-quiet quasars.   These jets are often relativistic flows, ejected
in a direction approximately perpendicular to the accretion disk (see Figure~\ref{fig:jet_disk}) by the effects of magnetic fields \citep{meier01,mckinney06}.  
If an AGN is radio-loud, and viewed at a small angle to the jet
axis (a `blazar'), then a third source of
optical continuum emission may be seen, in addition to the big blue bump and hot disk corona: emission from the relativistic jet beamed toward the observer.
This emission would be readily detectable by
SIM, because it would be offset with respect to the other sources of emission, and highly 
variable.  The \citet{konigl81} model for relativistic jets predicts
that the majority of the optical emission comes not from
synchrotron emission from the base of the jet but 
from synchrotron-self-Compton emission in the region of the jet where
the synchrotron emission peaks in the radio or millimeter \citep{hut86}.  A detailed application of this model to 3C\, 345, for example, predicts the optical emission to be offset $\sim 80 \, \muas$ from the center of mass of the system and nearly coincident with the 22 GHz radio emission which lies
$\sim 70 \, \muas$ or $0.4 \, {\rm pc}$ from the black hole \citep{unwin94}.  Not only is this readily detectable with SIM, but the vector position of the variability on the sky can be compared with the jet orientation seen on larger scales.

\subsection{Distinguishing AGN Models using Color-Dependent Astrometry
\label{qso-color}}

SIM can directly test the above modes by measuring a color-dependent shift in the astrometric position.  A displacement in optical photocenter 
between the red and blue ends of the passband can be readily measured, and is very
insensitive to systematic errors; measurements will be primarily limited by photon noise not instrument errors.  This color-dependent displacement, and its time-derivative in variable sources, is a vector quantity on the sky whose alignment can be compared to, say, the orientation of a radio jet imaged by VLBI or the VLA.

In radio-quiet quasars we do not expect to see a color shift, because of the absence of any contribution of a relativistic jet whose optical emission might introduce an astrometric asymmetry.   The red emission from the corona and the blue emission from the disk both should be coincident with the central black hole within $\sim 1\, \muas$.   Any color-dependent astrometric shift seen in radio-quiet quasars would challenge the current models of accreting systems in AGN.

By contrast, the astrometric position may be strongly color-dependent in any object with strong optical jet emission.  So  while the blue end of the spectrum
should be dominated by the thermal disk, the red region may be dominated
by the beamed relativistic jet. The relative contributions may vary as the activity level changes \citep[see][]{Ferrero2006}. Furthermore, we would expect any variability to be aligned on the sky with the direction of the color shift itself.

An example of a moderate-redshift jet-dominated quasar is 3C\,345 ($z = 0.6$) for which the red optical jet emission should be
roughly coincident with the 22 GHz radio emission, at about $80\, \muas$
from the black hole.  For 3C\,273 ($z = 0.16$), the separation may be as large as $300\, \muas$.  Not only is such a large shift readily detectable, but SIM could also detect variations in the offset with time. 

In the  nearby radio galaxy M\,87 we expect the  red optical emission should be
dominated by the accretion disk corona because its jets are not pointing
within a few degrees of our line of sight.  Therefore, SIM should
not see a significant color shift in this source.  However, M\,87
is so close that we might see an absolute position shift between
the measured radio photocenter and the overall optical photocenter.
This shift should be even larger for this low-redshift radio galaxy than
for 3C\,345 --- perhaps in the $1000\ \muas$ range.

\subsection{Finding Binary Black Holes}

Do the cores of galaxies harbor binary supermassive black holes
remaining from galaxy mergers?  This is a question of central importance
to understanding the onset and evolution of non-thermal activity in
galactic nuclei.  SIM can detect binary black holes in a manner analogous to
planet detection:  by measuring positional changes in the quasar optical
photocenters due to orbital motion.   If a quasar photocenter traces an elliptical path on the sky, then it harbors a binary black hole;  if the  motion is
random, or not detectable, then the quasar shows no evidence of binarity.  
If massive binary black holes are found, we will
have a new means of directly measuring their masses and estimating the
coalescence lifetimes of the binaries.

\begin{figure}[ht!]
\epsscale{1.12}
\plotone{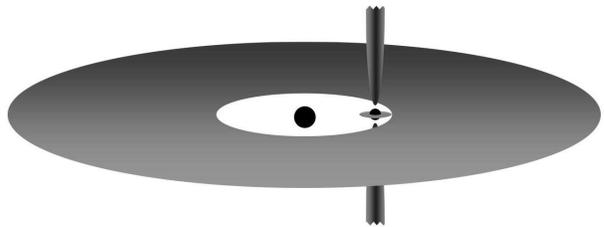}
\caption{Schematic diagram of a binary black hole system.  Binary separations of order $0.01 - 1$ pc may be expected. While the larger object will have a large disk, the
smaller will clear out the center of the disk, rendering the larger hole quiet, and leaving only the lower-mass black hole as a source of ionizing radiation and jets, as in Fig.~\ref{fig:jet_disk}.  The orbital shift of the secondary hole will be readily detectable by SIM especially for nearby quasars ($z \lesssim 0.2$). Expected
orbital periods for a $10^9\, $\msun primary range from $\sim 3$ yr for a 0.01 pc orbit to $\sim 90$ yr for a 0.1 pc orbit.
\vskip 2mm
\label{fig:binary}
}
\end{figure}

An AGN black hole system (see Fig.~\ref{fig:binary}) can occur near the
end of a galactic merger, when the two galactic nuclei themselves merge.
Time scales for the nuclei themselves to merge, and the black holes to
form a binary of $\sim1$ pc in size, are fairly short (on the order of several
million years) and significantly shorter
than the galaxy merger time (a few hundred million years).  Furthermore,
once the separation of the binary becomes smaller than 0.01 pc,
gravitational radiation also will cause the binary to coalesce in only
a few million years \citep{krolik99}.  However, the duration of the
`hard' binary phase (separation of 0.01 -- 1 pc) is largely unknown,
and depends critically on how much mass the binary can eject from the
nucleus as it interacts with ambient gas, stars, and other black holes
\citep{yu02,mer05}.  Depending on what processes
are at work, the lifetime in this stage can be longer than the age of the
universe, implying that binary black holes are numerous -- or as short as
the ``Salpeter'' time scale $\sim\, (M_1+M_2)/\dot{M_{Edd}}\ \sim 5\, \times 10^7
{\rm yr}$, implying that binaries might be rare.  Therefore, the search
for binary black holes in the nuclei of galaxies will yield important
information on their overall lifetime and on
the processes occurring in galaxies that affect black holes and quasars.

A promising candidate would be an object that looks like a quasar (with broad lines and optical-UV
continuum) but with absolute luminosity somewhat less than
10 \% of  the Eddington limit expected from the central black hole.
These might be objects with a large, but dark, primary central
black hole that is orbited by a secondary black hole of smaller mass;
the secondary would have cleared out the larger hole's accretion disk
interior, but still will be accreting prodigiously from the inner edge of
the primary's disk \citep{mil05}.  In this case, the
astrometric motion would indicate the full orbit of the secondary about
the primary, which could be a few to a few hundred $\muas$, depending
on the source distance. Figure \ref{fig:binary} is a simplified sketch of this situation.  OJ\,287 ($z = 0.3$) is a candidate, based on brightness 
variations with 12-year periodicity \citep{kidger00,valtonen06}, and we estimate 
that about 14 $\muas$ of orbital motion may be expected during a five-year span.

\section{Cosmology with SIM \label{CHAPTER14}}

In this section, we describe some contributions that SIM can make to
the fields of cosmology and dark energy (DE). There are at least three
distinct types of contributions: 1) the measurement of accurate
stellar ages which provide a robust lower bound for the age of the
Universe ($t_0$), 2) the estimation of the primordial Helium
abundance, and 3) the determination of percent-level distances to
several galaxies in the Local Group. Stellar ages and helium
abundances are best determined from spectrophotometry and accurate
parallaxes of eclipsing binaries \citep{L2000}.  For the present
purpose, we will assume that the Local-Group distances are equivalent
to determining the Hubble constant ($H_0$) itself. In fact, a robust
maximum value for the equation of state of dark energy derives from
the product $t_0$ and $H_0$ \citep{BHM2006}.

An error level for H$_0$ of one to two percent is optimal for
determining the EOS of Dark Energy and the total energy density of the
Universe. Errors of the order of 1\% and 0.1\% for $w$ and
$\Omega_{tot}$, respectively, are achievable with Planck-like CMB data
and a 2\% error on H$_0$. Assuming $\sqrt{N}$ statistics, 2\% distance
errors may be achieved with SIM observations of 43 stars in M\,31, 200
stars in M\,33 and 2,025 stars in the LMC.  Local-Group distances do
not determine $H_0$, rather they allow for a re-calibration of
secondary distance indicators that extend beyond the Local Group
such as Cepheids and the TRGB (see \S\ref{CHAPTER7}). The disks of these galaxies
encompass a wide range in metallicity, extinction and number density,
and thus provide ample opportunity for an accurate calibration of
methods such as the TRGB, Cepheids, RR~Lyrae, surface-brightness
fluctuations and so forth.

\subsection{The Extragalactic Distance Scale}
 \label{sec:The_Extra_Galactic_Distance_Scale}

When assuming a flat $\Lambda$\,CDM (cold Dark Matter) model, the fluctuations of the
cosmic microwave background (CMB) as observed by WMAP imply a value of
H$_0 = 73 \pm 3\, \kmsMpc$ \citep{WMAP03,WMAP07}; hereafter referred to
as WMAP07). This value is very close to the HST-derived value of $H_0$
= 74 $\pm$ 2 (random) $\pm$ 7 (systematic) $\kmsMpc$\ \citep{H0_HST}.
However, if the flatness assumption is abandoned, WMAP by itself
hardly constrains $H_0$ because WMAP measures the product of the
normalized matter density ($\Omega_m$) and $h^2$ (WMAP07, their Figure
20), as well as the baryon density ($\Omega_b h^2$). Here, $h \equiv
100 \, \kmsMpc / H_0$ and $\Omega_m h^2 \equiv \omega_m = 0.126 \pm\
0.009$. Thus, an independent and accurate determination of $H_0$ would
determine $\Omega_m$ and $\Omega_b$, and, if we assume a flat
universe, the dark-energy density ($\Omega_{DE}$).

While an accurate determination of $H_0$ has many advantages, here we
will concentrate on those with direct cosmological implications: 1)
determining the equation of state (EOS) defined as the ratio of pressure 
to density ($w\equiv p/\rho$) of dark energy (WMAP07), and 2) determining the
total density ($\Omega_{tot}$) of the universe. The equation of state
of dark energy tells us something about its nature: the cosmological
constant, strings, domain walls and so forth predict different values
for the equation of state \citep[e.g.,][]{PR2003}.

According to Hu (2005, hereafter H05), a change in
the EOS of Dark Energy by 30\% leads to a change in $H_0$ by 15\% if a constant
$w$ is assumed, and by about one-half that much if $w$ 
changes with redshift (H05, Figures 3a and 3b). Thus, an accuracy of
1\% in $H_0$ leads to a percent-level determination of the EOS of
dark energy. However, the absolute value of $w$ and its rate of change
can not be independently determined by fixing $H_0$. In fact, once
$H_0$ is fixed locally, the determination of $w$
and/or its slope and/or the curvature term requires sub-percent level
determination of $H_0$ at $z \sim 0.1 - 1.5$ (H05, Figures 3c and 3d).

As discussed above, if the CMB parameters are infinitely well known,
we expect that uncertainties in the EOS of Dark Energy remain at the level of
the uncertainty in $H_0$. Below, we estimate analytically the effects
of increasing both the accuracy of the CMB data and of $H_0$. 
The results of this exercise are consistent with the H05 findings. 
Because SIM is a targeted mission, a SIM-based
determination of $H_0$ can obtain, more or less, any required accuracy
if an appropriate amount of observing time is allocated.  Thus, our
analytical results are useful in quantifying more precisely the
resulting accuracy in the EOS of dark energy given a certain
expenditure of SIM time, and vice versa.

WMAP and other data currently constrain the EOS: $w \sim (-0.826
\pm 0.109) - (0.557 \pm 0.058) \, \omega_m \, h^{-2} \sim -0.95 \pm
0.11$, where we follow WMAP07 and assume a constant $w$, but allow for a 
non-zero curvature term \citep{O_RPDE_2007}. 
This follows from various relations between the
vacuum energy ($\Omega_\Lambda$), the matter density, the spatial
curvature of the universe ($\Omega_K$) and the Hubble
constant. Likewise, current data yield: $\Omega_{tot} =
\Omega_\Lambda + \Omega_m \sim (0.9438 \pm 0.0114) + 0.225 \,
\omega_m \, h^{-2} \sim 0.996 \pm 0.016$. Thus, current data allow
for the determination of the total density of the universe to plus or
minus one percent, while the EOS of dark energy is known to about
10\%.

To eliminate the uncertainty associated with $h$, one would want to
determine the Hubble constant via trigonometric parallaxes of nearby
galaxies. Unfortunately, this is not possible with foreseeable/planned
technology. However, the  `Rotational Parallax' method 
(see \S\,\ref{sec:Rotational_Parallax_Distances}) should be almost as good
\citep[][hereafter referred to as OP2000]{PS1997,OP2000}. 

\citet{O_RPDE_2007} estimates the effects of more accurate CMB data and
a better Hubble constant on the EOS of Dark Energy and finds that,
even at Planck accuracy, the errors on $w$ and $\Omega_{tot}$ are only
slightly smaller than the current values.  The accuracy of the EOS of
Dark Energy only improves significantly when the accuracy of $H_0$ is
improved.

\begin{figure}[th!]
\epsscale{1.17}
\plotone{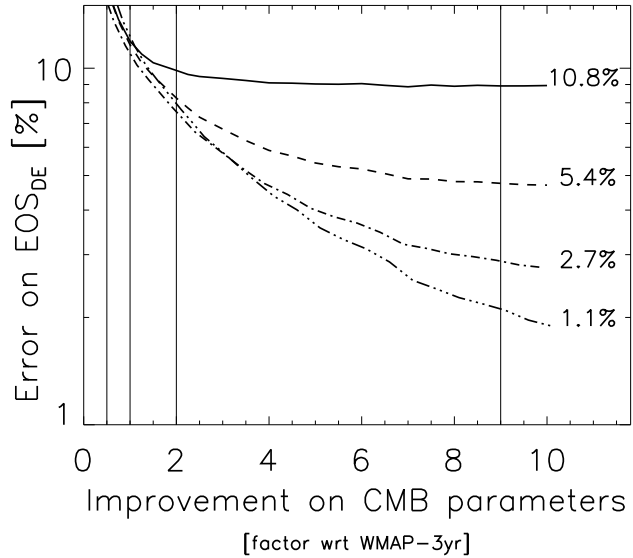}
\caption{The accuracy of the equation of state (EOS) of Dark
Energy $w$ as a function of the
uncertainty in the CMB data, essentially the factor by which 
the error on $\omega_m$ decreases, for four values of accuracy of the Hubble Constant (curves labeled  in \% accuracy on H$_0$).  
Vertical lines (from left to right) indicate the accuracies in the CMB parameters 
using WMAP 1-yr, 3-yr, and 8-yr data, and expected Planck data
\citep{PLANCK} normalized to the 3-year WMAP data.   
The resulting $w$ errors are 8.9\%, 4.8\%, 3.0\%
and 2.3\% for CMB accuracies at the level expected from Planck.  A significant improvement in $w$ requires both Planck CMB data {\it and} a more accurate Hubble Constant.
\label{fig:Cosmology}
} 
\end{figure}

The results are summarized in Figure~\ref{fig:Cosmology} which shows
the attainable accuracy on $w$ as a function of the improvement of our
knowledge of the CMB, with respect to the WMAP 3-yr data, for four values of $H_0$ accuracy. As
expected, the error on $w$ ($\epsilon_w$) decreases as knowledge of
the CMB parameters improves. However, these curves show that
$\epsilon_w$ approaches a limiting value. The curve that corresponds
approximately to the current error on $H_0$ ($\epsilon_{H_0} \approx $11\%,
top curve) indicates that an error of only about 9\% can be obtained
for $w$, even with Planck-like CMB accuracies \citep{PLANCK}. Note that this 
accuracy is only slightly better than our current knowledge as set by the WMAP 3-yr data. 
In order to significantly improve our knowledge of $w$, we
need Planck data and also need $H_0$ to much better accuracy.  
Achieving $\epsilon_w = 2.3 \%$, requires an accuracy of 1.1 \% on $H_0$.   
Errors on $\Omega_{tot}$ behave much like those on $w$, but with roughly ten
times smaller amplitude.

Further improvements in our knowledge of the Dark Energy EOS may come from 
improvements in observations other than $H_0$, though reducing the uncertainty in 
$H_0$ makes the biggest difference.  The other data sets used by WMAP07 are: large-scale structure
observations, galaxy redshift surveys, distant type-Ia Supernovae,
Big-Bang nucleosynthesis, Sunyaev-Zel'dovich fluctuations,
Lyman-$\alpha$ forest, and gravitational lensing. The Dark Energy Task
Force \citep{DETF2006} considers how future instruments and space missions might significantly reduce the error in $\epsilon_w$  through improvements in these datasets in four stages.  It defines Stage I to be the current state of the art, and Stage IV to comprise a Large Survey Telescope (LST),
and/or the Square Kilometer Array (SKA), and/or a Joint Dark Energy Mission (JDEM). Each higher stage represents an
improvement of $\epsilon_w$ by a factor of about three with respect to
the previous stage \citep{DETF2006}. It concludes that when these
data are of Stage IV quality, a reduction in the error on H$_0$ by a
factor of two matters at most 50\% for the accuracy of the Dark Energy EOS. 
However, \citet{O_RPDE_2007} argues that the effects of smaller
errors on $H_0$ are especially important at the early stages, estimating that a decrease of $\epsilon_{H_0}$ by a factor of ten yields an improvement of $\epsilon_w$ by factors of: 3.9, 3.0, 2.4 and 1.6 for Stages I, II, III and IV, respectively.

\subsection{Rotational Parallax Distances}
 \label{sec:Rotational_Parallax_Distances}

The method of Rotational Parallaxes (RP) combines proper motions
($\mu$) and radial velocities ($V_r$) of stars in external galaxies to
yield bias-free single-step distances with attainable accuracies of
the order of one percent. This method is analogous to the orbital
parallax technique \citep{Aea1992}.

For a nearby spiral galaxy at distance $D$ (in Mpc) which is inclined
by $i$ degrees and with a rotation speed of $V_c\, \kms$, the proper motion
due to rotation is ${V_c}/(\kappa D)\, \muasyr$, with $\kappa
\sim 4.74\ \kms$/(AU yr$^{-1}$). For M\,33, M\,31, and the LMC we find:
$\mu$ $\sim$24, $\sim$74 and $\sim$192 $\muasyr$, respectively, which
are easily measured by SIM.  A simplified RP method uses stars on the
major and minor axes at similar galactocentric distances of a galaxy
with a flat rotation curve.  The minor-axis proper motion measures the circular
velocity divided by $D$, while the ratio of minor to major-axis proper motions is simply  $\cos{i}$ \citep{PS1997}.  However, stars need to be close to the
principal axes, making it difficult to find enough targets.
Generalizing this method we find:

\begin{equation}
D = \frac{V_{\rm r}}{\kappa} \, 
   \sqrt{ \frac{-y'/\mu_{y'}}{x\mu_{x} +y' \mu_{y'}} }
\, ; \label{eqn:D_S}
\end{equation}

\noindent with $x$ and $y'$ the major-axis and minor-axis position of
individual stars in the target galaxy \citep[OP2000,][]{O_RPDE_2007}. The
attainable distance errors per star are estimated to be 13\%, 28\% and
90\% for M\,31, M\,33 and the LMC, respectively.
\vskip 7mm

\subsection{Realistic Rotational Parallaxes}
 \label{sec:Realistic_Rotational_Parallaxes}

The rotational parallax method allows distances to be made to 
M\,31, M\,33 and the LMC in a single step, to accuracies of about 0.92, 2.0 and 6.4\% respectively.  
Achieving these bias-free single-step distances requires careful modeling of non-circular motions (due to spiral-arm
streaming motions, perturbations from nearby galaxies, a bar, and warps) which could otherwise bias the distance determination (OP2000).   These
effects are likely to be significant for the LMC.  In order to achieve
errors of several percent, it will be necessary to correct for any
sizable deviations from circular motion.  OP2000 show that a
correction can be achieved with SIM-quality proper motions. For a disk geometry,
the combination of SIM proper motions and ground-based
radial velocities will yield four of the six phase space
coordinates for individual stars, where the missing components
can be chosen to be, for example, the vertical displacement ($z$) of
the star with respect to the galaxy plane and the average $z$
velocity \citep{O_RPDE_2007}. For M\,31 and M\,33, this
assumption is likely to be reasonable. Furthermore, we can make
different assumptions regarding the  fifth and sixth phase-space
parameters and require consistency between the results. Thus,
the rotational parallax method will yield very reliable
distances. This contrasts other techniques that claim to yield
extra-galactic distances at the few-percent level such as eclipsing
binaries, Cepheids, nuclear water masers and so forth. In fact, these
other techniques rely on additional assumption and/or inaccurate
slopes and/or zero points, albeit that many of those problems are
likely to be calibrated by SIM and Gaia \citep[see review in][]{O_RPDE_2007}.

Because non-circular motions can be correlated on large scales, a
substantial number of stars need to be observed (spread out over a
large area of the galaxy) to be able to apply reliable
corrections. OP2000 estimate that at least 200 stars are required to
achieve the 1\% distance error for M\,31 noted above. 
\vskip 7mm

\subsection{Other Local Group Distances}
 \label{sec:Other_Local_Group_Distances}

SIM can be used to apply the ``orbital parallax" technique to binary stars in
other nearby Local Group dwarf galaxies. This would increase the
number and the range of types of galaxies that can be used to
calibrate other rungs of the distance ladder. Let us estimate the
required accuracies by neglecting inclination effects and
eccentricity. Then, the orbital velocity, $v$, the semi-major axis,
$a$, and the period, $P$, for each of the components yield the
distance:

\begin{equation}
D_{100kpc} \hspace*{0em} =  \hspace*{0em} 
   10 \, \kappa \, \frac{ 2 \pi a_{\muas} } {P_{yr} \, v_{km/s}} 
   \,  \sim \, 
   298 \, \frac{ a_{\muas}}{ P_{yr} \, v_{km/s}}
   \label{eqn:D_orbital}
\end{equation}

\noindent where $v_{km/s} = \kappa 2\pi a_{AU}/P$ and
$a_{AU}=a_{\muas}/D_{Mpc}$. Because the distance error ($\Delta_D$)
scales according to $(\Delta_D/D)^2 = (\Delta_a/a)^2 +(\Delta_P/P)^2 +
(\Delta_v/v)^2$, a
1\% distance error requires that
the errors on semi-major axis ($\Delta_a/a$),  period ($\Delta_P/P$), and  radial velocity ($\Delta_v$) are all at the sub-percent level.

Given sufficient observing time, the period and the
orbital velocity and their errors can be determined from the ground
with the required accuracy. Also, short
period binaries are unlikely to have survived the expansion of the
primary, while long-period orbits will be poorly sampled during the
SIM mission and ground-based observations, so we will assume periods
between 2 and 10 years. We simulate a population of binaries at 100
kpc with a 1~M$_\sun$ primary. Secondaries are drawn from the stellar
initial mass function and are in circular orbits with a period
distribution according to \citet{DM1991}. We assume a 100\% binarity
rate. The result is that 1\% of stars lie in the required period range
(median is 7 yr) if $a_{\muas} \ge 12.5$. For these systems, the
median projected orbital velocity is $\overline{v}_{o;km/s} \sim
1.9$. To achieve the 1\% distance accuracy goal, SIM would need $(100/12.5)^2=64$
observations per star, while the radial-velocity program would need to
reach an accuracy of 19 m~s$^{-1}$. At 138 (200) kpc, only 0.4\%
(0.042\%) of binaries satisfy these criteria and have:
$\overline{P}_{yr} \sim$ 7.9 (9.3), $\overline{a}_{\muas} \sim$ 14.9
(13.2) and $\overline{v}_{o,km/s} \sim$ 1.9 (2.5), where 138 kpc
corresponds to the Fornax distance.

Excluding the Sagittarius dwarf and the Magellanic Clouds, we identify
five galaxies within 100 kpc in the \citet{Mateo98}
compilation of local group dwarfs, and seven within 200 kpc. If we make the optimistic
assumption that the total luminosity of these systems comes only
from stars 1 mag below the Tip of the Red Giant Branch (TRGB), then
these systems have about 10 TRGB binaries with the right properties
per galaxy.  These are all fainter than $V\sim18$ and would be fairly expensive in SIM observing time, but binaries in the Sagittarius dwarf, the Magellanic Clouds and
possibly the Sculptor and Fornax dwarfs are observable.


\section{Imaging with SIM \label{CHAPTER15}}

SIM will demonstrate synthesis imaging at optical wavelengths
in space, thereby showing the viability of this approach for the design of the
next generation of UV/Optical/IR imaging telescopes for space astronomy.
The telescopes we now have in space at these wavelengths (GALEX, HST, \& Spitzer) are all of the ``filled aperture'' type, which ultimately limits their angular resolution to that
dictated by the size of the largest rockets available for launching them. High
sensitivity is automatically provided with such high-resolution instruments, but
precision collecting area is expensive and may not always be required. Synthesis
imaging using phase-stable UV/Optical/IR interferometers provides, for the first
time, the possibility of separately choosing the  resolution and the collecting
area of space telescopes in order to provide a more cost-effective match to the
specific astrophysical problems to be addressed. Such flexibility is essential
for the future if instruments are to provide ever-increasing angular resolution
and still be affordable. SIM will break new ground by demonstrating these
imaging techniques at optical wavelengths in space.

\begin{figure}[ht!]

\vskip 5mm
\epsscale{1.1}
\plotone{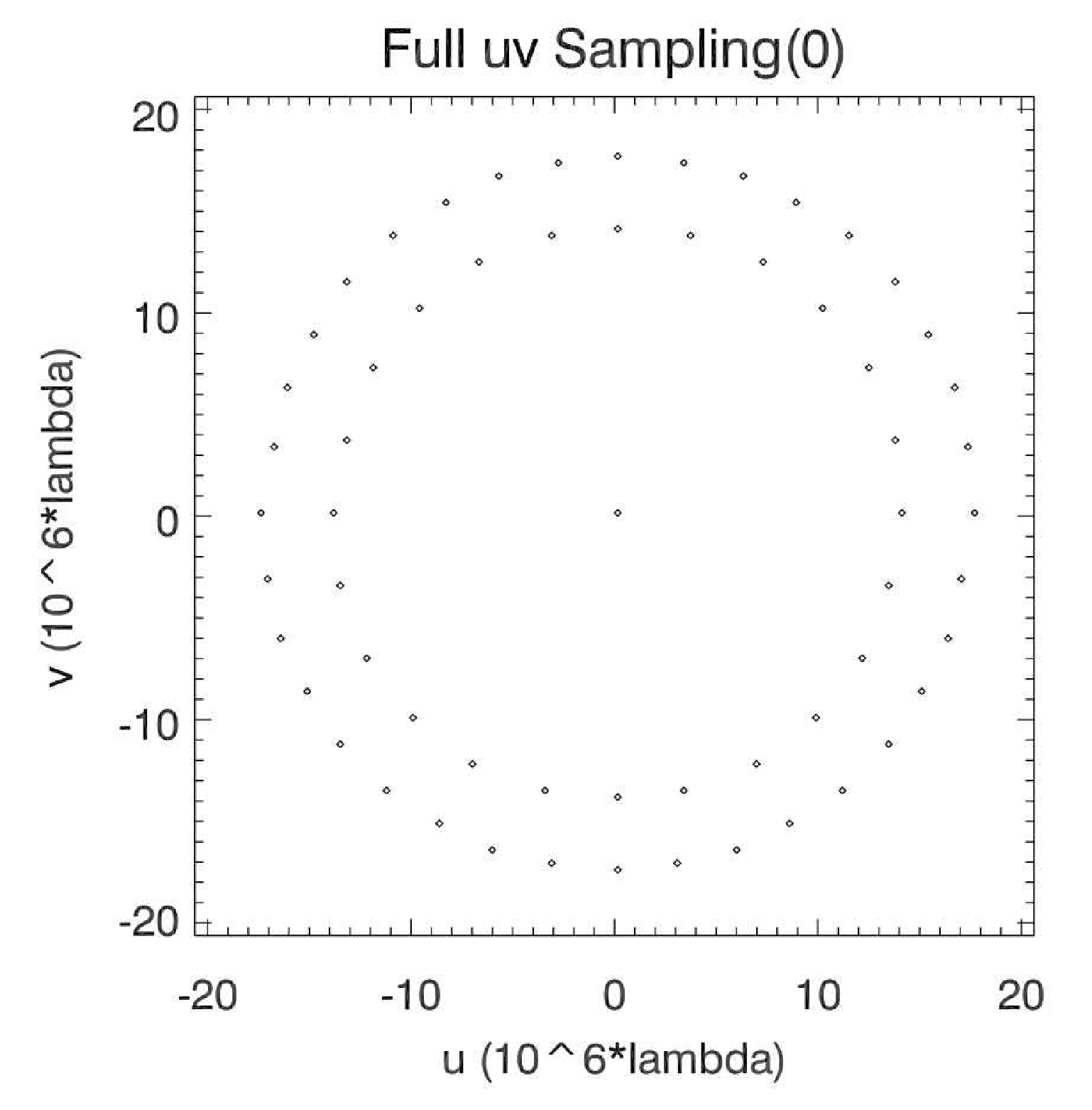}
\vskip 7mm
\epsscale{1.15}
\plotone{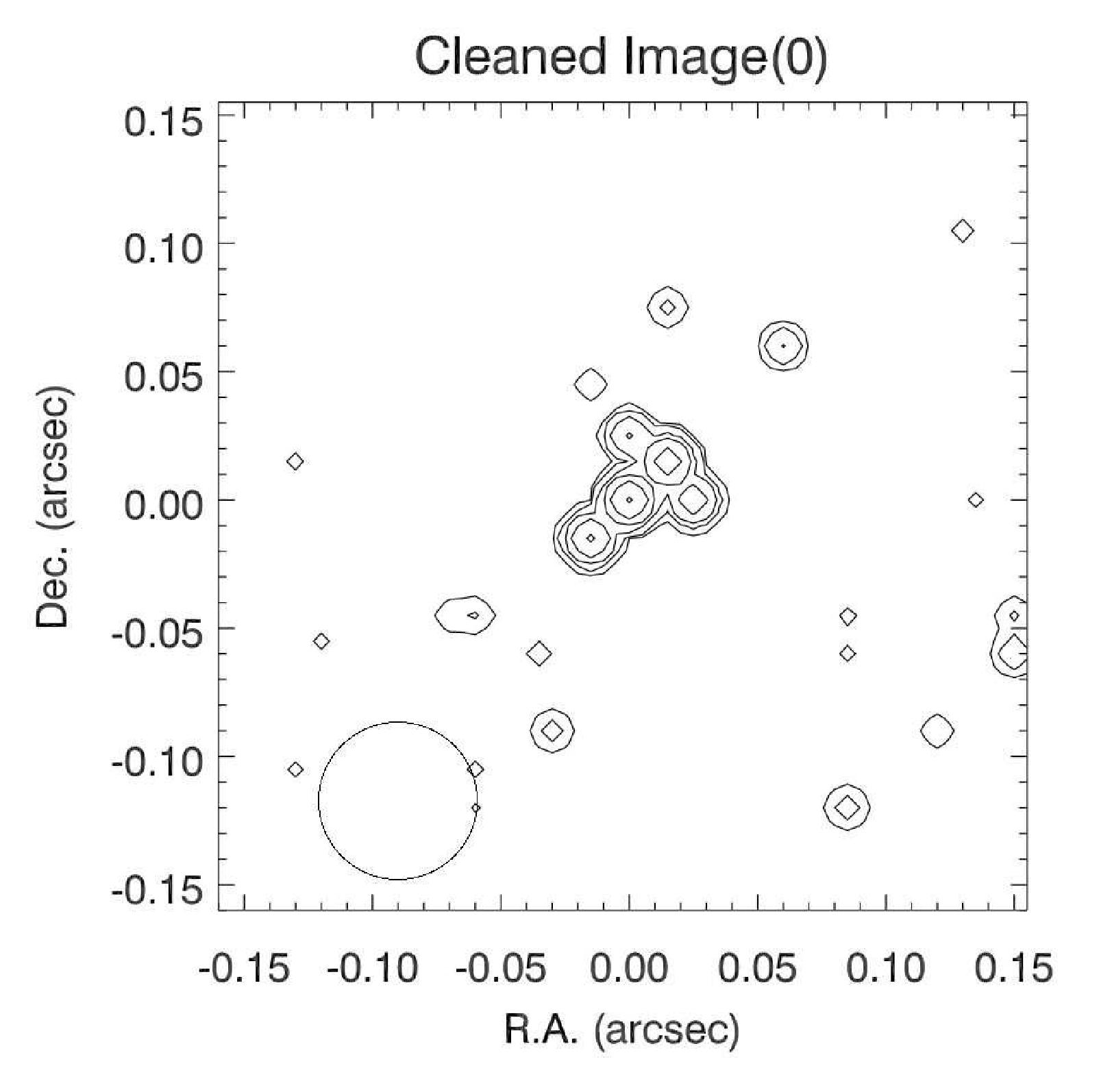}

\caption{Upper panel -- Example of the ($u,v$)-coverage for a single channel at mean
wavelength $\lambda = 500$ nm. Both the long ($\simeq 9.0$m) and short
($\simeq 7.2$m) baselines are included, and data taken at increments of
baseline orientations of $10^{\circ}$ and $15^{\circ}$, respectively.
Only half of the ($u,v$)-plane needs to be observed; the rest can be computed
using the fact that the field to be imaged is purely a real function. Note that SIM can provide its own zero-spacing data as well. 
The ($u,v$)-coverage for longer
wavelength channels would lie inside these circles, and the synthesized PSFs
would be proportionally wider. Lower panel -- A model cluster of 5 stars observed for ten seconds at each of 
the 30 ($u,v$) points shown in the left panel. The image has been CLEANed and restored with a gaussian beam of
FWHM 0.010 arcsec, and the residuals added in. Contours are logarithmic
at intervals of 1.25 mag. The brightest object (at 0,0) is
a 15th mag A0 star, the faintest member of the cluster (at 0, 0.025) is a 17th
mag G0 star. Other objects on this image are noise and artifacts of the restoration. The FWHM $\simeq 0.06$ arcsec of the HST/ACS/HRC camera PSF is drawn
in (the circle at the lower left) for comparison. 
\label{fig:imaging}
}
\end{figure}

Since SIM must have very high fringe phase stability over long periods of time
in order to accomplish its astrometric goals, it behaves at optical wavelengths
in much the same way as a ground-based radio interferometer. The results of
the observations can be put into the same general form of complex fringe
visibilities. Data can be obtained over a range of baseline orientations,
mimicking the rotation of the Earth for ground-based synthesis instruments. SIM
will also have two baselines available, one at 9.0-m and the other at 7.2-m, and
data for imaging observations can be taken at both baselines (although not
simultaneously). Figure \ref{fig:imaging}a shows a typical coverage of the
aperture ($u,v$) plane which can be achieved. A software simulator for this mode
of SIM is available, based on the initial work on this subject by \citet{ba99}.
The current version of the SIM imaging simulator, \textit{imSIM}, incorporates
the latest information on the expected performance parameters of SIM, and can generate simulated observations for a variety of source models. 

Figure~\ref{fig:imaging} shows the image which could be produced
by SIM on a (model) source field
consisting of 5 stars ranging in brightness from $V = 15 - 17$, and the ($u,v$) coverage of the simulated data. The stars were observed
for a total of 300 seconds of on-source integration time (30 points, 10 sec
per point). The point spread function of HST (convolved with the pixels of the HRC) is
shown for comparison.

For the imaging demonstration, targets of modest complexity but showing a wide
range of surface brightness will be chosen, consistent with the limited
number of physical baselines available in the SIM instrument. The observations
for this demonstration will be carried out soon after the launch of SIM during
the in-orbit checkout of the instrument. Subsequently, this capability will be
available for general use.

\subsection{Performance features of SIM as a synthesis imaging instrument}

In addition to its high resolution, SIM as an imager has several other novel
features. First, the instrument is a simple adding interferometer instead of
a correlation interferometer (common in radio astronomy), so that the total
flux of photons received in the FOV is retained in the measurement. If that
value is not significantly contaminated by photons from extraneous objects in
the FOV, it can provide an estimate of the ``zero-spacing" data, as indicated
by the dot at the origin in Figure \ref{fig:imaging}a. Second, and perhaps
most significant, images made synthetically with SIM will have exceptionally
high dynamic range. This is a result of the extraordinarily high phase
stability required of SIM in order to achieve its goal of \textit{repeatable}
microarcsecond astrometric precision. Since this stability means that the
theoretically-computed point-spread function (PSF) is an excellent approximation to the actual PSF, 
the former can be used to remove bright stars from the image, leaving only
photon noise and low-level artifacts from residual phase instability.
In marked contrast to the situation with filled apertures, these latter instabilities will be entirely uncorrelated from one measurement location
in the ($u,v$) plane to the next. The result is that the remaining low-level artifacts make equal contributions everywhere over the synthesized image
rather than piling up around the location of the bright stars. After PSF
subtraction, the remaining faint sources can be detected anywhere in the difference images with equal sensitivity, even as close as one interferometer
fringe ($\approx 11$ mas at $\lambda = 500$ nm) from a bright star.
Preliminary results from our simulations indicate that we can expect the
dynamic range (defined as the magnitude difference between
the peak value on the brightest star and the faintest detectable companion) on SIM images to routinely exceed 6 mag for target stars with V $\lesssim 13$, which is
comparable to that achieved at radio wavelengths with the VLA without any
additional data processing. However, in contrast to the case
with the VLA where values of dynamic range as high as 10 mag (factor $10^4$)
have been achieved by modeling the residual instrumental effects, the dynamic
range on SIM's synthetic images will ultimately be limited by the photon noise
from the stars in the FOV. With SIM's modest collecting area, it will be
difficult to significantly improve the dynamic range on SIM images much
further without large amounts of observing time.

\subsection{Synthetic imaging science with SIM}

Although intended primarily as a technology demonstration, the synthesis imaging
mode of SIM will also open new possibilities for science applications on targets
of high surface brightness, and where resolving structure beyond the limits of
the HST/ACS/HRC PSF would be significant to an understanding of their nature.
\citet{ba99} were the first to consider this question in some detail, and concluded
that targets as varied as the cores of dense Galactic globular
clusters, AGNs in galaxies, and dust disks around nearby stars were feasible.
Based partly on their analysis, a proposal to map the stellar distributions and
kinematics of the nuclear regions of M\,31 was included in the initial science
program for SIM \citep{ut2004}.  Since that time, instrument design changes have reduced the number of baselines available, and the nulling capability has been removed. These two de-scopes
of the SIM design have rendered the study of dust disks around nearby stars
significantly more difficult. However, the M\,31 project appears to be still
feasible, as are observations of compact high-surface-brightness targets in
general. The example of observations of a dense stellar cluster described in the previous section
makes it clear that SIM will have significant things to say about such targets.

The successful demonstration of synthesis imaging at optical wavelengths with
SIM is expected to have a significant impact on the design of future
space-based instruments for astronomical imaging. SIM imaging will also
offer new capabilities for the study of the structure of targets presently
only barely resolved by HST, or otherwise confused by the inability to
adequately remove diffraction spikes and scattered light from nearby bright
stars. The design of SIM is now quite stable, and many details of just how the
instrument will operate are now known. A new review of potential targets for
imaging science with SIM is under way.


\section{Measuring the PPN parameter $\gamma$\label{CHAPTER16}}

Our current lack of understanding of quantum gravity and dark energy demands new physics.  Searches for gravitational waves and attempts to understand the nature of dark matter are further motivated by the discovery of dark energy \citep{Turyshev-etal-2006}.  Promising theoretical models involve new gravitational effects that differ from General Relativity (GR), some of which  could manifest themselves as violations of the equivalence principle, variation of fundamental constants, modification of the inverse square law of gravity at short distances, Lorenz-symmetry breaking, and large-scale gravitational phenomena.  These effects are amenable to study with space-based experiments.  For example, scalar-tensor extensions of gravity \citep{Damour-Piazza-Veneziano-2002}, brane-world gravitational models \citep{Dvali-Gabadadze-Porrati-2000}, and modified gravity on large scales, motivate new searches for  deviations from a level of $10^{3-5}$ below the level currently tested by experiment \citep{Turyshev-etal-2004,Turyshev-etal-2006}. 

The parameterized-post-Newtonian (PPN) parameter $\gamma$ \citep{Will-2006} is currently measured to differ from unity by no more than  $(2.1 \pm 2.3)\times10^{-5}$, as obtained using radio-metric tracking data received from the Cassini spacecraft \citep{Bertotti-Iess-Tortora-2003} during a solar conjunction experiment.  This accuracy approaches the region where multiple tensor-scalar gravity models, consistent with the recent cosmological observations \citep{WMAP07}, predict a lower bound for the present value of this parameter at the level of $(1-\gamma) \sim 10^{-6}-10^{-7}$ \citep{Damour-Piazza-Veneziano-2002}, motivating the proposal of space-based experiments to improve the measurement of $\gamma$.

SIM will routinely operate at about this level of accuracy -- the Sun produces an astrometric bend of 4 mas at $90\arcdeg$, more than 1000 times the accuracy of individual SIM grid stars.  Gravitational effects in the Solar System  must therefore be included into the global astrometric model and the corresponding data analysis, formulated such that 
$\gamma$ is a measured parameter.  By performing differential astrometric measurements with an accuracy of a few $\muas$ over the instrument's FOV of 15$^\circ$ SIM will provide this precision as a by-product of its astrometric program and could measure the parameter $\gamma$ with accuracy of a few parts in $10^6$ \citep{turyshev2002}, a factor of 10 better than the Cassini result.  A precision SIM measurement would aid the search for cosmologically relevant scalar-tensor theories of gravity by looking for a remnant scalar field in today's Solar System.  

\begin{deluxetable*}{lccccc}[hb!]
\tablecaption{Strawman Key Project and Guest Observer Time Assignment
\label{obs-program}
} 
\tablewidth{170mm} 
\tablehead{
\colhead{Science Program}& \colhead{Number} & 
\colhead{Target {\it V}} & \colhead{Observing} & \colhead{Accuracy} &
\colhead{Mission}\\
\colhead{} & \colhead{of Targets} & \colhead{Magnitude} & Mode & 
($\muas$)\tablenotemark{a} & \colhead{Fraction}
}
\startdata

{\em Prime Mission} & & & & & 1.00\\
\\
Science Team Key Projects & $\sim10000$ & -1.4-20& Wide/Narrow & 4-25 & 0.36\\
Guest Observer Call 1  & $\sim75$ & 6-9 & Narrow & 0.6-1.2 & 
0.27\tablenotemark{b} \\
\ \ \ (Terrestrial planets)\\
Guest Observer Call 2 & $\sim10000$ & 12 & Wide & 8 & 0.05 \\
\ \ \ (Open)\\
Astrometric grid\tablenotemark{c} & 1302 & 9-10.5 & Wide & 3 & 0.24\\
Engineering\tablenotemark{d} & & & & & 0.08\\
\\

{\em Extended Mission} & & & & & 1.00\\
\\
Guest Observer Call 3  &  & & & & 0.7\\
\ \ \ (Sample programs)\\
\ \ \ \ GO Program 1 & $\sim180$ & 6-9 & Narrow & 1.5 & 0.2\tablenotemark{b} \\
\ \ \ \ GO Program 2 & $\sim5000$ & 12 & Wide & 5 & 0.1 \\
\ \ \ \ GO Program 3 & $\sim20000$ & 12 & Wide & 8 & 0.1 \\
\ \ \ \ GO Program 4 & $\sim12000$ & 14 & Wide & 10 & 0.1 \\
\ \ \ \ GO Program 5 & $\sim750$ & 18 & Wide & 12 & 0.1 \\
\ \ \ \ GO Program 6 & $\sim1500$ & 18 & Wide & 20 & 0.1 \\
Astrometric grid & 1302 & 9-10.5 & Wide & 2.5\tablenotemark{e} & 0.2\\
Engineering\tablenotemark{d} & & & & & 0.1\\

\enddata

\tablecomments{The SIM instrument and operations are designed for a 10-year total lifetime. Prime mission has a duration of 5 years; extended mission, a further 5 years.  Science Team Key Project time was assigned by NASA Announcement of Opportunity in 2000.  All Guest Observer (GO) Programs will be competed.  The study of terrestrial planets described in \S\,\ref{CHAPTER2} assumes that the first GO Call will be devoted to extending the target list for planet searches.  GO Call 3 is represented here by programs which span a range of magnitudes and accuracies, and which are purely intended to be illustrative. 
}
\tablenotetext{a}{In wide-angle mode, accuracy is mid-epoch position, one axis, at end of prime mission.  In narrow-angle mode, accuracy is for a single measurement in a local reference frame, one axis}
\tablenotetext{b}{Mission fraction is for 100 visits per target in each of two orthogonal axes} 
\tablenotetext{c}{Grid allocation also includes $\sim50$ quasars ($\sim1.5$\% mission fraction) to define the absolute reference frame (see Appendix \ref{APPC})}
\tablenotetext{d}{Includes instrument calibration and scheduling margin}
\tablenotetext{e}{Estimated accuracy of mid-epoch position, one axis, at end of extended mission
\vskip 2mm}

\end{deluxetable*}


\section{Conclusions \label{CHAPTER17}}

The Space Interferometry Mission (SIM) represents a 
revolutionary step forward in the development of astronomy's 
most ancient  measurement technique, determining  the positions 
of celestial objects. With an absolute precision of  $3\, \muas$ (microarcseconds) 
covering  a wide range of magnitudes (from $V<0$ mag to $V>19$ mag), 
SIM will measure positions, distances, and proper motions for almost 
every type of astronomical object with a parallax accuracy of 
10\% or  better across the entire Milky Way. Within a few parsecs,
the parallax precision will be better than 1\%. These precision 
measurements  will challenge our theories of stellar ($\S$~\ref{CHAPTER5},~\ref{CHAPTER6}) 
and galactic structure  ($\S$~\ref{CHAPTER7}), while  adding  to our understanding 
of dark matter and cosmology through refined knowledge of the distance
 to Cepheids ($\S$~\ref{CHAPTER8}), the value of the  Hubble Constant ($\S$~\ref{CHAPTER14}), 
the ages and distances of Globular Clusters ($\S$~\ref{CHAPTER9}), and the 
motions of  dwarf spheroidal and Local Group galaxies ($\S$~\ref{CHAPTER12}). 
Precision SIM astrometry even promises to probe small scale 
phenomena within the cores of quasars and active 
galactic nuclei, including  the possible presence of binary 
supermassive black holes ($\S$~\ref{CHAPTER13}).

With a differential astrometric  precision of 1 microarcsecond, 
SIM will measure the orbital motions of many classes of multiple objects, 
including every  sort of  normal and exotic star from low-mass 
stars ($\S$~\ref{CHAPTER5}) to X-ray binaries, evolved AGB stars, neutron star 
and  black hole binaries ($\S$~\ref{CHAPTER6}), 
and microlensing systems 
($\S$~\ref{CHAPTER10}). From these dynamical measurements, we will be able to 
determine masses for astronomical objects, perhaps the most 
fundamental single parameter in  understanding their physical 
nature. Coupling new physical  information from SIM with previously 
known characteristics of such systems, such  as ages and metallicity, 
will allow a new level of understanding of the evolution of 
astronomical objects.  

From a cultural standpoint and in the eyes of the public, as well as in the research community, SIM's 
most dramatic contribution  may come from opening up the next 
level of performance  in the reconnaissance of our nearest 
stellar neighbors by looking for planets with masses that are equal 
to or just a few times above the mass of our own Earth and located in 
the range of orbits conducive to development of habitable 
environments ($\S$~\ref{CHAPTER2}). As described in \S\,\ref{planet-yield}, 
within a deep 5 
year survey,  SIM could find planets with a  mass of 
$1\,\MEarth$ orbiting within the habitable zones of over 100 of 
the most favorable  stars.  SIM is the first step in a long 
term strategy for searching for other habitable worlds suitable 
for subsequent follow-up with direct detections of planet-light at 
visible or infrared wavelengths. SIM also offers dramatic new 
capabilities for finding planets orbiting stars not accessible  
to radial velocity or transit studies, such as young stars very 
massive stars, or stars with highly variable photospheres or 
weak spectral lines.  SIM will investigate the formation and 
migration of planets in  young planetary  systems, finding out 
whether Jupiter-mass planets are common or rare when orbiting stars 
younger than a few million years old ($\S$~\ref{CHAPTER3}).

As we noted in the \S\,\ref{CHAPTER1}, this paper is intended to highlight some of the many areas of astrophysics that precision astrometry will address, specifically those problems amenable to flexibly targeted observations at very high precision.  As a guide to possible opportunities, we show an overview of the assignment of SIM observing time in Table~\ref{obs-program}.  Key Project teams have the task of optimizing their science within their allocations, which requires a careful trade of the numbers of targets, their magnitudes, and the astrometric accuracy needed for the science objective.  We note that for faint targets, say $V>18$, accuracies of $< 10\muas$ require hours of mission time, and such targets will be selected with substantial care; many of the Sections in this paper discuss the considerations in target selection.   As this process of target definition for the Science Team is still underway, we show only the formal allocation to the Team.  The remaining time will be allocated via a future peer-reviewed Guest Observer (GO) program, and will be completely open with respect to science topic.  The number of selected programs will be set by NASA; it is expected that extended mission would be entirely open to competition.  We show a set of hypothetical, but representative, programs that together would complete the SIM time assignment. The first GO program will likely be assigned to further searches for terrestrial planets to supplement the NASA Key Projects (see \S\,\ref{CHAPTER2}).
The number of targets, magnitudes, and accuracies in Table~\ref{obs-program} should be viewed as illustrative, rather than the final word on target selection.  Observing time calculations depend on the science objective (e.g., parallax and proper motion require different optimizations). 

The technology pioneered by the SIM mission, the first long baseline 
Michelson Interferometer in space, represents an important 
investment for the future of space  astronomy. Visible- and 
infrared- imaging  on the milliarcsecond scale demands widely 
separated  apertures. SIM's long-baseline 
interferometric  capabilities ($\S$~\ref{CHAPTER15}) will make simple 
images with 10 milliarcsec resolution on $V=15$ objects. 
Future telescopes, potentially separated by hundreds of meters, 
will build upon the technology demonstrated by SIM.

The scientific measurements that SIM will provide cannot be 
duplicated by other means. As a pointed observatory, SIM will 
complement the  all-sky survey planned for the Gaia mission by 
enabling order-of-magnitude more precise observations of both very 
faint and very bright objects. The combination of SIM 
and Gaia will move astronomy and astrophysics into a new era of 
precision dynamical and kinematic knowledge unprecedented in the 2,000 year-old  history of astrometric measurements.

\section*{Acknowledgments}

The authors would like to thank their many colleagues, too numerous to mention, whose vision and determination over the course of more than a decade have brought SIM PlanetQuest to its current mature design.  Through their efforts, we are now poised, technology in hand, to build the next generation of space astrophysics instruments based on interferometry.
The research described in this paper was carried out at the Jet Propulsion
Laboratory, California Institute of Technology, under contract with the
National Aeronautics and Space Administration.  This research has made use of the NASA/IPAC Infrared Science Archive, which is operated by the Jet Propulsion Laboratory, California Institute of Technology, under contract with the National Aeronautics and Space Administration.  This work was supported by NSF grant AST-0307851 (RJP, SRM), NASA grant JPL 1228235 (SRM, CG, KVJ, RJP, ST), and NASA grant NAG5-9064 and NSF CAREER award AST-0133617 (KVJ).


\begin{appendix}

\section{Narrow Angle Astrometry \label{APPA}}

SIM is designed to perform narrow angle (relative) astrometry
in three different modes where it is capable of measuring periodic motions with a precision of 0.6 $\mu$as per axis in a narrow-angle frame.  The modes include the standard Narrow Angle (NA) mode, 
Gridless Narrow Angle Astrometry (GNAA) mode 
\citep{Shaklan03}, and Grid-based Differential Astrometry (GBDA) mode 
\citep{pan05}.
Because of its high efficiency, NA mode will be used for the vast majority of 
narrow angle observations.  Early mission results can be obtained with GNAA, while 
GBDA is useful once grid stars are known to the level of 0.1 arcsec.

NA mode is linked to the wide-angle grid campaign. It uses the end-of-mission 
grid star positions to determine the instrument baseline orientation and length at 
the time that a target and reference stars were observed.  Two roughly orthogonal baseline orientations are needed to make a 
2-dimensional measurement of the target star motion relative to the reference frame. 
Because all measurements are tied to the absolute grid, the target parallax and 
proper motion are determined with high precision and their errors do not contribute significantly to the single-measurement error.   A single visit typically requires $<$ 
30 min to achieve 0.6 $\mu$as precision.

GNAA is akin to traditional relative astrometry with a single telescope. Relative 
astrometry typically employs a photographic plate, a CCD, 
or Ronchi Ruling 
\citep{gatewood87}  to record the positions of a target star and a 
reference stars over several months to years. The reference stars are used 
to anchor a least-squares conformal transformation that matches the scale, rotation, 
translation, and potentially higher order field-dependent and time-dependent terms 
into a common reference frame, which is then applied to the target.  GNAA mode combines measurements of a target, at least four reference stars, and at least three baseline orientations \citep{Shaklan03}, and uses a conformal model to solve for instrument parameters (baseline length and orientation, and a phase-delay constant term).  Since the parallaxes and proper motions of the target and reference stars are not known to microarcsecond precision, this approach is not useful for long-period ($>$0.5 yr) signals.  

GBDA is a hybrid mode that uses coarse (early mission) grid star positions 
to solve for baseline parameters.  GBDA requires the measurement of several grid 
stars using two nearly-orthogonal baseline orientations in addition to the target 
and reference star measurements.  It is more efficient than GNAA (two vs. three 
baselines) but is still limited to the detection of short periods. In GBDA,  
grid catalog errors combine with stochastic baseline repointing errors to cause 
baseline orientation, length, and constant-term errors. These in turn mix with 
reference frame catalog errors resulting in random noise on the target star 
estimate. To keep the errors between the target and reference stars below a 
fraction of a microarcsecond, the baseline re-orientation requirements for 
subsequent visits are 0.1 degrees (around the line of sight) 
and 0.01 degrees (tilt toward the target).  These are easily met using star trackers.

\begin{figure}[ht!]
\epsscale{0.52}
\plotone{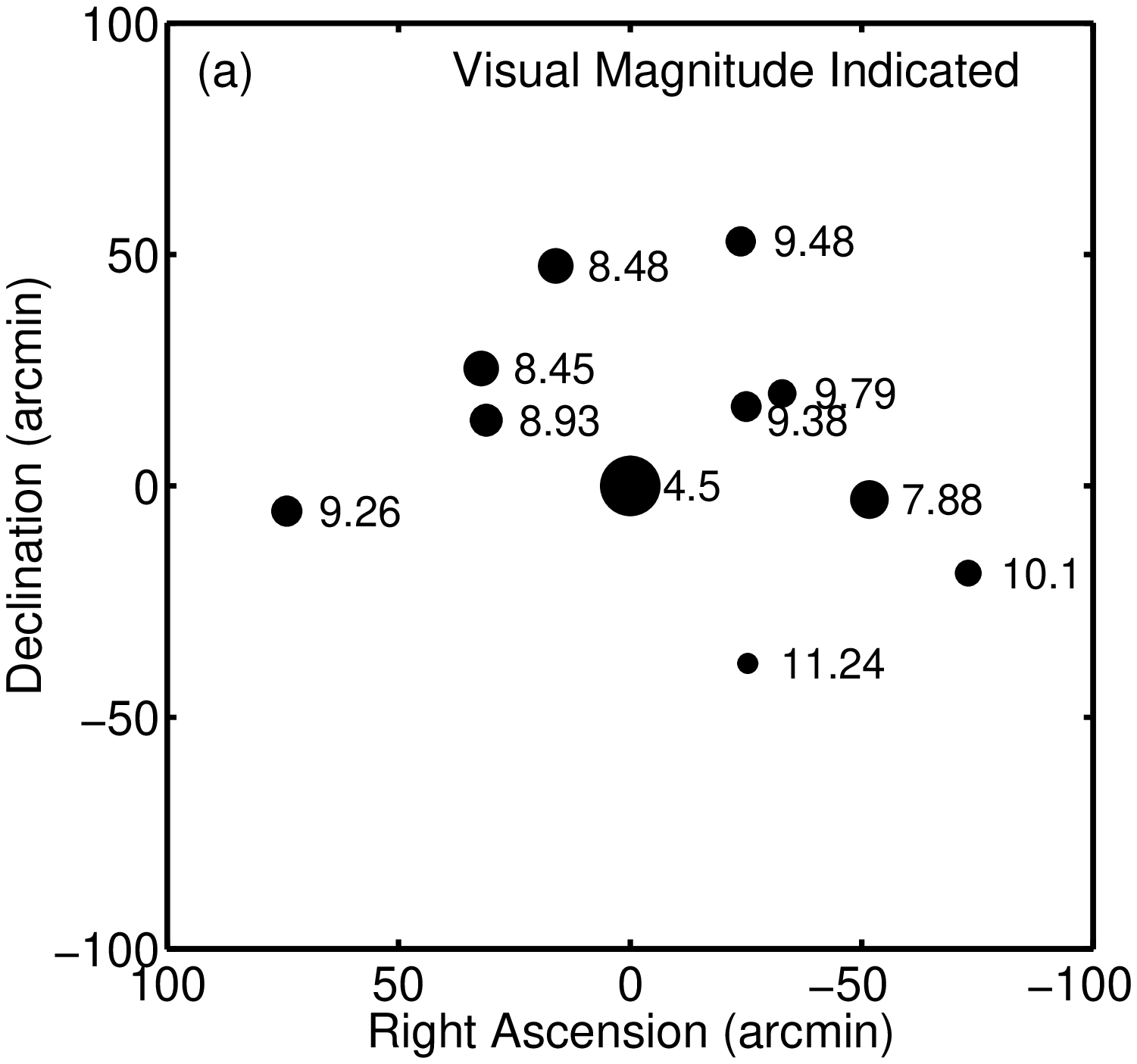}
\epsscale{0.6}
\plotone{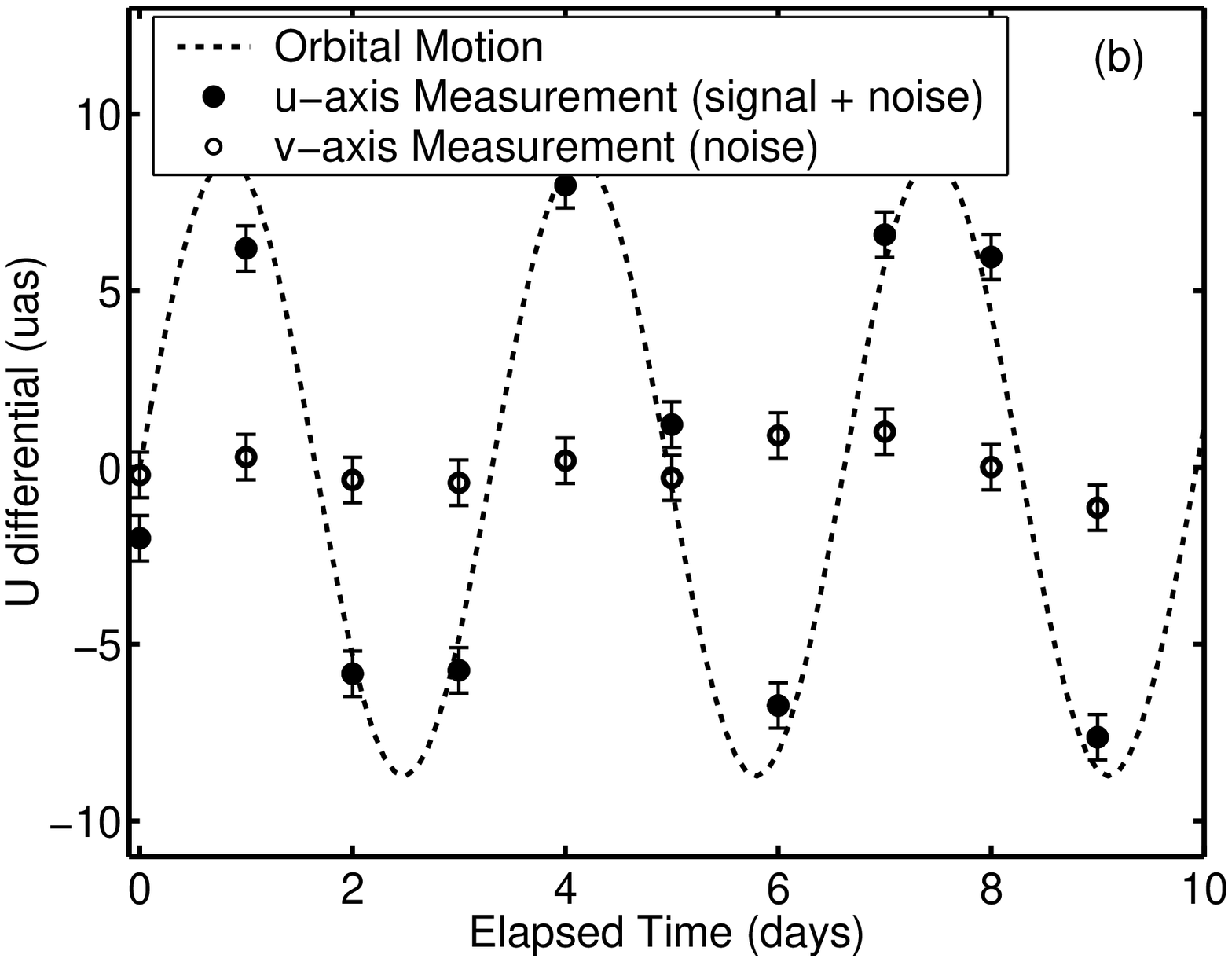}

\caption{Simulation of Grid-based Differential Astrometry (GBDA) with SIM of the 
$3.9 \MJup$ planet around Tau Boo.  (a) 10 brightest reference stars within 75 arcmin of Tau Boo.  
(b) Simulated measurement of the $9\,\muas$ amplitude signal (assumed to be along one axis). Residual proper motion and parallax are removed from the plot but were included as error sources in the astrometric model. 
\label{Appafig}
}
\end{figure}

To demonstrate the GBDA approach, we modeled the detection of the known planet orbiting Tau Boo \citep{butler97}.  The planet has a measured period of 3.3 days and (minimum) mass $3.9 \MJup$, from which we derive the astrometric amplitude is 9.0 $\mu$as.  In the model we assumed that the stellar reflex motion was on one axis in order to demonstrate that the magnitude of the astrometric errors is independent of the signal amplitude. Tau Boo's proper motion was assumed to be known to 1 mas\,yr$^{-1}$.
We modeled 20 GBDA measurements over a 10 day period.   We used realistic catalog errors for the reference and target star positions and proper motions. 
The NA reference frame consisted of the 10 brightest reference stars within a 75 
arcmin radius of Tau Boo (Fig.~\ref{Appafig}a).  We modeled instrument systematic errors based on the flight instrument design, and added signal-dependent noise added in quadrature to the target and reference star observations.  For our adopted observing scheme, this yields an astrometric error of 0.65 $\mu$as per axis per visit.  
Figure~\ref{Appafig}b shows the simulated signal and measurements, and that the  Tau Boo planet is easily detected by the GBDA technique.

\section{All-Sky Astrometric Grid\label{APPB}}

SIM  observes stars sequentially within a $15 \arcdeg$ patch of 
sky termed a `tile', while the instrument is held inertially stable by observing 
a pair of guide stars with guide interferometers.  One guide star lies close in angle to the target star; the other is roughly $90\arcdeg$ away. Details of the operation of SIM, and how  the guide stars are used to derive microarcsecond precisions on science 
targets, is discussed in, e.g., \citet{Laskin2006}.  Here
we describe the development of the astrometric grid of stars to which the 
science measurements are referenced, the astrophysical selection of those 
stars, and the `frame tie' of the grid to an inertial frame defined by distant quasars.

SIM will construct an all-sky astrometric grid including 1302 pre-selected stars and a smaller number of quasars.  The grid accuracy is expected to be about 3 $\muas$ in mean-epoch position, based on the current best estimate of instrument performance. Each grid object will be observed about 200 times during the nominal 5-year mission. The limited number of grid objects and the
moderate density of the grid is explained by its special role that is more utilitarian than research-driven. SIM will largely rely on self-calibration of the instrument and on the determination of the baseline orientation from its own interferometric
measurements of the grid objects interspersed with routine measurements of science targets. A number of key instrument parameters, for example, the baseline length, can not be determined to the required picometer-level
accuracy by external metrology techniques and, therefore, should be derived from observations of stars.  

Since the number of instrument and attitude parameters is large, 
it is necessary to observe a global grid of stars multiple times during the 
mission to construct an overdetermined system of linear equations.
The grid solution is a one-step direct Least Squares adjustment of a system of
$\sim\,300,000$  linear equations of $\sim\,160,000$ unknowns. Of the latter 
set, only some 6,500 unknown parameters are related to the grid objects (i.e., the mean positions, parallaxes and proper motions that
constitute the actual astrometric grid). An efficient and fast algorithm has been developed to solve the grid equations eliminating of the numerous attitude terms by QR factorization, described in \citet{mami}.

The astrometric accuracy of the grid can be directly evaluated from the global 
covariance, as far as random errors of measured positions are concerned. Systematic errors are not always possible to predict or to model, and are less amenable to straightforward covariance analysis. However, the manageable size of the reduced
design matrix allows us to employ rigorous mathematical analysis of {\it 
possible} systematic errors and compute the complete space of those perturbations that have a strong adverse  effect on the grid accuracy. As explained in more detail in \citet{mami}, measurement errors (accidental and systematic alike) propagate 
non-uniformly in different singular vectors
spanning the parameter space, which is likely to result in large-scale (so-called `zonal') correlated spatial errors in the grid.  They may be represented by spherical orthogonal functions for the parallax error distribution, and by vector spherical harmonics for the proper motion error distribution, which is a vector field on the
unit sphere \citep{koma}. The spatial power spectrum of grid
errors is `red', meaning the signal is larger at low frequencies.   

We explored the grid performance for a wide range of possible instrument systematic errors developed as a by-product of the detailed flight instrument design.  This allows us not only to evaluate the expected RMS grid performance and the properties of the zonal errors, but also to estimate confidence levels, for instance, how likely the actual SIM grid will be better than a given value.

An effective way to mitigate these zonal errors is to use a number of quasars as grid objects; quasar parallaxes are vanishingly small and can be constrained to zero in the global solution.  Numerical simulations and covariance analysis have shown that with only 25 additional optically bright quasars in the grid program, the grid parallax astrometric accuracy improves by $\sim$~28 \%, meeting the mission goal (4 $\uas$) with considerable margin.  Using the current best estimate for the instrument performance, the grid should achieve 3 $\muas$. 
Quasars also dramatically improve the grid confidence:
the 99\% confidence limit on grid parallax error drops from $6.2$ $\uas$ to $3.1$ $\uas$ (with the same simulated mission but with quasars included in the grid).

Analytical considerations showed that the astrometric accuracy on bright, frequently observed science program stars should be equal to or slightly surpassing that of grid stars.   Spreading observation times evenly across the mission time is not the best
strategy for parallax. Significantly better parallax can be obtained
for a given star by simply scheduling the chosen number of observation at the most favorable times, determined by a semi-empirical optimization algorithm implemented in the wide angle processing code. In a similar way, the observation schedule can be optimized for any object on proper motion performance, or a combination of all three types of astrometric parameters.   A realistic SIM schedule takes into account the desired science objectives of the measurements, and optimizes across the ensemble of all science targets both the integration time invested and the observation epochs.  While this ideal is likely impossible to realize in practice, observation planning will definitely play a major role in extracting the best astrometric performance from the instrument.

\begin{figure}[th]

\vskip -10mm
\epsscale{0.7}
\plotone{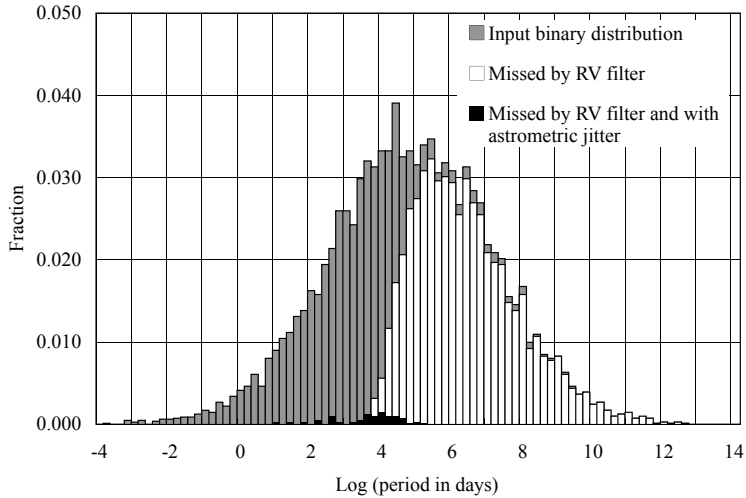}
\caption{Histogram of the simulated period distribution of stellar companions to candidate SIM grid stars, showing the effectiveness of radial velocity (RV) screening in eliminating binary stars with significant astrometric signatures from the grid. The binary mass and orbit distributions are assumed to be those of \citet{DM1991}.  Gray symbols represent the input period distribution.  RV filters virtually all binaries with periods less than $104$~days, leaving only the long-period companions (white symbols).  Very few of these remaining (i.e., missed) long-period companions produce astrometric motion detectable by SIM (jitter $> 4\, \muas$ RMS, shown in black).
\label{RV-screen}}
\end{figure}

\section{Astrometric Reference Frame\label{APPC}}

SIM will define a reference frame accurate to $3\, \muas$ using  50-100 quasars to `tie' the stellar grid to a presumed inertial frame. 
The SIM frame will complement the Gaia catalog, as the SIM frame will have much more precise positions than Gaia's, albeit for a smaller number of sources.  These distant  sources are assumed to have negligible parallaxes and proper motions, and as noted above, quasar observations will also reduce the amplitude of zonal errors in the astrometric grid.
Because of the physical effects discussed in \S\,\ref{CHAPTER13}, radio-quiet quasars are to be preferred, as there are less likely to be asymmetries, or worse, time-dependent position shifts, in the optical photocenters relative to the massive black holes that serve as inertial reference points (\S\,\ref{astrometric-shift}).  Most SIM science programs are not affected by frame rotation.   For instance, parallaxes are quite insensitive to a rotation rate.  Some topics require the frame to be inertial to $\sim 2\, \muasyr$.   Examples include galactic structure, including tidal tails 
(\S\,\ref{CHAPTER9}) and extragalactic rotational parallaxes ($\S$~\ref{CHAPTER13}).   This requirement should be easily met using 50-100 radio-quiet quasars, selected from bright quasar catalogs.  Note that a tie to the {\em radio} (ICRF) frame plays no direct role in these investigations.

Tying the SIM inertial frame (defined by radio-quiet quasars and grid stars) to the  ICRF serves two fundamental purposes.   First, future astrometric catalogs will require a registration with the ICRF since that frame is the basis for all current astrometric data.
Second, accurate registration between reference frames is important for science that combines astrometric data in two frames, for instance SIM and VLBI data.  This requires a tie to the ICRF, and hence SIM will include (radio-loud) quasars in the astrometric grid.  The study of non-thermal radio emission associated with stars will lead to a better understanding of the mechanisms giving rise to an individual source's spectral energy distribution (see \S\,\ref{masers}).   We showed in \S\,\ref{CHAPTER13} that astrometry is a powerful tool for probing the structure of blazars. A tie to radio (VLBI) images is important for a full understanding of the physical processes in quasars that manifest themselves as astrometric motion.

The tie must of course be done using radio loud quasars, since the ICRF is fundamentally a radio reference frame.  The accuracy of the radio-optical frame tie  will therefore depend on the quasars in common between the two frames.  Since the SIM reference frame will be much more accurate than the ICRF, errors in tying the frames are likely to be dominated by the VLBI observations that define the ICRF.
Quasars will be chosen from the  212 ICRF sources; there is approximately one ICRF quasar per $15 \arcdeg$ diameter SIM tile, with fewer in the southern hemisphere and in the plane of the Milky Way Galaxy.  By establishing an accurate link between the optical SIM frame and the radio ICRF, high resolution imaging data at these disparate
wavelengths can be accurately aligned to allow absolute positional correlation.

Quasars with radio emission may not make ideal astrometric reference targets, due to variability.  It is not yet clear how the known structure changes seen at milliarcsecond levels (from VLBI imaging) in the most variable radio-loud objects 
translate into astrometric shifts in the optical.  Section \ref{CHAPTER13} discusses the physical models that allow us to make estimates of the effects, prior to SIM launch.
Although there is a lot of experience with the ICRF defined by radio observations, the physics is not understood well enough to make a simple prediction.  Indeed, we fully expect that SIM will provide key insights into this problem.

The quality of the radio-optical frame tie can be improved in a number of ways.   First, the quasars in common should be selected to avoid the most highly variable targets.  Second, repeated astrometric measurements taken during the SIM 5-year mission can be checked for consistency and obvious outliers rejected.   Third, the number of quasars in common should be large to allow us to average out the offsets, which will be uncorrelated from target to target.  Since the investment of SIM observing time may be traded between number of quasars and accuracy of the astrometric measurements, intrinsic astrometric variability would lead one to prefer to increase the number of quasars in common; this would also make the identification of outliers easier.   Those objects which show astrometric signatures  as large as perhaps $15-20\, \uas$ will be easily eliminated from the reference frame program but will make for very interesting astrophysical studies.  Finally, we note that detailed studies of the physical processes in quasars, as probed by the astrometric program described in  \S\,\ref{CHAPTER13}, can provide insights into the most effective selection of quasars to use in the end-of-mission solution for the reference frame, allowing some quasars to be selectively omitted from the solution.


\section{Selection of SIM Grid Stars\label{APPD}}

The SIM grid comprises 1302 K giant stars spread quasi-uniformly over
the sky. K giant stars were chosen for several reasons. K giant stars are
numerous and located at all galactic latitudes. K giant stars are intrinsically
much brighter than dwarf stars.  Compared to similarly numerous F and G dwarfs (the
most common dwarfs in a $V \approx 9-12$ magnitude limited sample), K giants
are intrinsically brighter by 4 to 5 magnitudes at $V$. Thus, for a given
brightness, K giant stars are $5-10$ times more distant than F and G dwarfs.
Astrometric motions induced by unseen stellar and planetary companions are of course
minimized by the increased distance \citep[e.g.,][]{gould01,frink01,plz02}.

Candidate K giant stars were selected from the Tycho-2 and 2MASS catalogs based
upon a compiled catalog of photometric colors ($BVJHK_s$), proper motions, and
distances (if known) of the stars.  A near-infrared reduced proper motion
diagram was constructed to separate candidate K giant stars from dwarf stars of
similar color \citep{gm03}.  The visual extinction for each candidate K giant
star was estimated to produce extinction corrected magnitudes and colors (i.e.,
spectral types) and distance estimates \citep{ciardi04}.  The Tycho-2 catalog
giant star extraction was supplemented with a ground-based survey, sparsely covering the whole sky with a roughly uniform pattern of `bricks', each covering 
$\approx 0.5$ sq-deg per brick \citep{patterson01}. The survey
utilized the Washington $M$, $T_2$, and DDO51 filter system to identify
candidate low metallicity K giant stars \citep{Majewski2000}. Together
the two methods provided a total of $\sim170,000$ K giant stars from which
the grid star candidates could be selected.  For each `brick', six candidate grid
stars were selected for a total of 7812 candidate grid stars.  A minimum
distance of 500 pc was required for candidate selection.  The candidate grid
stars have a median visual magnitude of $V=9.9$ mag, with 90\% of the stars
between $V=9.0 - 10.5$ mag.  The median distance of the grid stars is $d=600$
pc, with 90\% of the stars between $d = 500 - 1000$ pc.

\begin{figure}[th!]

\vskip -8mm
\begin{center}
\includegraphics[scale=0.4,angle=90]{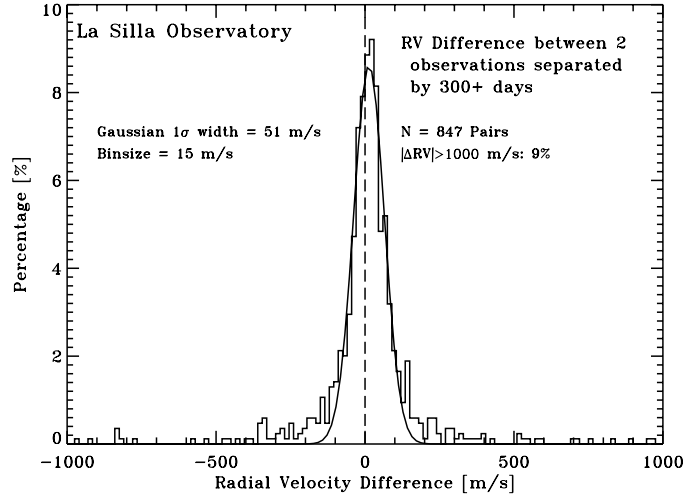}
\end{center}
\vskip -10mm

\caption{Frequency distribution of the difference between consecutive
radial velocity measurements separated by more than the 300 days.  Data taken at La Silla Observatory as part of the SIM grid star RV monitoring program.
\label{LAS-fig}}
\end{figure}


Detailed simulations of the effectiveness of radial velocity (RV) `screening' of grid star candidates has been performed by \citet{Cat2004}, using reasonable assumptions about the population of binary companions that induce astrometric perturbations in the primary star.  The results show that without RV screening, the grid is likely to be significantly contaminated, but that screening to a fairly modest precision of $50\,\mse$, with 3 observations spread over at least 3 years, are sufficient to reduce the contamination to a very low level (Fig.~\ref{RV-screen}).  RV screening mostly fails to reject brown dwarfs and giant planets, but most of the companions remaining produce small astrometric signatures, and have a negligible effect on the grid as a whole.
A pilot study of the RV stability of K giant stars by
\citet{frink01} showed that early K giant stars ($B-V \approx 0.8 - 1.1$ mag)
were intrinsically stable at a level of $20-30\,\mse$.  Analytic studies and Monte
Carlo simulations \citep{gould01, frink01, plz02} indicated that a modest
RV screening program (30-50 \mse) would reduce the fraction of grid
stars with unmodeled astrometric motions detectable by SIM to less than
$1-10\%$. 
In September 2004, a high precision ($<50$ \mse) RV
monitoring campaign was begun to remove those stars from the candidate list
with unseen stellar companions. Observations are currently being performed at
the 1.2m Euler Telescope at La Silla.
The RV program spans four years and is structured such that half
of the candidates (3 per brick) are observed twice in years 1 and 2 with a
minimum separation between observations of 9 months. Those stars which display
large RV excursions ($\chi^2_\nu > 4$) are removed from the
candidate list.  The removed stars are replaced and are observed during years 3
and 4, along with the surviving candidates from years 1 and 2.  The program has
finished its second year.

The effective sensitivity of the radial velocity program to detect
companions is a convolution of the stellar atmospheric jitter and the
instrumental jitter.  Thus, the effective precision of the RV
measurements is given by
$\sigma_{total} = \sqrt{(\sigma_{inst}^2 + \sigma_\star^2)}$
where $\sigma_{inst}$ is the RV precision of the instrument/data
reduction, and $\sigma_\star$ is the intrinsic stellar atmospheric jitter.   In
Figure \ref{LAS-fig}, we present the frequency distribution of the difference
between two consecutive RV measurements separated by 300$+$ days
for 847 candidate grid stars. The precision for each RV
measurement is approximately $\sigma_{inst}=20-30$ \mse.  The intrinsic stellar
jitter of early K giants, as found by \citet{frink01}, is $\sigma_\star \approx
25$ \mse, yielding an expected distribution width (in the absence of
companions) of $\sigma_{total} \approx 30-35$ \mse. The central portion of the
distribution is well modeled with a gaussian of width $\sigma_{total} = 51$ \mse, slightly larger than the expected width, indicating the presence (not
unexpectedly) of binaries in the sample or a slightly higher atmospheric jitter
(40 vs. 25 \mse). Additionally, the high velocity wings ($|\Delta {\rm RV}|
> 100$ \mse) are likely the result of the presence of unseen companions to the
grid stars. Using a reduced chi-square of $\chi^2_\nu < 4$ as the cut-off for
acceptance, approximately 60\% of the stars in this sub-sample would pass
through for further radial velocity vetting in years 3 and 4 of the program.

\end{appendix}


{}

\end{document}